\RequirePackage{lineno}
\documentclass[12pt, letter
  , twoside
]{article} 
\usepackage{geometry} 
\usepackage{fancyhdr} 
\usepackage[compact,noindentafter]{titlesec}
\usepackage{natbib}

\bibpunct{(}{)}{;}{author-year}{}{,} 
\usepackage{amsmath
, amsthm
, amssymb
}
\usepackage{MnSymbol} 
\usepackage{bm}
\usepackage{booktabs} 
\usepackage{threeparttable} 
\usepackage{array} 
\usepackage{paralist} 
\usepackage{verbatim} 
\usepackage{subfigure} 
\usepackage{topcapt} 
\usepackage{graphicx}
\usepackage[font={small},labelsep=quad,labelfont={small,bf}]{caption}
\usepackage{rotating}
\usepackage{float}
\usepackage{setspace}
\usepackage{multirow}
\usepackage{multicol}
\usepackage{umoline}
\usepackage[lowtilde]{url}
\usepackage{color}
	
\definecolor{mygreen}{rgb}{0, 0.5, 0}	
\definecolor{myblue}{rgb}{0, 0, 0.75}		
\definecolor{myred}{rgb}{0.75, 0, 0}		
\definecolor{myrev1}{rgb}{0, 0, 0}
\definecolor{myrev2}{rgb}{0, 0, 0}
\definecolor{todocol}{rgb}{1, 0.4, 0}		
\definecolor{temphidden}{rgb}{0, 0, 0}		
\definecolor{mycorr1}{rgb}{0, 0, 1}		

\newcommand{\rev}[1] {\textcolor{myrev1}{ #1}}
\newcommand{\revv}[1] {\textcolor{myrev2}{ #1}}

\newfloat{file}{thp}{lofi}
\floatname{file}{File}

\newcommand{\textuperscript}[1]{$^{\textnormal{#1}}$}

\newcommand{\aloc}{\mathsf{A}}
\newcommand{\bloc}{\mathsf{B}} 
\newcommand{\cloc}{\mathsf{C}}

\newcommand{\qeqb}{{\hat{q}_{\bloc}}} 
\newcommand{\qeqbc}{{\hat{\tilde{q}}_{\bloc}}} 

\newcommand{\sbt}{\,\begin{picture}(-1,1)(-1,-1.625)\circle*{1.5}\end{picture}\ }

\newcommand{\toneQLE}{t_{1,\mathrm{QLE}}(p; p_0)} 
\newcommand{\ttwoQLE}{t_{2,\mathrm{QLE}}(p; p_0)} 
\newcommand{\tbarQLE}{\bar{t}_{\mathrm{QLE}}} 

\newcommand{\tonetilQLE}{\tilde{t}_{1,\mathrm{QLE}}(p; p_0)} 
\newcommand{\ttwotilQLE}{\tilde{t}_{2,\mathrm{QLE}}(p; p_0)} 
\newcommand{\tbartilQLE}{\tilde{\bar{t}}_{\mathrm{QLE}}} 

\newcommand{\toneQLErho}{t_{1,\mathrm{QLE},\rho\gg0}(p; p_0)} 
\newcommand{\ttwoQLErho}{t_{2,\mathrm{QLE},\rho\gg0}(p; p_0)} 
\newcommand{\tbarQLErho}{\bar{t}_{\mathrm{QLE},\rho \gg 0}} 

\newcommand{\tonetilQLErho}{\tilde{t}_{1,\mathrm{QLE},\rho\gg0}(p; p_0)} 
\newcommand{\ttwotilQLErho}{\tilde{t}_{2,\mathrm{QLE},\rho\gg0}(p; p_0)} 
\newcommand{\tbartilQLErho}{\tilde{\bar{t}}_{\mathrm{QLE},\rho \gg 0}} 

\newcommand{\toneOLM}{t_{1,\mathrm{OLM}}(p; p_0)} 
\newcommand{\ttwoOLM}{t_{2,\mathrm{OLM}}(p; p_0)} 

\newcommand{\tonetilOLM}{\tilde{t}_{1,\mathrm{OLM}}(p; p_0)} 
\newcommand{\ttwotilOLM}{\tilde{t}_{2,\mathrm{OLM}}(p; p_0)} 
\newcommand{\tbartilOLM}{\tilde{\bar{t}}_{\mathrm{OLM}}} 

\newcommand{\configL}{\textsf{C}$-$\textsf{A}$-$\textsf{B}}
\newcommand{\configM}{\textsf{A}$-$\textsf{C}$-$\textsf{B}}
\newcommand{\configR}{\textsf{A}$-$\textsf{B}$-$\textsf{C}}

\newcommand*\patchAmsMathEnvironmentForLineno[1]{%
  \expandafter\let\csname old#1\expandafter\endcsname\csname #1\endcsname
  \expandafter\let\csname oldend#1\expandafter\endcsname\csname end#1\endcsname
  \renewenvironment{#1}%
     {\linenomath\csname old#1\endcsname}%
     {\csname oldend#1\endcsname\endlinenomath}}%
\newcommand*\patchBothAmsMathEnvironmentsForLineno[1]{%
  \patchAmsMathEnvironmentForLineno{#1}%
  \patchAmsMathEnvironmentForLineno{#1*}}%
\AtBeginDocument{%
\patchBothAmsMathEnvironmentsForLineno{equation}%
\patchBothAmsMathEnvironmentsForLineno{align}%
\patchBothAmsMathEnvironmentsForLineno{flalign}%
\patchBothAmsMathEnvironmentsForLineno{alignat}%
\patchBothAmsMathEnvironmentsForLineno{gather}%
\patchBothAmsMathEnvironmentsForLineno{multline}%
}

\titlespacing\section{0pt}{10pt plus 4pt minus 2pt}{0pt plus 2pt minus 2pt}
\titlespacing\subsection{0pt}{10pt plus 4pt minus 2pt}{0pt plus 2pt minus 2pt}
\titlespacing\subsubsection{0pt}{10pt plus 4pt minus 2pt}{0pt plus 2pt minus 2pt}

\title{The effect of linkage on establishment and survival\\of locally beneficial mutations}

\author{Simon Aeschbacher$^{\ast,1,2}$, Reinhard B\"{u}rger$^{1}$}

\begin{document}

\maketitle

\begin{center}
\noindent 1 Department of Mathematics, University of Vienna, 1090 Vienna, Austria\\
2 Department of Evolution and Ecology, University of California, Davis, CA 95616
\\
$\ast$ E-mail: saeschbacher@mac.com
\end{center}

\vspace{1cm}

\begin{abstract}
	\noindent
We study invasion and survival of weakly beneficial mutations arising in linkage to an established migration--selection polymorphism. Our focus is on a continent--island model of migration, with selection at two biallelic loci for adaptation to the island environment. Combining branching and diffusion processes, we provide the theoretical basis for understanding the evolution of islands of \rev{divergence}, the genetic architecture of locally adaptive traits, and the importance of so-called `divergence hitchhiking' relative to other mechanisms, such as `genomic hitchhiking', chromosomal inversions, or translocations. We derive approximations to the invasion probability and the extinction time \rev{of a de-novo mutation}. Interestingly, the invasion probability is maximised at a non-zero recombination rate if the focal mutation is sufficiently beneficial. If a proportion of migrants carries a beneficial background allele, the mutation is less likely to become established. Linked selection may increase the survival time by several orders of magnitude. \rev{By altering the time scale of stochastic loss, it can therefore affect the dynamics at the focal site to an extent that is of evolutionary importance, especially in small populations}. We derive an effective migration rate experienced by the weakly beneficial mutation, which accounts for the reduction in gene flow imposed by linked selection. Using the concept of the effective migration rate, we also quantify the long-term effects on neutral variation embedded in a genome with arbitrarily many sites under selection. Patterns of neutral diversity change qualitatively and quantitatively as the position of the neutral locus is moved along the chromosome. This will be \revv{useful} for population-genomic inference. \rev{Our results strengthen the emerging view that physically linked selection is biologically relevant if linkage is tight or if selection at the background locus is strong.}
\end{abstract}

\clearpage


\section{Introduction}
Adaptation to local environments may generate a selective response at several loci, either because the fitness-related traits are polygenic, or because multiple traits are under selection. However, populations adapting to spatially variable environments often experience gene flow that counteracts adaptive divergence. The dynamics of polygenic adaptation is affected by physical linkage among selected genes, and hence by recombination \citep{Barton:1995fk}. \rev{Recombination allows contending beneficial mutations to form optimal haplotypes, but it also breaks up existing beneficial associations \citep{Fisher:1930fk,Muller:1932zr,Hill:1966fk,Lenormand:2000uq}.} On top of that, finite population size causes random fluctuations of allele frequencies that may lead to fixation or loss. Migration and selection create statistical associations even among physically unlinked loci.

The availability of genome-wide marker and DNA-sequence data has  spurred both empirical and theoretical work on the interaction of selection, gene flow, recombination, and genetic drift. Here, we study the stochastic fate of a locally beneficial mutation that arises in linkage to an established migration--selection polymorphism. We also investigate the long-term effect on linked neutral variation of adaptive divergence with gene flow.

Empirical insight on local adaptation with gene flow emerges from studies of genome-wide patterns of genetic differentiation between populations or species. Of particular interest are studies that have either related such patterns to function and fitness \citep[e.g.][]{Nadeau:2012vn,Nadeau:2013vn}, or detected significant deviations from neutral expectations \citep[e.g.][]{Karlsen:2013vn}, thus implying that some of this divergence is adaptive. One main observation is that in some organisms putatively adaptive differentiation (e.g.\ measured by elevated $F_{\mathrm{ST}}$) is clustered at certain positions in the genome \citep[][and references therein]{Nosil:2012fk}. This has led to the metaphor of genomic islands of \rev{divergence} or speciation \citep{Turner:2005fk}. Other studies did not identify such islands, however \cite[see][ for a review of plant studies]{Strasburg:2012uq}.

These findings have stimulated theoretical interest in mechanistic explanations for the presence or absence of genomic islands. 
Polygenic local adaptation depends crucially on the genetic architecture of the selected traits, but, in the long run, local adaptation may also lead to the evolution of this architecture. Here, we define genetic architecture as the number of, and physical distances between, loci contributing to local adaptation, and the distribution of selection coefficients of established mutations.

Using simulations, \citet{Yeaman:2011fk} have shown that mutations contributing to adaptive divergence in a quantitative trait may physically aggregate in the presence of gene flow. In addition, these authors reported cases where the distribution of mutational effects changed from many divergent loci with mutations of small effect to few loci with mutations of large effect.
Such clustered architectures reduce the likelihood of recombination breaking up locally beneficial haplotypes and incorporating maladaptive immigrant alleles. This provides a potential explanation for genomic islands of divergence. However, it is difficult to explain the variability in the size of \rev{empirically observed islands of divergence}, especially the existence of very long ones. Complementary mechanisms have been proposed, such as the accumulation of adaptive mutations in regions of strongly reduced recombination (e.g.\ at chromosomal inversions; \citealt{Guerrero:2012uq,McGaugh:2012ve}), or the assembly of adaptive mutations by large-scale chromosomal rearrangements (e.g.\ transpositions of loci under selection; \citealt{Yeaman:2013uq}).

It is well established that spatially divergent selection can cause a reduction in the effective migration rate \citep{Charlesworth:1997vn,Kobayashi:2008fk,Feder:2010fk}. 
This is because migrants tend to carry combinations of alleles that are maladapted, such that selection against a locally deleterious allele at one locus also eliminates incoming alleles at other loci. The effective migration rate can either be reduced by physical linkage to a gene under selection, or by statistical associations among physically unlinked loci.
Depending on whether physical or statistical linkage is involved, the process of linkage-mediated differentiation with gene flow has, by some authors, been called `divergence hitchhiking' or `genomic hitchhiking', respectively \citep{Nosil:2012fk,Feder:2012dq,Via:2012uq}. These two processes are not mutually exclusive, and, recently, there has been growing interest in assessing their relative importance in view of explaining observed patterns of divergence. If not by inversions or translocations, detectable islands of divergence are expected as a consequence of so-called divergence hitchhiking, but not of genomic hitchhiking. This is because physical \rev{linkage} reduces the effective migration rate only locally (i.e.\ in the neighbourhood of selected sites), whereas statistical \rev{linkage} \revv{may reduce} it \revv{across the whole genome}. Yet, if there are many loci under selection, it is unlikely that all of them are physically unlinked \citep{Barton:1983fk}, and so the two sources of \rev{linkage disequilibrium} may be confounded.

A number of recent studies have focussed on the invasion probability of neutral or locally beneficial de-novo mutations in the presence of divergently selected loci in the background \citep{Feder:2010fk,Feder:2012dq,Flaxman:2013fk,Yeaman:2011uq,Yeaman:2013uq}. They showed that linkage elevates invasion probabilities only over very short map distances, implying that physical \rev{linkage} provides an insufficient explanation for both the abundance and size of islands of divergence. Such conclusions hinge on assumptions about the distribution of effects of beneficial mutations, the distribution of recombination rates along the genome, and the actual level of geneflow. \rev{T}hese studies were based on time-consuming simulations \citep{Feder:2010fk,Feder:2012dq,Flaxman:2013fk,Yeaman:2013uq} or heuristic ad-hoc aproximations \citep{Yeaman:2011uq,Yeaman:2013uq} that provide limited understanding. Although \rev{crucial}, invasion probabilities on their own might not suffice to gauge the importance of physical linkage in creating observed patterns of divergence. In finite populations, the time to extinction of adaptive mutations is \rev{also relevant}. It codetermines the potential of synergistic interactions among segregating adaptive alleles.

Here, we fill a gap in existing theory to understand the role of physical linkage in creating observed patterns of divergence with gene flow. First, we provide numerical and analytical approximations to the invasion probability of locally beneficial mutations arising in linkage to an existing migration--selection polymorphism. This sheds light on the ambiguous role of recombination and allows for an approximation to the distribution of fitness effects of successfully invading mutations. Second, we obtain a diffusion approximation to the proportion of time the beneficial mutation segregates in various frequency ranges (the sojourn-time density), and the expected time to its extinction (the mean absorption time). From these, we derive an invasion-effective migration rate experienced by the focal mutation. Third, we extend existing approximations of the effective migration rate at a neutral site linked to two migration--selection polymorphisms \citep{Buerger:2011uq} to an arbitrary number of such polymorphisms. These formulae are used to predict the long-term footprint of polygenic local adaptation on linked neutral variation. We extend some of our analysis to the case of standing, rather than de-novo, adaptive variation at the background locus.

\section{Methods}

\subsection{Model}
We consider a discrete-time version of a model with migration and selection at two biallelic loci \citep{Buerger:2011uq}. Individuals are monoecious diploids and reproduce sexually. Soft selection occurs at the diploid stage and then a proportion $m$ ($0 < m < 1$) of the island population is replaced by immigrants from the continent \citep{Haldane:1930fk}.
Migration is followed by gametogenesis, recombination with probability $r$ ($0 \le r \le 0.5$), and random union of gametes including population regulation. 
Generations do not overlap.

We denote the two loci by $\aloc$ and $\bloc$ and their alleles by $A_1$ and $A_2$, and $B_1$ and $B_2$, respectively. Locus $\aloc$ is taken as the focal locus and locus $\bloc$ as background locus. The four haplotypes 1, 2, 3, and 4 are $A_1B_1$, $A_1B_2$, $A_2B_1$, and $A_2B_2$.
On the island, the frequencies of $A_1$ and $B_1$ are $p$ and $q$, and the linkage disequilibrium is denoted by $D$ (see section 1 in File \ref{txt:SI} for details).

\subsection{Biological scenario}
We assume that the population on the continent is fixed for alleles $A_2$ and $B_2$. The island population is of size $N$ and initially fixed for $A_2$ at locus $\aloc$. At locus $\bloc$, the locally beneficial allele $B_1$ has arisen some time ago and is segregating at migration--selection balance. Then, a weakly beneficial mutation occurs at locus $\aloc$, resulting in a single copy of $A_1$ on the island. Its fate is jointly determined by direct selection on locus $\aloc$, linkage to the selected locus $\bloc$, migration, and random genetic drift. If $A_1$ occurs on the beneficial background ($B_1$), the fittest haplotype is formed and invasion is likely unless recombination transfers $A_1$ to the deleterious background ($B_2$). If $A_1$ initially occurs on the $B_2$ background, a suboptimal haplotype is formed ($A_1B_2$; Eq.\ \ref{eq:fitMatAdd} below) and $A_1$ is doomed to extinction unless it recombines  onto the $B_1$ background early on. These two scenarios occur proportionally to the marginal equilibrium frequency $\qeqb$ of $B_1$. Overall, recombination is therefore expected to play an ambiguous role.

Two aspects of genetic drift are of interest: random fluctuations when $A_1$ is initially rare
, and random sampling of alleles between successive generations. 
In the first part of the paper, we focus exclusively on the random fluctuations when $A_1$ is rare, assuming that $N$ is so large that the dynamics is almost deterministic after an initial stochastic phase.
In the second part, we allow for small to moderate population size $N$ on the island. The long-term invasion properties of $A_1$ are expected to differ in the two cases \cite[p.\ 167--171]{Ewens:1979hc}. With $N$ sufficiently large and parameter combinations for which a fully-polymorphic internal equilibrium exists under deterministic dynamics, the fate of $A_1$ is decided very early on. If it survives the initial phase of stochastic loss, it will reach the (quasi-) deterministic equilibrium frequency and stay in the population for a very long time \citep{Petry:1983uq}. This is what we call invasion, or establishment. Extinction will finally occur, because migration introduces $A_2$, but not $A_1$. Yet, extinction occurs on a time scale much longer than is of interest for this paper. For small or moderate $N$, however, genetic drift will cause extinction of $A_1$ on a much shorter time scale, even for moderately strong selection. In this case, stochasticity must be taken into account throughout, and interest shifts to the expected time $A_1$ spends in a certain range of allele frequencies (sojourn \revv{time}) and the expected time to extinction (absorption time).

As an extension of this basic scenario, we allow the background locus to be polymorphic on the continent. Allele $B_1$ is assumed to segregate at a constant frequency $q_c$. This reflects, for instance, a polymorphism maintained at drift--mutation or mutation--selection balance. \rev{It could also apply to the case where the continent is a meta-population or  receives migrants from other populations.} A proportion $q_c$ of haplotypes carried by immigrants \rev{to the focal island} will then be $A_2B_1$, \rev{and a proportion $1-q_c$ will be $A_2B_2$}.

\subsection{Fitness and evolutionary dynamics}
We define the relative fitness of a genotype as its expected relative contribution to the gamete pool from which the next generation of zygotes is formed. 
We use $w_{ij}$ for the relative fitness of the genotype composed of haplotypes $i$ and $j$ ($i,j \in \{1,2,3,4\}$). Ignoring parental and position effects in heterozygotes, we distinguish nine genotypes. We then have $w_{ij} = w_{ji}$ for all $i \neq j$ and $w_{23} = w_{14}$.

The extent to which analytical results can be obtained for general fitnesses is limited \citep{Ewens:1967fk,Karlin:1968fk}. Unless otherwise stated, we therefore assume absence of dominance and epistasis, i.e., allelic effects combine additively within and between loci. The matrix of relative genotype fitnesses $w_{ij}$ (Eq.\ \ref{eq:fitMat} in File \ref{txt:SI}) may then be written as
\begin{equation}
	\label{eq:fitMatAdd}
	\bordermatrix{
			~ & B_1B_1 & B_1B_2 & B_2B_2 \cr
                  		A_1A_1 & 1+a+b & 1+a & 1+a-b \cr
                  		A_1A_2 & 1+b & 1 & 1-b \cr
			A_2A_2 & 1-a+b & 1-a & 1-a-b \cr 
			},
\end{equation}
where $a$ and $b$ are the selective advantages on the island of alleles $A_1$ and $B_1$ relative to $A_2$ and $B_2$, respectively. To enforce positive fitnesses, we require that $0<a,b<1$ and $a+b<1$. We assume that selection in favour of $A_1$ is weaker than selection in favour of $B_1$ ($a < b$). Otherwise, $A_1$ could be maintained in a sufficiently large island population independently of $B_1$, whenever $B_1$ is not swamped by gene flow \citep{Haldane:1930fk}. As our focus is on the effect of linkage on establishment of $A_1$, this case is not of interest.

The deterministic dynamics of the haplotype frequencies are given by the recursion equations in Eq.\ \eqref{eq:recEqFull} in File \ref{txt:SI} (see also File \ref{prc:determDiscrNB}).
A crucial property of this dynamics is the following. Whenever a marginal one-locus migration--selection equilibrium $E_{\bloc}$ exists such that the background locus \textsf{B} is polymorphic and locus \textsf{A} is fixed for allele $A_2$, this equilibrium is asymptotically stable. After occurrence of $A_1$, $E_{\bloc}$ may become unstable, in which case a fully-polymorphic (internal) equilibrium emerges and is asymptotically stable, independently of whether the continent is monomorphic ($q_c = 0$) or polymorphic ($0 < q_c < 1$) at the background locus. Therefore, in the deterministic model, invasion of $A_1$ via $E_{\bloc}$ is always followed by an asymptotic approach towards an internal equilibrium (see File \ref{txt:SI}, sections 3 and 6). 

Casting our model into a stochastic framework is difficult in general. By focussing on the initial phase after occurrence of $A_1$, the four-dimensional system in Eq.\ \eqref{eq:recEqFull} can be simplified to a two-dimensional system (Eq.\ \ref{eq:recEqApprox} in File \ref{txt:SI}). This allows for a branching-process approach as described in the following.

\subsection{Two-type branching process}
As shown in section 2 in File \ref{txt:SI}, for rare $A_1$, we need to follow only the frequencies of haplotypes $A_1B_1$ and $A_1B_2$. This corresponds to $A_1$ initially occurring on the $B_1$ or $B_2$ background, respectively, and holds as long as $A_1$ is present in heterozygotes only. Moreover, it is assumed that allele $B_1$ is maintained constant at the marginal one-locus migration--selection equilibrium $E_{\bloc}$ of the dynamics in Eq.\ \eqref{eq:recEqFull}. At this equilibrium, the frequency of $B_1$ is
\begin{equation}
	\label{eq:FreqB1OneLoc}
	\qeqb = \frac{b - m(1 - a)}{b(1+m)}
\end{equation}
for a monomorphic continent (see section 3 in File \ref{txt:SI} for details, and Eq.\ \ref{eq:FreqB1OneLocPolymCont} for a polymorphic continent).

To model the initial stochastic phase after occurrence of $A_1$ for large $N$, we employed a two-type branching process in discrete time \citep{Harris:1963dq}. 
We refer to haplotypes $A_1B_1$ and $A_1B_2$ as type 1 and 2, respectively. They are assumed to propagate independently and contribute offspring to the next generation according to type-specific distributions. We assume that the number of $j$-type offspring produced by an $i$-type parent is Poisson-distributed with parameter $\lambda_{ij}$ ($i \in \{1,2\}$). 
Because of independent offspring distributions, the probability-generating function (pgf) for the number of offspring of any type produced by an $i$-type parent is $f_{i}(s_1, s_2) = \prod_{j=1}^{2} f_{ij}(s_{j})$, where $f_{ij}(s_j) = e^{-\lambda_{ij}(1-s_j)}$ for $i,j \in \{1,2\}$ \rev{(section 4 in File \ref{txt:SI})}. The $\lambda_{ij}$ depend on fitness, migration, and recombination, and are derived from the deterministic model (Eq.\ \ref{eq:recGEntries} in File \ref{txt:SI}). The matrix $\mathbf{L} = (\lambda_{ij})$, ${i,j \in \{1,2\}}$, is called the mean matrix. Allele $A_1$ has a strictly positive invasion probability if $\nu > 1$, where $\nu$ is the leading eigenvalue of $\mathbf{L}$. The branching process is called supercritical in this case.

We denote the probability of invasion of $A_1$ conditional on initial occurrence on background $B_1$ ($B_2$) by $\pi_1$ ($\pi_2$), and the corresponding probability of extinction by $Q_1$ ($Q_2$). The latter are found as the smallest positive solution of
\begin{subequations}
	\label{eq:bpSol}
		\begin{align}
			f_1(s_1, s_2) = s_1\\
			f_2(s_1, s_2) = s_2
		\end{align}
\end{subequations}
such that $s_i < 1$ ($i \in \{ 1, 2 \}$). Then, $\pi_1 = 1-Q_1$ and $\pi_2=1-Q_2$ \citep{Haccou:2005bh}. The overall invasion probability of $A_1$ is given as the weighted average of the two conditional probabilities,
\begin{equation}
	\label{eq:invProbOverall}
	\bar{\pi} = \qeqb \pi_1 + (1-\qeqb) \pi_2
\end{equation}
\cite[cf.][]{Kojima:1967fk, Ewens:1967fk, Ewens:1968fk}. Section 4 in File \ref{txt:SI} gives further details and explicit expressions for additive fitnesses. 

\subsection{Diffusion approximation}
The branching process described above models the initial phase of stochastic loss and applies as long as the focal mutant $A_1$ is rare. To study long-term survival of $A_1$, we employ a diffusion approximation. We start from a continuous-time version of the deterministic dynamics in Eq.\ \eqref{eq:recEqFull}, assuming additive fitnesses as in Eq.\ \eqref{eq:fitMatAdd}. For our purpose, it is convenient to express the dynamics in terms of the allele frequencies ($p$, $q$) and the linkage disequilibrium ($D$), as given in Eq.\ \eqref{eq:diffEqs} in File \ref{txt:SI}. Changing to the diffusion scale, we measure time in units of $2N_e$ generations, where $N_e$ is the effective population size.

We introduce the scaled selection coefficients $\alpha = 2N_e a$ and $\beta = 2N_e b$, the scaled recombination rate $\rho = 2 N_e r$, and the scaled migration rate $\mu = 2N_e m$. As it is difficult to obtain analytical results for the general two-locus diffusion problem \citep{Ewens:1979hc,Ethier:1980dq,Ethier:1988bh,Ethier:1989qf}, we assume that recombination is much stronger than selection and migration. Then, linkage disequilibrium decays on a faster time scale, whereas allele frequencies evolve on a slower one under quasi-linkage equilibrium (QLE) \citep{Kimura:1965fk,Nagylaki:1999uq,Kirkpatrick:2002uq}. In addition, we assume that the frequency of the beneficial background allele $B_1$ is not affected by establishment of $A_1$ and stays constant at $q = \qeqbc$. Here, $\qeqbc$ is the frequency of $B_1$ at the one-locus migration--selection equilibrium when time is continuous, $\tilde{E}_{\bloc}$ (section 6 in File \ref{txt:SI}). As further shown in section 6 of File \ref{txt:SI}, these assumptions lead to a one-dimensional diffusion process. The expected change in $p$ per unit time is
\begin{equation}
	\label{eq:Mp}
	M(p) =  \alpha p(1-p) - \mu p + \frac{\mu(\beta-\mu)}{\beta - \mu - \alpha(1-2p) + \rho}\,p
\end{equation}
if the continent is monomorphic. The first term is due to direct selection on the focal locus, the second reflects migration, and the third represents the interaction of all forces.

For a polymorphic continent, $M(p)$ is given by Eq.\ \eqref{eq:MpPolymCont} in File \ref{txt:SI}, and the interaction term includes the continental frequency $q_c$ of $B_1$. In both cases, assuming random genetic drift according to the Wright--Fisher model, 
the expected squared change in $p$ per unit time is $V(p) = p(1-p)$ \citep{Ewens:1979hc}. We call $M(p)$ the \emph{infinitesimal mean} and $V(p)$ the \emph{infinitesimal variance} \cite[p.\ 159]{Karlin:1981fk}.

Let the initial frequency of $A_1$ be $p_0$. We introduce the \emph{sojourn-time density} (STD) $t(p; p_0)$ such that the integral $\int_{p_1}^{p_2} t (p; p_0) dp$ approximates the expected time $A_1$ segregates at a frequency between $p_1$ and $p_2$ before extinction, conditional on $p_0$. 
Following \citet[Eqs.\ 4.38 and 4.39]{Ewens:1979hc}, we define
\begin{equation}
	\label{eq:STDQLEDefMain}
	t_{\mathrm{QLE}}(p; p_0) = \left\{ \begin{array}{ll}
		\toneQLE & \textrm{if $0 \le p \le p_0$}\\
		\ttwoQLE & \textrm{if $p_0 \le p \le 1$}
	\end{array} \right.,
\end{equation}
with subscript QLE for the assumption of quasi-linkage equilibrium. The densities $t_{i,\mathrm{QLE}}(p; p_0)$ are
\begin{subequations}
	\label{eq:STDQLE}
		\begin{align}
			\toneQLE & = \frac{2}{V(p)\psi(p)} \int_{0}^{p}\psi(y)dy \label{eq:STD1QLE}, \\
			\ttwoQLE & = \frac{2}{V(p)\psi(p)} \int_{0}^{p_0}\psi(y)dy \label{eq:STD2QLE},
		\end{align}
\end{subequations}
where $\psi(p) = \exp\left[-2\int_{0}^{p} \frac{M(z)}{V(z)} dz\right]$. Integration over $p$ yields the expected time to extinction,
\begin{equation}
	\label{eq:meanAbsTimeQLE}
	\tbarQLE = \int_{0}^{p_0} \toneQLE dp + \int_{p_0}^{1} \ttwoQLE dp,
\end{equation}
or the \emph{mean absorption time}, in units of $2N_e$ generations. A detailed \rev{exposition} is given in section 7 of File \ref{txt:SI}. 

\subsection{Simulations}
We \rev{conducted} two types of simulation, one for the branching-process regime and another for a finite island population with Wright--Fisher random drift. In the branching-process regime, we simulated the absolute frequency of the two types of interest ($A_1B_1$ and $A_1B_2$) over time. Each \rev{run} was initiated with a single individuum and its type determined according to Eq.\ \eqref{eq:FreqB1OneLoc}. Every generation, each individual produced \rev{a} Poisson-distributed \rev{number} of offspring of either type (see above). \rev{We performed $n = 10^6$ runs. Each run was terminated if either the mutant population went extinct (no invasion), reached a size of $500/(2a)$ (invasion), or survived for more than $5 \times 10^4$ generations (invasion). We estimated the invasion probability from the proportion $\hat{\pi}$ of runs that resulted in invasion, and its standard error as $\sqrt{\hat{\pi}(1-\hat{\pi})/n}$.}

In the Wright--Fisher type simulations, each generation was initiated by zygotes built from gametes of the previous generation. Viability selection, migration, and gamete production including recombination (meiosis) were implemented according to the deterministic recursions for the haplotype frequencies in Eq.\ \eqref{eq:recEqFull}. Genetic drift was simulated through the formation of $N_e$ (rather, the nearest integer) zygotes for the next generation by random union of pairs of gametes. Gametes were sampled with replacement from the gamete pool in which haplotypes were represented according to the deterministic recursions. Replicates were terminated if either allele $A_1$ went extinct or a maximum of $10^9$ generations was reached. \rev{Unless otherwise stated, for each parameter combination we performed 1000 runs, each with 1000 replicates. Replicates within a given run provided one estimate of the mean absorption time, and runs provided a distribution of these estimates.}
Java source code and JAR files are available from the corresponding author. 

\section{Results}
\subsection{Establishment in a large island population} 
We first describe the invasion properties of the beneficial mutation $A_1$, which arises in linkage to a migration-selection polymorphism at the background locus $\bloc$. Because we assume that the island population is large, random genetic drift is ignored after $A_1$ has overcome the initial phase during which stochastic loss is likely. Numerical and analytical results were obtained from the two-type branching process and confirmed by simulations (see Methods). We will turn to the case of small to moderate population size further below.
%

\subsubsection{Conditions for the invasion of $A_1$}
Mutation $A_1$ has a strictly positive invasion probability whenever
\begin{equation}
	\label{eq:condNonExtGener}
	r w_{14} \left[ \frac{\bar{w}}{1-m} - \qeqb w_1 - (1-\qeqb) w_2 \right]  < - \left( \frac{\bar{w}}{1-m} - w_1 \right) \left( \frac{\bar{w}}{1-m} - w_2 \right)
\end{equation}
(section 4 in File \ref{txt:SI}; File \ref{prc:stochDiscrNB}).
Here, $w_i$ is the marginal fitness of type $i$ and $\bar{w}$ the mean fitness of the resident population (see Eqs.\ \ref{eq:margFitA1rare} and \ref{eq:meanFitA1rare} in File \ref{txt:SI}). 
Setting $m=0$, we recover the invasion condition obtained by \citet{Ewens:1967fk} for a panmictic population in which allele $B_1$ is maintained at frequency $\qeqb$ by overdominant selection.
%
%
All remaining results in this subsection assume additive fitnesses as in Eq.\ \eqref{eq:fitMatAdd}.

For a monomorphic continent ($q_c=0$), it follows from Eq.\ \eqref{eq:condNonExtGener} that $A_1$ can invade only if $m < m^{\ast}$, where
\begin{equation}
	\label{eq:mCrit5Main}
	m^{\ast} = \frac{a(b-a+r)}{(a-r)(a-b)+r(1-a)}.
\end{equation}
In terms of the recombination rate, $A_1$ can invade only if $r < r^{\ast}$, where
\begin{equation}
	\label{eq:rCrit}
	r^{\ast} = \left\{ \begin{array}{ll}
 	\frac{1}{2} & \textrm{if $m \leq \frac{a}{1-2a+b}$},\\
 	\frac{a(a-b)(1+m)}{a(1+2m)-(1+b)m} & \textrm{otherwise}
  	\end{array} \right.
\end{equation}
(see sections 3 and 4 in File \ref{txt:SI}, and \rev{File \ref{prc:determDiscrNB}} for details).

For a polymorphic continent ($0< q_c<1$), $A_1$ has a strictly positive invasion probability whenever $r$ and $q_c$ are below the critical values $r^{\ast}$ and $q_c^{\ast}$ derived in section 3 of File \ref{txt:SI} (cf.\ File \ref{prc:stochCompareJacobianVsMeanMatrixNB}). \rev{In this case, we could not determine the critical migration rate $m^{\ast}$ explicitly.}
%

\subsubsection{Invasion probability}
\begin{figure}[!ht]
\begin{center}
\includegraphics[width=1\textwidth]{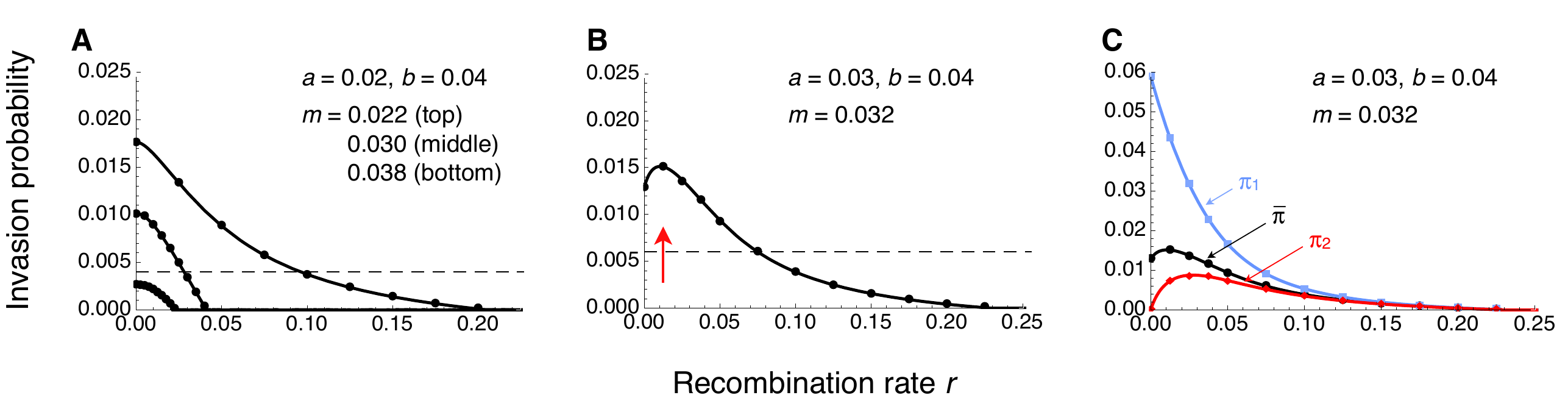}
\end{center}
\caption{
{\textrm{Invasion probability of $A_1$ as a function of the recombination rate for a monomorphic continent.}} \textbf{(A)} and \textbf{(B)} show the weighted average invasion probabilitiy $\bar{\pi}$ across the two genetic backgrounds $B_1$ and $B_2$ (Eqs.\ \ref{eq:FreqB1OneLoc} and \ref{eq:invProbOverall}). 
For comparison, horizontal dashed lines give 10\% of Haldane's \citeyearpar{Haldane:1927fk} approximation $2a$, valid for $m=0$ and $r=0$.
\textbf{(B)} The optimal recombination rate $r_{\mathrm{opt}}$, defined as the recombination rate at which $\bar{\pi}$ is maximised (red arrow), is non-zero. \textbf{(C)} Same as in \textbf{(B)}, but in addition to the weighted average, the invasion probabilities of $A_1$ conditional on initial occurrence on the $B_1$ or $B_2$ background are shown in blue or red, respectively. Note the difference in the scale of the vertical axis between \textbf{(B)} and \textbf{(C)}. In all panels, curves show exact numerical solutions to the branching process. \rev{Dots represent the point estimates across $10^{6}$ simulations under the branching-process assumptions (see Methods). Error bars span twice the standard error on each side of the point estimates, but are too short to be visible}.}
\label{fig:invProbFuncR}
\end{figure}

We obtained exact conditional invasion probabilities, $\pi_1$ and $\pi_2$, of $A_1$ by numerical solution of the pair of transcendental equations in Eq.\ \eqref{eq:bpSol}. From these, we calculated the average invasion probability $\bar{\pi}$ according to Eq.\ \eqref{eq:invProbOverall}, with $\qeqb$ as in Eq.\ \eqref{eq:FreqB1OneLoc} (Figures \ref{fig:invProbFuncR} and \ref{fig:invProbFuncM} for a monomorphic continent). \citet{Haldane:1927fk} approximated the invasion probability without migration and linked selection by $2a$, i.e.\ twice the selective advantage of $A_1$ in a heterozygote. With linked selection, the map distance over which $\bar{\pi}$ is above, say, 10\% of $2a$ can be large despite gene flow (Figures \ref{fig:invProbFuncR}A and \ref{fig:invProbFuncR}B).

\begin{figure}[!ht]
\begin{center}
\includegraphics[width=1\textwidth]{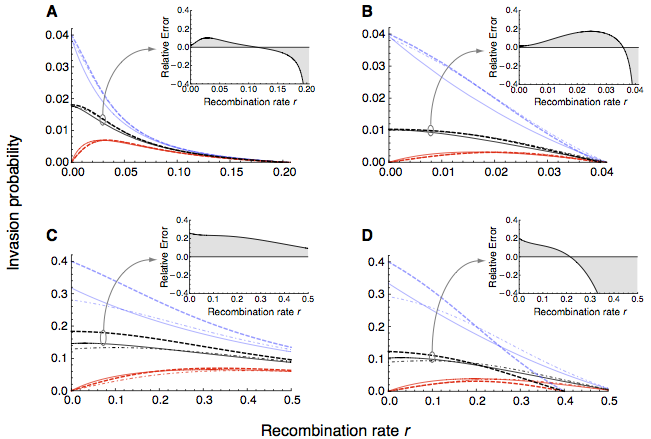}
\end{center}
\caption{
{\textrm{Approximation to the invasion probability of $A_1$ for a monomorphic continent.}} Invasion probabilities are shown for $A_1$ initially occurring on the beneficial background $B_1$ (blue), on the deleterious background $B_2$ (red), and as a weighted average across backgrounds (black). 
Analytical approximations assuming a slightly supercritical branching process (dot-dashed curves) and, in addition, weak evolutionary forces (Eq.\ \ref{eq:approxInvProbAdd}; thick dashed curves) are compared to the exact numerical branching-process solution (solid curves). Inset figures show the error of the analytical approximation $\bar{\tilde{\pi}}(\zeta)$ (thick dashed black curve) relative to $\bar{\pi}$ (solid black curve), $\bar{\tilde{\pi}}(\zeta) / \bar{\pi} - 1$.  \textbf{(A)} $a = 0.02$, $b = 0.04$, $m = 0.022$. \textbf{(B)} $a = 0.02$, $b = 0.04$, $m = 0.03$. \textbf{(C)} $a = 0.2$, $b = 0.4$, $m = 0.22$. \textbf{(D)} $a = 0.2$, $b = 0.4$, $m = 0.3$. As expected, the analytical approximations are very good for weak evolutionary forces (top row), but less so for strong forces (bottom row).}
\label{fig:invProbFuncRApprox}
\end{figure}

Analytical approximations were obtained by assuming that the branching process is slightly supercritical, i.e., that the leading eigenvalue of the mean matrix $\mathbf{L}$ is of the form $\nu = 1 + \xi$, with $\xi > 0$ small. We denote these approximations by $\pi_{1}(\xi)$ and $\pi_{2}(\xi)$. The expressions are long (File \ref{prc:stochDiscrSlightlySupercritBPNB}) and not shown here. 
For weak evolutionary forces ($a,b,m,r \ll 1$), $\pi_{1}(\xi)$ and $\pi_{2}(\xi)$ can be approximated by
\begin{subequations}
	\label{eq:approxInvProbAdd}
	\begin{align}
		\tilde{\pi}_{1}(\xi) = & \max\left[0, \frac{a\left(b + r + \sqrt{R_2}\right) - 2 m r}{\sqrt{R_2}}\right],\\
		\tilde{\pi}_{2}(\xi) = & \max\left[0, \frac{b^2 - 2 m r + b\left(r - \sqrt{R_2} \right) - a\left(b - r - \sqrt{R_2} \right)}{\sqrt{R_2}}\right],
	\end{align}
\end{subequations}
where $R_2 = b^2 + 2br - 4mr + r^2$ and $\xi \approx \frac{1}{2}\left(2a - b - r + \sqrt{R_2} \right)$. The approximate average invasion probability $\bar{\tilde{\pi}}(\xi)$ is obtained according to Eq.\ \eqref{eq:invProbOverall}, with $\qeqb$ as in Eq.\ \eqref{eq:FreqB1OneLoc}. 
Formally, these approximations are justified if $\xi$ is much smaller than 1 (section 4 in File \ref{txt:SI}). Figure \ref{fig:invProbFuncRApprox} suggests that the assumption of weak evolutionary forces is more crucial than $\xi$ small, and that if it is fulfilled, the approximations are very good (compare Figures \ref{fig:invProbFuncRApprox}A and \ref{fig:invProbFuncRApprox}B to \ref{fig:invProbFuncRApprox}C and \ref{fig:invProbFuncRApprox}D).

For a polymorphic continent, exact and approximate invasion probabilities are derived in Files \ref{prc:stochDiscrNB} and \ref{prc:stochDiscrSlightlySupercritBPNB} (see also section 4 in File \ref{txt:SI}).
The most important, and perhaps surprising, effect is that the average invasion probability 
decreases with increasing continental frequency $q_c$ of the beneficial background allele $B_1$ (Figure \ref{fig:invProbFuncRPolymCont}). As a consequence, invasion requires tighter linkage if $q_c > 0$. This is because the resident island population has a higher mean fitness when a proportion $q_c > 0$ of immigrating haplotypes carry the $B_1$ allele, which makes it harder for $A_1$ to become established. Competition against fitter residents therefore compromises the increased probability of recombining onto a beneficial background ($B_1$) when $A_1$ initially occurs on the deleterious background ($B_2$). However, a closer look suggests that if $A_1$ is sufficiently beneficial and recombination sufficiently weak ($r \ll a$), there are cases where the critical migration rate below which $A_1$ can invade is maximised at an intermediate $q_c$ (Figure \ref{fig:invProbFuncMPolymCont}, right column). In other words, for certain combinations of $m$ and $r$, the average invasion probability as a function of $q_c$ is maximised at an intermediate (non-zero) value of $q_c$ (Figure \ref{fig:invProbFuncqCPolymCont}).

For every combination of selection coefficients ($a$, $b$) and recombination rate ($r$), the mean invasion probability 
decreases as a function of the migration rate $m$. This holds for a monomorphic and a polymorphic continent (Figures \ref{fig:invProbFuncM} and \ref{fig:invProbFuncMPolymCont}, respectively). In both cases, migrants carry only allele $A_2$ and, averaged across genetic backgrounds, higher levels of migration make it harder for $A_1$ to invade \cite[cf.][]{Buerger:2011uq}.

\subsubsection{Optimal recombination rate}
Deterministic analysis showed that $A_1$ can invade if and only if recombination is sufficiently weak; without epistasis, large $r$ is always detrimental to establishment of $A_1$ (\citeauthor{Buerger:2011uq} \citeyear{Buerger:2011uq}; section 3 in File \ref{txt:SI}). 
In this respect, stochastic theory is in line with deterministic predictions. However, considering the average invasion probability $\bar{\pi}$ as a function of $r$, we could distinguish two qualitatively different regimes. In the first one, $\bar{\pi}(r)$ decreases monotonically with increasing $r$ (Figure \ref{fig:invProbFuncR}A). In the second one, $\bar{\pi}(r)$ is maximised at an intermediate recombination rate $r_{\mathrm{opt}}$ (Figure \ref{fig:invProbFuncR}B). A similar dichotomy was previously found for a panmictic population in which the background locus is maintained polymorphic by heterozygote superiority \citep{Ewens:1967fk}, and has recently been reported in the context of migration and selection in \rev{simulation studies \citep{Feder:2010fk,Feder:2012dq}}.
As shown in section 5 of File \ref{txt:SI}, 
$r_{\mathrm{opt}}>0$ holds in our model whenever
\begin{equation}
	\label{eq:aCritOptRecombRateGeneric}
	w_1 - w_2 > \bar{w}\ \frac{\pi_{1}^{\circ}}{(1-m)(1-\pi_{1}^{\circ})},
\end{equation}
where $w_1$ ($w_2$) is the marginal fitness of type 1 (2) and $\bar{w}$ the mean fitness of the resident population (defined in Eqs.\ \ref{eq:margFitA1rare} and \ref{eq:meanFitA1rare} in File \ref{txt:SI}). 
Here, $\pi_{1}^{\circ}$ is the invasion probability of $A_1$ conditional on background $B_1$ and complete linkage ($r=0$). Setting $m=0$, we recover Eq.\ (36) of \citet{Ewens:1967fk} for a panmictic population with overdominance at the background locus.

\begin{figure}[!ht]
\begin{center}
\includegraphics[width=0.5\textwidth]{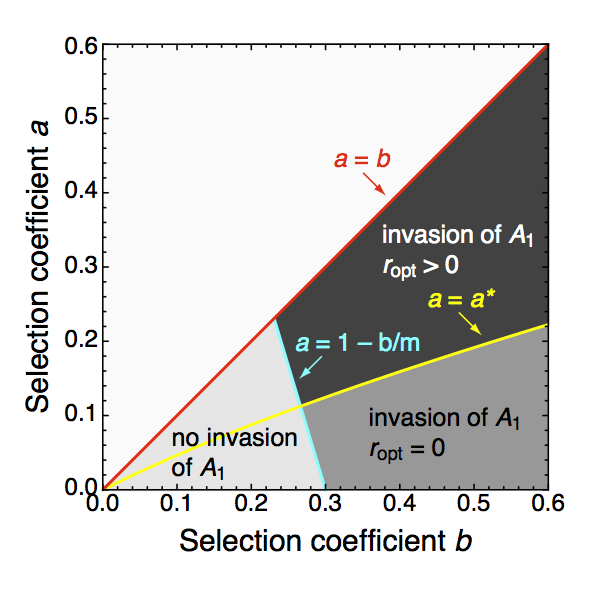}
\end{center}
\caption{
{\textrm{Optimal recombination rate and regions of invasion.}} The dark grey area indicates where 
the optimal recombination rate $r$ is 
positive ($r_{\mathrm{opt}} > 0$; cf.\ Figure \ref{fig:invProbFuncR}B). The medium grey area shows the parameter range for which 
$r_{\mathrm{opt}} = 0$ (cf.\ Figure \ref{fig:invProbFuncR}A). Together, these two areas indicate where $A_1$ can invade via the marginal one-locus migration--selection equilibrium $E_{\bloc}$ if $r$ is sufficiently small. The light grey area shows where $E_{\bloc}$ does not exist and $A_1$ cannot invade via $E_{\bloc}$. The area above $a=b$ is not of interest, as we focus on mutations that are weakly beneficial compared to selection at the background locus ($a<b$). The critical selection coefficient $a^{\ast}$ is given in Eq.\ \eqref{eq:aCritOptRecombRate} and the migration rate is $m = 0.3$ (other values of $m$ yield qualitatively similar diagrams). The continent is monomorphic ($q_c = 0$).}
\label{fig:optRecombRate}
\end{figure}

Inequality \eqref{eq:aCritOptRecombRateGeneric} is very general. In particular, it also holds with epistasis or dominance. However, explicit conclusions require calculation of $\pi_{1}^{\circ}$, $\bar{w}$, and $w_{i}$, which themselves depend on $\qeqb$ and hence on $m$ (cf.\ Eq.\ \ref{eq:FreqB1OneLoc}).
For mathematical convenience, we resorted to the assumption of additive fitnesses (Eq.\ \ref{eq:fitMatAdd}). For a monomorphic continent, $\pi_{1}^{\circ} \approx 2 a (1+m)/(1+b)$ to first order in $a$. Moreover, we found that
\begin{equation}
	\label{eq:aCritOptRecombRate}
	a > a^{\ast}
\end{equation}
is a necessary condition for $r_{\mathrm{opt}}>0$, where
\begin{equation*}
	a^{\ast} = \frac{1}{2}\left\{ 1 + b(2+m) - \sqrt{1 + 2b(1+m) + b^2\left[ 2 + m(4+m) \right]} \right\}
\end{equation*}
(section 5 in File \ref{txt:SI}). Thus, $A_1$ must be sufficiently beneficial for $r_{\mathrm{opt}} > 0$ to hold. Figure \ref{fig:optRecombRate} shows the division of the parameter space where $A_1$ can invade into two areas where $r_{\mathrm{opt}} = 0$ or $r_{\mathrm{opt}} > 0$ holds.

The two regimes $r_{\mathrm{opt}} = 0$  and $r_{\mathrm{opt}} > 0$ arise from the ambiguous role of recombination. On the one hand, when $A_1$ initially occurs on the deleterious background ($B_2$), some recombination is needed to transfer $A_1$ onto the beneficial background ($B_1$) and rescue it from extinction. This is reminiscent of \citeauthor{Hill:1966fk}'s \citeyearpar{Hill:1966fk} result that recombination improves the efficacy of selection in favour of alleles that are partially linked to other selected sites \citep{Barton:2010vn}. On the other hand, when $A_1$ initially occurs on the beneficial background, recombination is always deleterious, as it breaks up the fittest haplotype on the island ($A_1B_1$). This interpretation is confirmed by considering $\pi_1$ and $\pi_2$ separately as functions of $r$ (Figure \ref{fig:invProbFuncR}C). Whereas $\pi_1(r)$ always decreases monotonically with increasing $r$, $\pi_2(r)$ is always 0 at $r=0$ (section 5 in File \ref{txt:SI}) 
and then increases to a maximum at an intermediate recombination rate (compare blue to red curve in Figure \ref{fig:invProbFuncR}C). As $r$ increases further, $\pi_1(r)$ and $\pi_2(r)$ both approach 0. We recall from Eq.\ \eqref{eq:invProbOverall} that the average invasion probability $\bar{\pi}$ is given by $\qeqb \pi_1 + (1-\qeqb) \pi_2$. Depending on $\qeqb$, either $\pi_1$ or $\pi_2$ make a stronger conbribution to $\bar{\pi}$, which then leads to either $r_{\mathrm{opt}} > 0$ or $r_{\mathrm{opt}} = 0$.

A more intuitive interpretation of Eq.\ \eqref{eq:aCritOptRecombRate} is as follows. If $A_1$ conveys a weak advantage on the island ($a < a^{\ast}$), it will almost immediately go extinct when it initially arises on background $B_2$. Recombination has essentially no opportunity of rescuing $A_1$, even if $r$ is large. Therefore, $\pi_2$ contributes little to $\bar{\pi}$. If $A_1$ is sufficiently beneficial on the island ($a > a^{\ast}$), however, it will survive for some time even when arising on the deleterious background. Recombination now has time to rescue $A_1$ if $r$ is sufficiently different from 0 (but not too large). In this case, $\pi_2$ makes an important contribution to $\bar{\pi}$ and leads to $r_{\mathrm{opt}} > 0$. 
For a polymorphic continent, $r_\mathrm{opt} > 0$ may also hold (section 5 in File \ref{txt:SI}). However, in such cases, $r_{\mathrm{opt}}$ approaches zero quickly with increasing $q_c$ (File \ref{prc:stochDiscrDerivativeZeroRecombRateNB} and Figure \ref{fig:invProbFuncRPolymCont}).
%

\subsubsection{Distribution of fitness effects of successful mutations}
\begin{figure}[!ht]
\begin{center}
\includegraphics[width=0.5\textwidth]{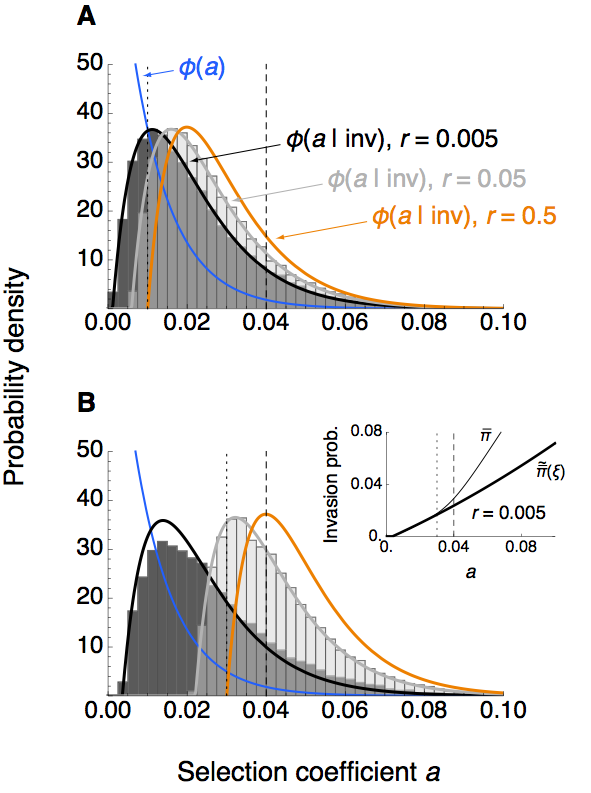}
\end{center}
\caption{
{\textrm{Distribution of fitness effects (DFE) of successfully invading mutations for a monomorphic continent.}} The DFE of successfully invading mutations was obtained as $\phi\left(a \mid \mathrm{inv}\right) = \phi\left(\mathrm{inv} \mid a\right) \phi(a) / \int_{0}^{\infty} \phi\left(\mathrm{inv} \mid a\right) \phi(a)\, da$, where $\phi\left(\mathrm{inv} \mid a\right) = \bar{\tilde{\pi}}(\xi) = \qeqb \tilde{\pi}_{1}(\xi) + (1-\qeqb) \tilde{\pi}_{2}(\xi)$, with $\qeqb$ and $\tilde{\pi}_{i}(\xi)$ as in Eqs.\ \eqref{eq:FreqB1OneLoc} and \eqref{eq:approxInvProbAdd}, respectively. The mutational input distribution was assumed to be exponential, $\phi(a) = \lambda e^{-a \lambda}$ (blue). Vertical lines denote $a=m$ (dotted) and $a=b$ (dashed). 
Histograms were obtained from simulations under the branching-process assumptions \rev{(intermediate grey shading indicates where histograms overlap)}. Each represents $2.5 \times 10^4$ realisations in which $A_1$ successfully invaded (see Methods). As a reference, the one-locus model (no linkage) is shown in orange. \textbf{(A)} Relatively weak migration: $b = 0.04$, $m = 0.01$. \textbf{(B)} Migration three times stronger: $b = 0.04$, $m = 0.03$. In both panels, $\lambda = 100$ and $\phi\left(a \mid \mathrm{inv}\right)$ is shown for a recombination rate of $r=0.005$ (black) and $r = 0.05$ (grey). The inset in panel \textbf{(B)} shows why the fit is worse for $r = 0.005$: in this case, $\bar{\tilde{\pi}}(\xi)$ underestimates the exact invasion probability $\bar{\pi}$ (Eq.\ \ref{eq:invProbOverall}) for large $a$. }
\label{fig:DFEmain}
\end{figure}

Using Eq.\ \eqref{eq:approxInvProbAdd} we can address the distribution of fitness effects (DFE) of successfully invading mutations. This distribution depends on the distribution of selection coefficients $a$ of novel mutations \citep{Kimura:1979lq}, which in general is unknown \citep{Orr:1998ul}. In our scenario, the island population is at the marginal one-locus migration--selection equilibrium $E_\bloc$ before the mutation $A_1$ arises. Unless linkage is very tight, the selection coefficient $a$ must be above a threshold for $A_1$ to effectively withstand gene flow (this threshold is implicitly defined by Eq.\ \ref{eq:mCrit5Main}). Therefore, we assumed that $a$ is drawn from the tail of the underlying distribution, which we took to be exponential \citep{Gillespie:1983pd,Gillespie:1984uq,Orr:2002gf,Orr:2003kx,Barrett:2006fk,Eyre-Walker:2007fk} \rev{\cite[for alternatives, see][]{Cowperthwaite:2005vn,Barrett:2006fk,Martin:2008fk}}.
We further assumed that selection is directional with a constant fitness gradient (Eq.\ \ref{eq:fitMatAdd}).
We restricted the analysis to the case of a monomorphic continent. As expected, linkage to a migration--selection polymorphism shifts the DFE of successfully invading mutations towards smaller effect sizes (Figure \ref{fig:DFEmain}). Comparison to simulated histograms in Figure \ref{fig:DFEmain} suggests that the approximation based on Eq.\ \eqref{eq:approxInvProbAdd} is very accurate.

\subsection{Survival in a finite island population}
We now turn to island populations of small to moderate size $N$. In this case, genetic drift is strong enough to cause extinction on a relevant time scale even after successful initial establishment. Our focus is on the sojourn-time density and the mean absorption time of the locally beneficial mutation $A_1$ (see Methods). We also derive an approximation to the effective migration rate experienced by $A_1$.

\subsubsection{Sojourn-time density}
A general expression for the sojourn-time density (STD) was given in Eq.\ \eqref{eq:STDQLE}. Here, we describe some properties of the exact numerical solution and will then discuss analytical approximations (see also File \ref{prc:stochDiffApproxQLENB}). Because $A_1$ is a de-novo mutation, it has an initial frequency of $p_0 = 1/(2N)$. For simplicity, we assumed that the effective population size on the island is equal to the actual population size, i.e.\ $N_e = N$ (this assumption will be relaxed later). As $p_0 = 1/(2N)$ is very close to zero in most applications, we used $\ttwoQLE$ as a proxy for $t_{\mathrm{QLE}}(p; p_0)$ (cf.\  Eq.\ \ref{eq:STDQLEDefMain}).

\begin{figure}[!ht]
\begin{center}
\includegraphics[width=1\textwidth]{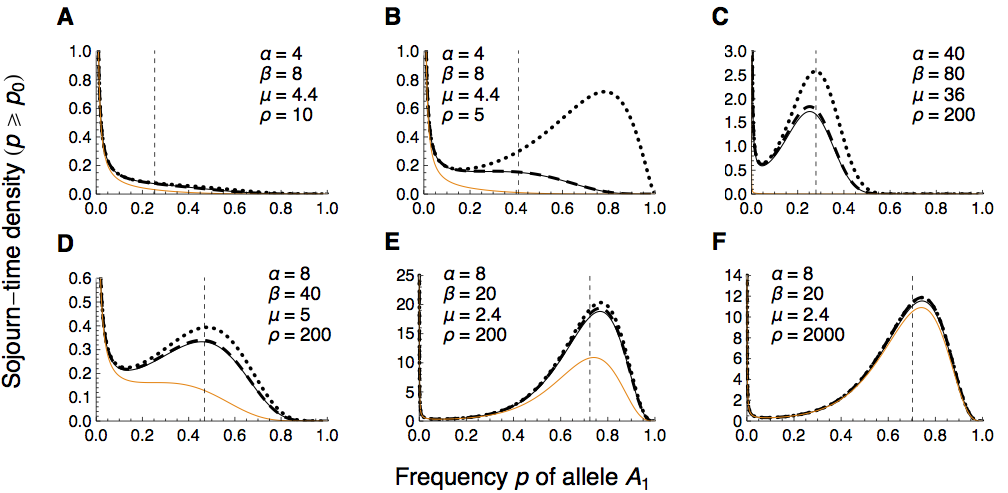}
\end{center}
\caption{
{\textrm{Diffusion approximation to the sojourn-time density of $A_1$ under quasi-linkage equilibrium for a monomorphic continent.}} Comparison of the sojourn-time density (STD) $\ttwoQLE$ (thin black, Eq.\ \ref{eq:STD2QLE}) to the approximation valid for small $p_0$, $\ttwotilQLE$ (dashed black, Eq.\ \ref{eq:STD2QLEApprox} in File \ref{txt:SI}), 
and the one based on the additional assumption of $\rho \gg \max(\alpha, \beta, \mu)$, $\ttwotilQLErho$ (dotted, Eq.\ \ref{eq:STD2QLEApproxRhoLarge}). The STD for the one-locus model, $\ttwotilOLM$, is shown in orange as a reference. Vertical lines give the deterministic frequency $\hat{p}_{+}$ of $A_1$ at the fully-polymorphic equilibrium (computed in File \ref{prc:stochDiffApproxQLENB}). \textbf{(A)} Weak evolutionary forces relative to genetic drift.
\textbf{(B)} As in \textbf{(A)}, but with half the scaled recombination rate $\rho$. 
The assumption of $\rho \gg \max(\alpha, \beta, \mu)$ is violated and hence $\ttwotilQLErho$ is a poor approximation of $\ttwoQLE$. \textbf{(C)} Strong evolutionary forces relative to genetic drift.
The STD has a pronounced mode different from $p = 0$, but $\tilde{t}_{2,\mathrm{QLE},\rho \gg 0}(p; p_0)$ overestimates $\ttwoQLE$ considerably. 
\textbf{(D)} Strong assymmetry in selection coefficients and moderate migration.
As in \textbf{(C)}, the STD has a pronounced mode different from $p = 0$, but $\ttwotilQLErho$ now approximates $\ttwoQLE$ better. \textbf{(E)} Recombination ten times stronger than selection at locus $\bloc$.
\textbf{(F)} As in \textbf{(E)}, but with recombination 100 times stronger than selection at locus $\bloc$.
In \textbf{(A)}--\textbf{(F)}, $p_0=0.005$, which corresponds to an island population of size $N=100$ and a single initial copy of $A_1$. 
}
\label{fig:stdQLE}
\end{figure}

The STD always has a peak at $p = 0$, because most mutations go extinct after a very short time (Figure \ref{fig:stdQLE}). However, for parameter combinations favourable to invasion of $A_1$ (migration weak relative to selection, or selection strong relative to genetic drift), the STD has a second mode at an intermediate allele frequency $p$. Then, allele $A_1$ may spend a long time segregating in the island population before extinction. The second mode is usually close to -- but slightly greater than -- the corresponding deterministic equilibrium frequency (solid black curves in Figures \ref{fig:stdQLE}C--\ref{fig:stdQLE}F for a monomorphic continent).  The peak at this mode becomes shallower as the continental frequency $q_c$ of $B_1$ increases (Figure \ref{fig:stdQLEPolym}).

The effect on the STD of linkage 
is best seen from a comparison to the one-locus model (OLM), for which the STD is given by \cite{Ewens:1979hc} as
\begin{equation}
	\label{eq:STDOLM}
	t_{\mathrm{OLM}}(p; p_0) = \left\{ \begin{array}{ll}
		\toneOLM = 2 e^{2 p \alpha}(1-p)^{2\mu - 1} & \textrm{if $0 \le p \le p_0$},\\
		\ttwoOLM = 2 p_0 e^{2 p \alpha}p^{-1}(1-p)^{2\mu - 1} & \textrm{if $p_0 \le p \le 1$}.
	\end{array} \right.
\end{equation}
If invasion of $A_1$ is unlikely without linkage, but selection at the background locus is strong, even loose linkage has a large effect and causes a pronounced second mode in the STD (compare orange to black curves in Figure \ref{fig:stdQLE}C). In cases where $A_1$ can be established without linkage, the STD of the one-locus model also shows a second mode at an intermediate allele frequency $p$. Yet, linkage to a background polymorphism leads to a much higher peak, provided that selection at the background locus is strong and the recombination rate not too high (Figures \ref{fig:stdQLE}D--\ref{fig:stdQLE}F). Specifically, comparison of Figures \ref{fig:stdQLE}E with \ref{fig:stdQLE}F suggests that the effect of linkage becomes weak if the ratio of the (scaled) recombination rate to the (scaled) selection coefficient at the background locus, $\rho/\beta$, becomes much larger than $\sim$10. In other words, for a given selective advantage $b$ of the beneficial background allele, a weakly beneficial mutation will profit from linkage if it occurs within about $b\times10^3$ map units (centimorgans) from the background locus. This assumes that one map unit corresponds to $r = 0.01$.

An analytical approximation of the STD can be obtained under two simplifying assumptions. The first is that the initial frequency $p_0$ of $A_1$ is small ($p_0$ on the order of $1/(2N_e) \ll 1$). The second concerns the infinitesimal mean $M(p)$ of the change in the frequency of $A_1$:\ assuming that recombination is much stronger than selection and migration, we may approximate Eq.\ \eqref{eq:Mp} by
\begin{equation}
	\label{eq:MpRhoLarge}
	M_{\rho\gg0}(p) = \alpha p(1-p) - \mu p + \frac{\mu(\beta - \mu)}{\rho} p
\end{equation}
for a monomorphic continent. 
The STDs in Eq.\ \eqref{eq:STDQLE} can then be approximated by
\begin{subequations}
	\label{eq:STDQLEApproxRhoLarge}
	\begin{align}
		\tonetilQLErho & = 2 e^{2p\alpha}(1-p)^{\frac{2\mu(\mu - \beta + \rho)}{\rho} - 1} \label{eq:STD1QLEApproxRhoLarge},\\
		\ttwotilQLErho & = 2p_0 e^{2p\alpha} p^{-1}(1-p)^{\frac{2\mu(\mu - \beta + \rho)}{\rho} - 1}. \label{eq:STD2QLEApproxRhoLarge}
	\end{align}
\end{subequations}
Here, we use `$\sim$' to denote the assumption of $p_0$ small, and a subscript `$\rho \gg 0$' for the assumption of $\rho \gg \max(\alpha, \beta, \mu)$.
For a polymorphic continent, expressions analogous to Eqs.\ \eqref{eq:MpRhoLarge} and \eqref{eq:STDQLEApproxRhoLarge} are given in Eqs.\ \eqref{eq:MpPolymContApprox} and \eqref{eq:STDQLEApproxRhoLargePolymCont} \revv{in} File \ref{txt:SI}.

Better approximations than those in Eqs.\ \eqref{eq:STDQLEApproxRhoLarge} and \eqref{eq:STDQLEApproxRhoLargePolymCont} are obtained by making only one of the two assumptions above.
We denote by $\tilde{t}_{1, \mathrm{QLE}}(p; p_0)$ and $\tilde{t}_{2, \mathrm{QLE}}(p; p_0)$ the approximations of the STD in Eq.\ \eqref{eq:STDQLE} based on the assumption $p_0 \ll 1$ (Eqs.\ \ref{eq:STD1QLEApprox} and \ref{eq:STD2QLEApprox} in File \ref{txt:SI}). Alternatively, the approximations obtained from the assumption $\rho \gg \max(\alpha, \beta, \mu)$ in $M(p)$ are called $\toneQLErho$ and $\ttwoQLErho$ (Eq.\ \ref{eq:STDQLERhoLarge}).

In the following, we compare the different approximations to each other and to stochastic simulations.
Conditional on $p_0 = 1/(2N)$, the approximation $\ttwotilQLE$ (Eqs.\ \ref{eq:STD2QLEApprox}) 
is indeed very close to the exact numerical value $\ttwoQLE$ from Eq.\ \eqref{eq:STD2QLE}.
This holds across a wide range of parameter values, as is seen from comparing solid to dashed curves in Figure \ref{fig:stdQLE} (monomorphic continent) and Figure \ref{fig:stdQLEPolym} (polymorphic continent). The accuracy of the approximation 
$\tilde{t}_{2,\mathrm{QLE}, \rho \gg 0}(p; p_0)$ from Eq.\ \eqref{eq:STD2QLEApproxRhoLarge} is rather sensitive to violation of the assumption $\rho \gg \max{(\alpha, \beta, \mu)}$, however (dotted curves deviate from other black curves in Figures \ref{fig:stdQLE}B and \ref{fig:stdQLE}C). The same applies to a polymorphic continent, but the deviation becomes smaller as $q_c$ increases from zero (Figure \ref{fig:stdQLEPolym}A).

Comparison of the diffusion approximation $\ttwotilQLE$ 
to sojourn-time distributions obtained from stochastic simulations shows a very good agreement, except at the boundary $p = 0$. There, the continuous solution of the diffusion approximation is known to provide a suboptimal fit to the discrete distribution (Figures \ref{fig:STDCompSimMonomCont} and \ref{fig:STDCompSimPolymCont}).

Based on the analytical approximations above, we may summarise the effect of weak linkage relative to the one-locus model as follows. For a monomorphic continent, the ratio of $\ttwotilQLErho$ to $\ttwoOLM$ is $\tilde{R} = (1-p)^{-\gamma}$, where $\gamma = 2\mu\left(\beta - \mu\right)/\rho$. The exponent $\gamma$ is a quadratic function of $\mu$ and linear in $\beta$. For weak migration, $\tilde{R} \approx 1-2 (\beta\mu/\rho) \ln(1-p)$, suggesting the following rule of thumb. For the focal allele to spend at least the $\tilde{R}$-fold amount of time at frequency $p$ compared to the case without linkage, 
we require
\begin{equation}
	\label{eq:ruleOfThumb}
		\frac{\beta \mu}{\rho} >\frac{\tilde{R}-1}{-2\ln(1-p)}.
\end{equation}
For example, allele $A_1$ will spend at least twice as much time at frequency $p = 0.5$ (0.8) if $\beta \mu \gtrsim 0.72 \rho$ ($0.31 \rho$). Because we assumed weak migration and QLE, we conducted numerical explorations to check when this rule is conservative, meaning that it does not predict a larger effect of linked selection than is observed in simulations. We found that, first, genetic drift must not dominate, i.e.\ $1 < \alpha, \beta, \mu, \rho$ holds. Second, migration, selection at the background locus, and recombination should roughly satisfy $\mu < \beta/4 < 0.1\rho$. 
\rev{This condition applies only to the validity of Eq.\ \eqref{eq:ruleOfThumb}, which is based on $\ttwotilQLErho$ in Eq.\ \eqref{eq:STD2QLEApproxRhoLarge}. It does not apply to $\ttwotilQLE$, which fits simulations very well if $\rho$ is as low as $1.25\beta$ (Figure \ref{fig:STDCompSimMonomCont}D).} For related observations in different models, see \citet{Slatkin:1975fk} and \citet{Barton:1983fk}.

\subsubsection{Mean absorption time}
\begin{figure}[!ht]
\begin{center}
\includegraphics[width=1\textwidth]{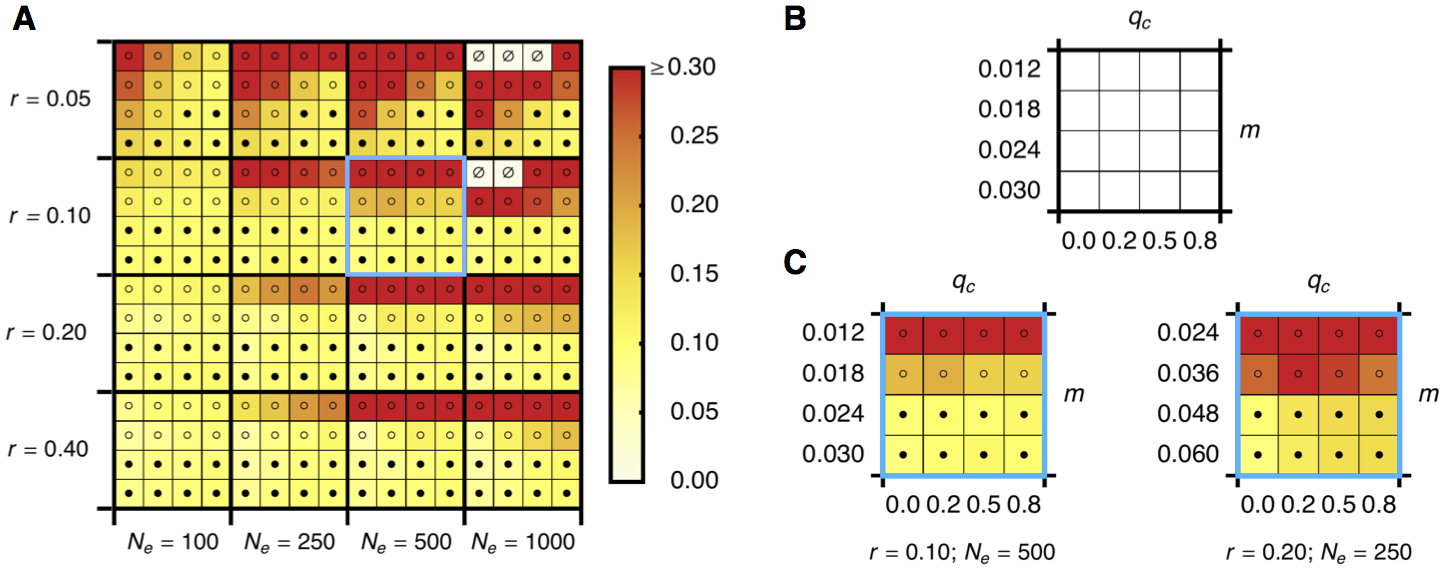}
\end{center}
\caption{
{\textrm{Relative error of  the diffusion approximation to the mean absorption time of $A_1$.}} \textbf{(A)} \rev{The error of $\tbarQLE$ from Eq.\ \eqref{eq:meanAbsTimeQLE} relative to simulations for various parameter combinations.} Squares bounded by bold lines delimit combinations of values of the recombination rate $r$ and the effective population size $N_e$. \rev{Within each of them, values of the migration rate $m$ and the continental frequency $q_c$ of $B_1$ are as shown in \textbf{(B)}}. No negative relative errors were observed. For better resolution, we truncated values above \rev{0.30} (the maximum was 3.396 for $N_e=1000$, $r=0.05$, $m=0.018$, $q_c=0.0$). Empty (filled) circles indicate that the marginal one-locus equilibrium $\tilde{E}_\bloc$ is unstable (stable) and $A_1$ can (not) be established under deterministic dynamics. Parameter combinations for which simulations were too time-consuming are \rev{indicated} by $\emptyset$. Selection coefficients are $a=0.02$ and $b=0.04$. \rev{\textbf{(C)} The left plot corresponds to the square in \textbf{(A)} that is framed in blue. The right plot shows the fit of the diffusion approximation to simulations conducted with unscaled parameters twice as large, and $N_e$ half as large, as on the left side. Scaled paramters are equal on both sides. As expected, the diffusion approximation is worse on the right side.} Simulations were as described in Methods. See Table \ref{tab:summaryTableSTDAndMAT} for numerical values.
}
\label{fig:meanabstimeQLEWF}
\end{figure}

The mean absorption time is obtained by numerical integration of the STD as outlined in Methods. Comparison to stochastic simulations shows that the diffusion approximation $\tbarQLE$ from Eq.\ \eqref{eq:meanAbsTimeQLE} is fairly accurate: the absolute relative error is less than 15\%, provided that the QLE assumption is not violated and migration not too weak (Figure \ref{fig:meanabstimeQLEWF}). 

Given the approximations to the STD derived above, various degrees of approximation are available for the mean absorption time, too. Their computation is less prone to numerical issues than that of the exact expressions. 
%
%
Extensive numerical computations showed that if $p_0 = 1/(2N)$ and $N_e = N$, the approximations based on the assumption of $p_0$ small ($\tbartilQLE$ and $\tilde{\bar{t}}_{\mathrm{QLE}, \rho \gg 0}$ as given in Eqs.\ \ref{eq:meanAbsTimeQLEApprox} and \ref{eq:meanAbsTimeQLEApproxRhoLarge}) provide an excellent fit to their more exact counterparts ($\tbarQLE$ and $\bar{t}_{\mathrm{QLE}, \rho \gg 0}$ in Eqs.\ \ref{eq:meanAbsTimeQLE} and \ref{eq:meanAbsTimeQLERhoLarge}, respectively). Across a wide range of parameter values, the absolute relative error never exceeds 1.8\% (Figures \ref{fig:relErrDiffApproxComb}A and \ref{fig:relErrDiffApproxComb}C).
In contrast, the approximation based on the assumption of $\rho \gg 0$, $\bar{t}_{\mathrm{QLE}, \rho \gg 0}$, is very sensitive to violations of this assumption. For large effective population sizes and weak migration, the relative error becomes very high if recombination is not strong enough (Figure \ref{fig:relErrDiffApproxComb}B). 

\begin{figure}[!ht]
\begin{center}
\includegraphics[width=1\textwidth]{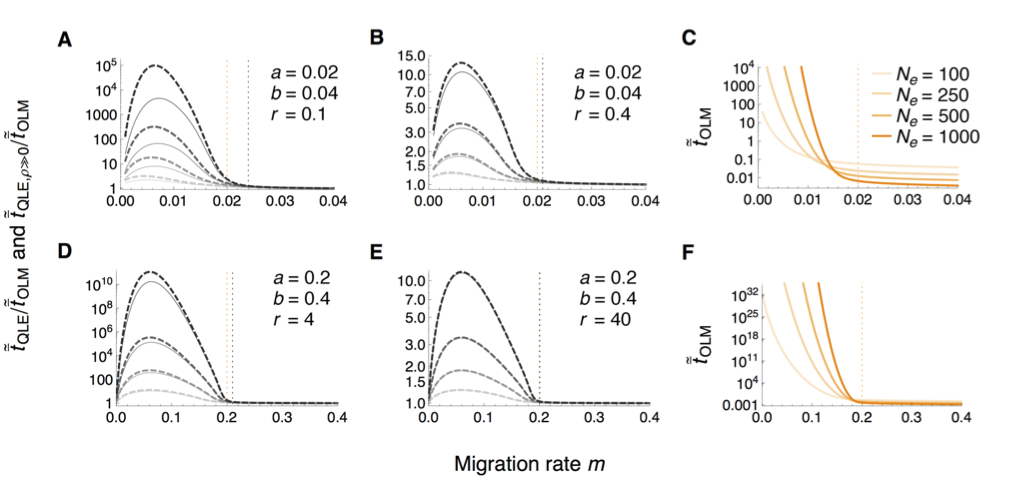}
\end{center}
\caption{
{\textrm{Mean absorption time of $A_1$ under quasi-linkage equilibrium relative to the one-locus model (OLM).}} In panels \textbf{(A)}, \textbf{(B)}, \textbf{(D)} and \textbf{(E)}, thin solid curves show the ratio $\tbartilQLE / \tbartilOLM$ and thick dashed curves $\tbartilQLErho / \tbartilOLM$, as a function of the migration rate $m$. The effective population size $N_e$ increases from light to dark grey, taking values of 100, 250, 500, and 1000. Vertical lines denote the migration rate below which $A_1$ can invade in the deterministic one-locus (orange) and two-locus (black) model. \textbf{(A)} Recombination is too weak for the assumption $\rho \gg \max{(\alpha, \beta, \mu)}$ to hold. 
\textbf{(B)} As in \textbf{(A)}, but with recombination four times stronger.
\textbf{(D)} Evolutionary forces -- other than drift -- are ten times stronger than in \textbf{(B)}. 
\textbf{(E)} As in \textbf{(D)}, but with recombination ten times stronger. Panels \textbf{(C)} and \textbf{(F)} show the mean absorption time \rev{(in multiples of $2N_e$)} under the one-locus model for the respective row. 
For $m$ close to 0, numerical procedures are unstable and we truncated the curves. As $m$ converges to 0, $\tbartilQLE / \tbartilOLM$ and $\tbartilQLErho / \tbartilOLM$ are expected to approach unity, however.
}
\label{fig:meanabstimeQLEOLM}
\end{figure}

The effect of linkage is again demonstrated by a comparison to the one-locus model. If selection is strong relative to recombination, 
the mean absorption time with linkage, $\tbartilQLE$, is increased by several orders of magnitude compared to the one-locus case, $\tilde{\bar{t}}_{\mathrm{OLM}}$ (Figures \ref{fig:meanabstimeQLEOLM}A and \ref{fig:meanabstimeQLEOLM}D; Table \ref{tab:ratioTAbsQLEToTAbsOLM}). The effect is reduced, but still notable, when the recombination rate becomes substantially higher than ten times the strength of selection in favour of the beneficial background allele, i.e.\ $\rho/\beta \gg 10$ (Figures \ref{fig:meanabstimeQLEOLM}B and \ref{fig:meanabstimeQLEOLM}E). Importantly, large ratios of $\tbartilQLE/\tilde{\bar{t}}_{\mathrm{OLM}}$ are not an artefact of $\tilde{\bar{t}}_{\mathrm{OLM}}$ being very small, as Figures \ref{fig:meanabstimeQLEOLM}C and \ref{fig:meanabstimeQLEOLM}F confirm. Moreover, $\tbartilQLE/\tilde{\bar{t}}_{\mathrm{OLM}}$ is maximised at intermediate migration rates: for very weak migration, $A_1$ has a fair chance of surviving for a long time even without linkage ($\tilde{\bar{t}}_{\mathrm{OLM}}$ is large); for very strong migration, $\tilde{\bar{t}}_{\mathrm{OLM}}$ and $\tbartilQLE$ both tend to zero and $\tbartilQLE/\tilde{\bar{t}}_{\mathrm{OLM}}$ approaches unity. 

As expected from deterministic theory (\citeauthor{Buerger:2011uq} \citeyear{Buerger:2011uq}; see also section 3 in File) 
and invasion probabilities calculated above, the mean absorption time decreases as a function of the migration rate $m$ (Figure \ref{fig:meanAbsTimeQLEWFrLargeCombM}). A noteworthy interaction exists between $m$ and the effective population size $N_e$.
For small $m$, the mean absorption time increases with $N_e$, whereas for large $m$, it decreases with $N_e$.
Interestingly, the transition occurs at a value of $m$ lower than the respective critical migration rate below which $A_1$ can invade in the deterministic model (Figure \ref{fig:meanAbsTimeQLEWFrLargeCombM}). Hence, there exists a small range of intermediate values of $m$ for which deterministic theory suggests that $A_1$ will invade, but the stochastic model suggests that survival of $A_1$ lasts longer in island populations of small rather than large effective size. Similar, but inverted, relations hold for the dependence of the mean absorption time on the selective advantage $a$ of allele $A_1$ and $N_e$ (Figure \ref{fig:meanAbsTime_effectOfA_variousqC}).

For the parameter ranges we explored, the mean absorption time decreases with increasing continental frequency $q_c$ of $B_1$. 
As for the invasion probabilities, competition against a fitter resident population has a negative effect on maintenance of the focal mutation $A_1$. For a given recombination rate, the effect depends on the relative strength of migration and selection, though: increasing $q_c$ from 0 to 0.8 will decrease the mean absorption time by a considerable amount only if $m$ is low or $a$ is large enough; otherwise, genetic drift dominates (Figure \ref{fig:meanAbsTime_effectOfMandA_variousqC}). This effect is more pronounced for weak than for strong recombination (Figure \ref{fig:meanAbsTimeQLEWFrLargeqCComb}). 

So far, we assumed that the initial frequency of $A_1$ is small, i.e.\ $p_0 = 1/(2N)$, and that $N_e = N$. In many applications, $N_e < N$ holds and hence $1/(2N) < 1/(2N_e)$. Approximations based on the assumption of $p_0$ being small, i.e.\ on the order of $1/(2N_e)$ or smaller, will then cause no problem. However, $N_e > N$ may hold in certain models, e.g.\ with spatial structure \citep{Whitlock:1997qf}, and $p_0 = 1/(2N)$ may be much greater than $1/(2N_e)$. We therefore investigated the effect of violating the assumption of $p_0 \le 1/(2N_e)$. For this purpose, we fixed the initial frequency at $p_0 = 0.005$ (e.g.\ a single copy of $A_1$ in a population of actual size $N = 100$) and then assessed the relative error of our approximations for various $N_e \ge 100$. As expected, the approximate mean absorption times based on the assumption of $p_0$ small ($\tbartilQLE$ and $\tilde{\bar{t}}_{\mathrm{QLE}, \rho \gg 0}$) deviate further from their exact conterparts ($\tbarQLE$ and $\bar{t}_{\mathrm{QLE}, \rho \gg 0}$, respectively) as $N_e$ increases from 100 to $10^4$ (Figures \ref{fig:relErrDiffApproxComb}D and \ref{fig:relErrDiffApproxComb}F). 
For strong migration, the relative error tends to be negative, while it is positive for weak migration (blue versus red boxes in Figures \ref{fig:relErrDiffApproxComb}D and \ref{fig:relErrDiffApproxComb}F). The assumption of $\rho \gg \max{(\alpha, \beta, \mu)}$ in $M(p)$ does not lead to any further increase of the relative error, though (Figure \ref{fig:relErrDiffApproxComb}E). 
Moreover, violation of $p_0 \le 1/(2N_e)$ has almost no effect on the ratio of the two-locus to the one-locus absorption time, $\tbarQLE/\bar{t}_{\mathrm{OLM}}$ (compare Table \ref{tab:ratioTAbsQLEToTAbsOLMp0Fixed} to Table \ref{tab:ratioTAbsQLEToTAbsOLM}).

\subsubsection{Invasion-effective migration rate}
Comparison of the sojourn-time densities given in Eqs.\ \eqref{eq:STDOLM} and \eqref{eq:STDQLEApproxRhoLarge} suggests that if $\mu$ in the one-locus model is replaced by $\mu_{e} = \mu(\mu - \beta + \rho)/\rho$, one obtains the STD for the two-locus model. Hence, $\mu_e$ denotes the scaled migration rate in a one-locus model such that allele $A_1$ has the same sojourn properties as it would have if it arose in linkage (decaying at rate $\rho$) to a background polymorphism maintained by selection against migration at rate $\mu$. \rev{In other words, if the assumptions stated above hold, we may use single-locus migration--selection theory, with $\mu$ replaced by $\mu_e$, to describe two-locus dynamics.} Transforming from the diffusion to the natural scale, we therefore defined an invasion-effective migration rate as
\begin{equation}
	\label{eq:mEffQLE}
	m_e = m \frac{m + r - b}{r},
\end{equation}
which, for small $m$, is approximately
\begin{equation}
	\label{eq:mEffQLEweakMig}
	\tilde{m}_e = m\,\left(1-\frac{b}{r}\right)
\end{equation}
(Figure \ref{fig:mEffSojourn}A). Note that $m_e$ and $\tilde{m}_e$ are non-negative only if $r \ge b - m$ and $r \ge b$, respectively. As we assumed quasi-linkage equilibrium 
in the derivation, these conditions do not impose any further restriction.

\citet{Petry:1983uq} previously derived an effective migration rate for a neutral site linked to a selected site. In our notation, it is given by
\begin{equation}
	\label{eq:mEffPetry}
	m_{e}^{(P)} = m\left(1+\frac{b}{r}\right)^{-1} = m \frac{r}{b + r}
\end{equation}
(see \citealt{Bengtsson:1985fk}, and \citealt{Barton:1986fk} for an extension of the concept). Petry obtained this approximation
by comparing the moments of the stationary allele-frequency distribution for the two-locus model to those for the one-locus model. He assumed that selection and recombination are strong relative to migration and genetic drift. To first order in $r^{-1}$, i.e.\ for loose linkage, Petry's $m_e^{(P)}$ is equal to our $\tilde{m}_e$ in Eq.\ \eqref{eq:mEffQLEweakMig}.
As we derived $\tilde{m}_e$ under the assumption of QLE, 
convergence of $m^{(P)}_{e}$ to $\tilde{m}_e$ is reassuring.
Effective gene flow decreases with the strength of background selection $b$, but increases with the recombination rate $r$ (Figures \ref{fig:mEffSojourn}B and \ref{fig:mEffSojourn}C).

\subsection{Long-term effect on linked neutral variation}
Selection maintaining genetic differences across space impedes the homogenising effect of gene flow at closely linked sites \citep{Bengtsson:1985fk, Barton:1986fk}. This has consequences for the analysis of sequence or marker data, as patterns of neutral diversity may reveal the action of recent or past selection at nearby sites \citep{Maynard-Smith:1974kx,Barton:1998fk,Kaplan:1989ly,Takahata:1990ys}. 
We investigated the impact of a two-locus polymorphism contributing to local adaptation on long-term patterns of linked genetic variation. For this purpose, we included a neutral locus \textsf{C} with alleles $C_1$ and $C_2$. Allele $C_1$ segregates on the continent at a constant frequency $n_c$ ($0 \le n_c < 1$), for example at drift--mutation equilibrium. This may require that the continental population is very large\rev{, such that extinction or fixation of $C_1$ occur over sufficiently long periods of time compared to the events of interest on the island}. 
The neutral locus is on the same chromosome as \textsf{A} and \textsf{B}, either to the left (\configL), in the middle (\configM) or to the right (\configR) of the two selected loci (without loss of generality, \textsf{A} is to the left of \textsf{B}). We denote the recombination rate between locus $X$ and $Y$ by $r_{XY}$, where $r_{XY} = r_{YX}$, and assume that the recombination rate is additive. For example, if the configuration is \configM, we set $r_{\aloc\bloc} = r_{\aloc\cloc} + r_{\cloc\bloc}$.

Unless linkage to one of the selected loci is complete, under deterministic dynamics, allele $C_1$ will reach \rev{the equilibrium frequency} $\hat{n} = n_c$ on the island, independently of its initial frequency on the island. Recombination affects only the rate of approach to this equilibrium, not its value. 
We focus on the case where the continent is monomorphic at locus \textsf{B} ($q_c = 0$). Selection for local adaptation acts on loci \textsf{A} and \textsf{B}, and migration--selection equilibrium will be reached at each of them (section 6 in File \ref{txt:SI}). 
Gene flow from the continent will be effectively reduced in their neighbourhood on the chromosome. 
Although the expected frequency of $C_1$ remains $n_c$ throughout, drift will cause variation around this mean to an extent that depends on the position of \textsf{C} on the chromosome. It may take a long time for this drift--migration equilibrium to be established, but the resulting signal should be informative for inference.

To investigate the effect of selection at two linked loci, we employed the concept of an effective migration rate according to \citet{Bengtsson:1985fk}, \citet{Barton:1986fk}, and \citet{Kobayashi:2008fk}. As derived in section 8 of File \ref{txt:SI} and File \ref{prc:determEffMigRate}, 
for continuous time and weak migration, the effective migration rates for the three configurations are
\begin{subequations}
	\label{eq:effMigRatesNeutr}
	\begin{align}
		m_{e}^{\cloc\aloc\bloc} & = m \frac{r_{\cloc\aloc} \left(a + r_{\cloc\bloc}\right)}{\left(a + r_{\cloc\aloc}\right)\left(a + b + r_{\cloc\bloc}\right)}\label{eq:effMigRateNeutrL}, \\
		m_{e}^{\aloc\cloc\bloc} & = m \frac{r_{\aloc\cloc}  r_{\cloc\bloc}}{\left(a + r_{\aloc\cloc}\right)\left(b + r_{\cloc\bloc}\right)}\label{eq:effMigRateNeutrM}, \\
		m_{e}^{\aloc\bloc\cloc} & = m \frac{r_{\bloc\cloc} \left(b + r_{\aloc\cloc}\right)}{\left(b + r_{\bloc\cloc}\right)\left(a + b + r_{\aloc\cloc}\right)}\label{eq:effMigRateNeutrR}.
	\end{align}
\end{subequations}
We note that $m_{e}^{\aloc\cloc\bloc}$ has been previously derived \cite[Eq.\ 4.30]{Buerger:2011uq}. From Eq.\ \eqref{eq:effMigRatesNeutr}, we defined the effective migration rate experienced at a neutral site as
\begin{equation}
	\label{eq:effMigNeutr}
	m_e^{(n)} = \left \{
		\begin{array}{ll}
			m_{e}^{\cloc\aloc\bloc} & \textrm{if\ \configL\ holds}, \\
			m_{e}^{\aloc\cloc\bloc} & \textrm{if\ \configM\ holds}, \\
			m_{e}^{\aloc\bloc\cloc} & \textrm{if\ \configR\ holds}.
		\end{array}
	\right.
\end{equation}

Equation \eqref{eq:effMigNeutr} subsumes the effect on locus \textsf{C} of selection at loci \textsf{A} and \textsf{B}. It can be generalised to an arbitrary number of selected loci. Let $\aloc_i$ $(i = 1\dots I)$ and $\bloc_j$ $(j = 1\dots J)$ be the $i$th and $j$th locus to the left and right of the neutral locus, respectively. \rev{We} find that the effective migration rate at the neutral locus is
\begin{equation}
	\label{eq:effMigNeutrGeneral}
	m_e^{(n)} = m \left [ \prod_{i=1}^{I} \left( 1 + \frac{a_i}{\sum_{k=1}^{i-1} a_k + r_{\aloc_{i}}} \right)^{-1} \right]  \times \left [ \prod_{j=1}^{J} \left( 1 + \frac{b_j}{\sum_{k=1}^{j-1} b_k + r_{\bloc_{j}}} \right)^{-1} \right],
\end{equation}
where $a_i$ ($b_j$) is the selection coefficient at locus $\aloc_i$ ($\bloc_j$), and $r_{\aloc_{i}}$ ($r_{\bloc_{j}}$) the recombination rate between the neutral locus and $\aloc_i$ ($\bloc_j$).
Each of the terms in the round brackets in Eq.\ \eqref{eq:effMigNeutrGeneral} is reminiscent of \citeauthor{Petry:1983uq}'s \citeyearpar{Petry:1983uq} effective migration rate for a neutral linked site (Eq.\ \ref{eq:mEffPetry}). For weak linkage, these terms are also similar to the invasion-effective migration rate experienced by a weakly beneficial mutation (Eq.\ \ref{eq:mEffQLEweakMig}). This suggests that the effective migration rate experienced by a linked neutral site is approximately the same as that experienced by a linked weakly beneficial mutation, which corroborates the usefulness of Eq.\ \eqref{eq:effMigNeutrGeneral}. 
In the following, we study different long-term properties of the one-locus drift-migration model by substituting effective for actual migration rates.

\subsubsection{Mean absorption time}
Suppose that $C_1$ is absent from the continent ($n_c = 0$), but present on the island as a de-novo mutation. Although any such mutant allele is doomed to extinction, recurrent mutation may lead to a permanent influx and, at mutation--migration equilibrium, to a certain level of neutral differentiation between the continent and the island. Here, we ignore recurrent mutation and focus on the fate of a mutant population descending from a single copy of $C_1$. We ask how long it will survive on the island, given that a migration--selection polymorphism is maintained at equilibrium at both selected loci in the background (\textsf{A}, \textsf{B}). Standard diffusion theory 
predicts that the mean absorption (extinction) time of $C_1$ is approximately $\tilde{\bar{t}}_{\mathrm{neut}} = N_e^{-1}\int_{1/(2N)}^{1} n^{-1}(1-n)^{2\mu - 1}dn$ \cite[pp.\ 171--175]{Ewens:1979hc}. We replace the scaled actual migration rate $\mu$ by $\mu_e^{(n)} = 2 N_e m_{e}^{(n)}$, with $m_{e}^{(n)}$ from Eq.\ \eqref{eq:effMigNeutr}. This assumes that the initial frequency of $C_1$ on the island is $n_0 = 1/(2N)$ and that $N_e = N$. 
For moderately strong migration ($\mu_{e}^{(n)} \approx 1$), $\tilde{\bar{t}}_{\mathrm{neut}}$ is of order $\log(2 N_e)$, meaning that $C_1$ will on average remain in the island population for a short time. However, if locus \textsf{C} is tightly linked to one of the selected loci, or if configuration \configM\ applies and \textsf{A} and \textsf{B} are sufficiently close, 
the mean absorption time of $C_1$ is strongly elevated (Figure \ref{fig:meanabstimeNeutr}).

\subsubsection{Stationary distribution of allele frequencies}
\begin{figure}[!ht]
\begin{center}
\includegraphics[width=1\textwidth]{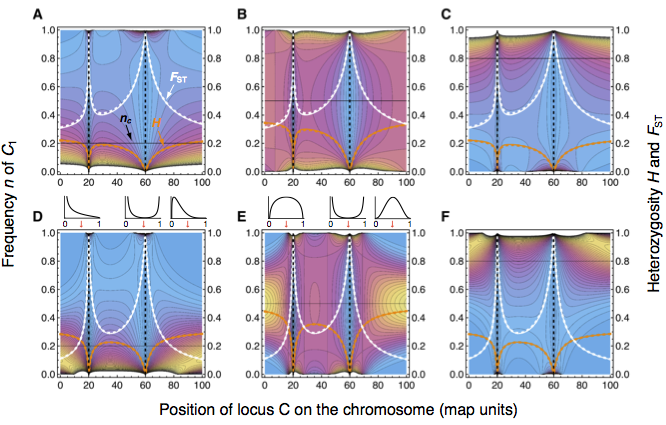}
\end{center}
\caption{
{\textrm{The effect of linked selection on neutral diversity and population divergence.}} Shown are top views of the stationary allele frequency distribution on the island for a neutral biallelic locus \textsf{C} linked to two selected sites at 20 (\textsf{A}) and 60 (\textsf{B}) map units from the left end of the chromosome. Density increases from light blue to yellow (high peaks were truncated for better resolution). Orange and white curves show the expected diversity (heterozygosity $H$) and population divergence ($F_{\mathrm{ST}}$) as a function of the position of the neutral site. Solid curves use exact, numerically computed values of the effective migration rate and dashed ones use the approximations given in Eq.\ \eqref{eq:effMigNeutr}. One map unit (cM) corresponds to $r=0.01$ and the effective size of the island population is $N_e = 100$. The continental frequency $n_c$ of allele $C_1$ is indicated by a horizontal black line and, from left to right, equal to 0.2, 0.5 and 0.8. \textbf{(A)}--\textbf{(C)} Relatively strong drift, and weak migration compared to selection: $\alpha = 4$, $\beta = 80$, $\mu = 2$. \textbf{(D)}--\textbf{(F)} Relatively weak drift, and migration on the same order of magnitude as selection at locus \textbf{A}: $\alpha = 40$, $\beta = 800$, $\mu = 48$. Note that $H$ is sensitive to $n_c$, whereas  $F_{\mathrm{ST}}$ is not. On top of panels \textbf{(D)} and \textbf{(E)} allele frequency distributions are shown that result from taking vertical slices at positions indicated by red arrows (15, 59 and 90 map units).
}
\label{fig:statAllDistNeutr}
\end{figure}

In contrast to above, assume that $C_1$ is maintained at a constant frequency $n_c \in (0,1)$ on the continent. Migrants may therefore carry both alleles, and genetic drift and migration will lead to a stationary distribution of allele frequencies given by
\begin{equation*}
	\phi(n) = \frac{\Gamma\left(2\mu\right)}{\Gamma\left(2\mu n_c\right) \Gamma\left(2\mu\left[1-n_c\right]\right)}\; n^{2\mu n_c -1}(1-n)^{2\mu (1-n_c) -1},
\end{equation*}
where $\Gamma(x)$ is the Gamma function \cite[p.\ 239--241]{Wright:1940fk}. As above, we replace $\mu$ by $\mu_e^{(n)}$ to account for the effect of linked selection. The mean of the distribution $\phi(n)$ is $n_c$, independently of $\mu_e$, whereas the stationary variance is $\mathrm{var}(n) = n_c (1-n_c)/(1+2\mu_e)$ \citep{Wright:1940fk}. 
The expected 
heterozygosity is $H = 4\mu_e n_c (1-n_c)/ (1 + 2\mu_e)$, and the divergence from the continental population is $F_{\mathrm{ST}} = \mathrm{var}(n)/\left[ n_c (1-n_c) \right] = 1/(1+2\mu_e)$ (see File \ref{prc:stochDiffNeutrVar} for details). Depending on the position of the neutral locus, $\phi(n)$ may change considerably in shape, for example, from L- to U- to bell-shaped (Figure \ref{fig:statAllDistNeutr}). The pattern of $\phi(n)$, $H$ and $F_{\mathrm{ST}}$ along the chromosome reveals the positions of the selected loci, and their rate of change per base pair contains information about the strength of selection if the actual migration rate is known.

\subsubsection{Rate of coalescence}
As a third application, we study the rate of coalescence for a sample of size two taken from the neutral locus \textsf{C}, assuming that migration--selection equilibrium has been reached a long time ago at the selected loci \textsf{A} and \textsf{B}. We restrict the analysis to the case of strong migration compared to genetic drift, for which results by \citet{Nagylaki:1980uq} (forward in time) and \citet{Notohara:1993kx} (backward in time) apply \cite[see][ for a detailed review]{Wakeley:2009kx}. The strong-migration limit follows from a separation of time scales: going back in time, migration spreads the lineages on a faster time scale, whereas genetic drift causes lineages to coalesce on a slower one.

For a moment, let us assume that there are two demes of size $N_1$ and $N_2$, and denote the total number of diploids by $\bar{N} = N_1 + N_2$. We define the relative deme size $c_i = N_i / \bar{N}$ and let the backward migration rates $m_1$ and $m_2$ denote the fractions of individuals in deme 1 and 2 in the current generation that were in deme 2 and 1 in the previous generation, respectively. The strong-migration limit then requires that $N_i m_i = \bar{N} c_i m_i$ is large \cite{Wakeley:2009kx}. Importantly, the relative deme sizes $c_i$ are constant in the limit of $\bar{N} \rightarrow \infty$. Under these assumptions, it can be shown that the rate of coalescence for a sample of two is independent of whether the two lineages were sampled from the same or different demes. The rate of coalescence is given by
\begin{equation}
	\label{eq:coalRateNeutr}
	G = \frac{m_2^2}{(m_1 + m_2)^2}\,\frac{1}{c_1} + \frac{m_1^2}{(m_1 + m_2)^2}\, \frac{1}{c_2}
\end{equation}
\cite[p.\ 193]{Wakeley:2009kx}. The coalescent-effective population size is defined as the actual total population size times the inverse of the rate of coalescence, $N_{e}^{\mathrm{(coal)}} = \bar{N}/G$ \citep{Sjodin:2005ys}.

In our context, we substitute $m_e^{(n)}$ from Eq.\ \eqref{eq:effMigNeutr} for $m_1$ in $G$. To be consistent with the assumption of continent--island migration -- under which we studied the migration--selection dynamics at \textsf{A} and \textsf{B} -- we require $N_2 \gg N_1$ and $m_2 \ll m_1$. This way, the assumptions of $N_1 m_1$ and $N_2 m_2$ being large can still be fulfilled. However, note that $m_2 \ll m_1$ does not automatically imply $m_2 \ll m_e^{(n)}$; depending on the strength of selection and recombination, $m_e^{(n)}$ may become very small. Hence, in applying the theory outlined here, one should bear in mind that the approximation may be misleading if $m_e^{(n)}$ is small (for instance, if locus \textsf{C} is tightly linked to either \textsf{A} or \textsf{B}). 
The neutral coalescent rate $G$ is strongly increased in the neighbourhood of selected sites; accordingly, $N_e^{\mathrm{(coal)}}$ is increased (Figure \ref{fig:coalRateAndNeNeutr}). Reassuringly, this pattern parallels those for linked neutral diversity and divergence in Figure \ref{fig:statAllDistNeutr}.

\section{Discussion}

We have provided a comprehensive analysis of the fate of a locally beneficial mutation that arises in linkage to a previously established migration--selection polymorphism. 
In particular, we obtained explicit approximations to the invasion probability. These reveal the functional dependence on the key parameters and substitute for time-consuming simulations. Further, we found accurate approximations to the mean extinction time, showing that a unilateral focus on invasion probabilities yields an incomplete understanding of the effects of migration and linkage. Finally, we derived the effective migration rate experienced by a neutral or weakly beneficial mutation that is linked to arbitrarily many migration--selection polymorphisms. This opens up a genome-wide perspective of local adaptation and establishes a link to inferential frameworks.

\subsection{Insight from stochastic modelling}
Previous theoretical studies accounting for genetic drift in the context of polygenic local adaptation with gene flow were mainly simulation-based \citep{Yeaman:2011fk,Feder:2012dq,Flaxman:2013fk} or did not model recombination explicitly (\citealt{
Lande:1984ul,Lande:1985bh,Barton:1987kx, Rouhani:1987ly, Barton:1991ys}; but see \citealt{Barton:1986fk}). 
Here, we used stochastic processes to model genetic drift and to derive explicit expressions that provide an alternative to simulations. We distinguished between the stochastic effects due to initial rareness of a de-novo mutation on the one hand and the long-term effect of finite population size on the other.

For a two-locus model with a steady influx of maladapted genes, we found an implicit condition for invasion of a single locally beneficial mutation linked to the background locus (Eq.\ \ref{eq:condNonExtGener}). This condition is valid for arbitrary fitnesses, i.e.\ any regime of dominance or epistasis. It also represents an extension to the case of a panmictic population in which the background polymorphism is maintained by overdominance, rather than migration--selection balance \citep{Ewens:1967fk}. Assuming additive fitnesses, we derived simple explicit conditions for invasion in terms of a critical migration or recombination rate (Eqs.\ \ref{eq:mCrit5Main} or \ref{eq:rCrit}, respectively). Whereas these results align with deterministic theory \citep{Buerger:2011uq}, additional quantitative and qualitative insight 
emerged from studying invasion probabilities and extinction times. Specifically, invasion probabilities derived from a two-type branching process (Eqs.\ \ref{eq:bpSol} and \ref{eq:approxInvProbAdd}) capture the ambiguous role of recombination breaking up optimal haplotypes on the one hand and creating them on the other. Diffusion approximations to the sojourn and mean absorption time shed light on the long-term effect of finite population size. A comparison between the dependence of invasion probabilities and extinction times on migration and recombination rate revealed important differences (discussed further below). Deterministic theory fails to represent such aspects, and simulations only provide limited understanding of functional relationships.


Recently, \citet{Yeaman:2013uq} derived an ad-hoc approximation of the invasion probability, using the so-called `splicing approach' \citep{Yeaman:2011uq}. There, the leading eigenvalue of the 
appropriate Jacobian  \citep{Buerger:2011uq} is taken as a proxy for the selection coefficient and inserted into  \citeauthor{Kimura:1962fk}'s \citeyearpar{Kimura:1962fk} formula for the one-locus invasion probability in a panmictic population. \rev{\citeauthor{Yeaman:2013uq}'s \citeyearpar{Yeaman:2013uq} method provides a fairly accurate approximation to the invasion probability if $A_1$ initially occurs on the beneficial background $B_1$ (at least for tight linkage). However, it does not describe the invasion probability of an average mutation (Figure \ref{fig:invProbCompAnalytApprox}), and hence does not predict the existence of a non-zero optimal recombination rate. As a consequence, \citeauthor{Yeaman:2013uq}'s \citeyearpar{Yeaman:2013uq} conclusion that physically linked selection alone is of limited importance for the evolution of clustered architectures is likely conservative, because it is based on an approximation that inflates the effect of linked selection.}


\subsection{Non-zero optimal recombination rate}
We have shown that the average invasion probability of a linked beneficial mutation can be maximised at a non-zero recombination rate ($r_{\mathrm{opt}}>0$). Equation \eqref{eq:aCritOptRecombRateGeneric} provides a general condition for when this occurs. With additive fitnesses, the local advantage of the focal mutation must be above a critical value (Eq.\ \ref{eq:aCritOptRecombRate}). Otherwise, the invasion probability is maximised at $r_{\mathrm{opt}} = 0$.

Existence of a non-zero optimal recombination rate in the absence of epistasis and dominance is noteworthy. For a panmictic population in which the polymorphism at the background locus is maintained by overdominance, \citet{Ewens:1967fk} has shown that the optimal recombination rate may be non-zero, but this requires epistasis. In the context of migration, the existence of $r_{\mathrm{opt}}>0$ has been noticed \rev{and discussed} in a simulation study \cite[Figure 5 of][]{Feder:2012dq}, but no analytical approximation or explanation has been available that captures this feature. In principle, $r_{\mathrm{opt}} > 0$ suggests that the genetic architecture of polygenic adaptation may evolve such as to optimise the recombination rates between loci harbouring adaptive mutations. 
Testing this prediction requires modifier-of-recombination theory \cite[e.g.][]{Otto:1997fk,Martin:2006zr,Roze:2006uq,Kermany:2012ys}. While we expect evolution towards the optimal recombination rate in a deterministic model \citep{Lenormand:2000uq}, it will be important to determine if and under which conditions this occurs in a stochastic model, and what the consequences for polygenic adaptation are. 

For instance, in a model with two demes and a quantitative trait for which the fitness optima are different in the two demes, \citet{Yeaman:2011fk} have shown that mutations contributing to adaptive divergence in the presence of gene flow may cluster with respect to their position on the chromosome. Moreover, architectures with many weakly adaptive mutations tended to become replaced by architectures with fewer mutations of larger effect. Although our migration model is different, existence of a non-zero optimal recombination rate suggests that there might be a limit to the degree of clustering of locally adaptive mutations. \rev{It is worth recalling that our result of $r_{\mathrm{opt}}>0$ applies to the \emph{average} invading mutation (black curves in Figures \ref{fig:invProbFuncR} and \ref{fig:invProbFuncRApprox}). For any \emph{particular} mutation that arises on the beneficial background, $r=0$ is (almost) always optimal (blue curves in Figures \ref{fig:invProbFuncR} and \ref{fig:invProbFuncRApprox}; see Figure \ref{fig:invProbFuncM}D for an exception).}


\subsection{Long-term dynamics of adaptive divergence}
Finite population size on the island eventually leads to extinction of a locally beneficial mutation even after successful initial establishment. This is accounted for neither by deterministic nor branching-process theory. 
Employing a diffusion approximation, we have shown that linkage of the focal mutation to a migration--selection polymorphism  can greatly increase the time to extinction and thus alter the long-term evolutionary dynamics. In such cases, the time scale of extinction may become similar to that on which mutations occur. This affects the rate at which an equilibrium between evolutionary forces is reached. 
We provided a rule of thumb for when the time spent by the focal allele at a certain frequency exceeds a given multiple of the respective time without linkage. Essentially, the product of the background selection coefficient times the migration rate must be larger than a multiple of the recombination rate (Eq.\ \ref{eq:ruleOfThumb}). 

The effect of linked selection can also be expressed in terms of an invasion-effective migration rate (Eqs.\ \ref{eq:mEffQLE} and \ref{eq:mEffQLEweakMig}). 
Both our rule of thumb and the formula for the effective migration rate provide a means of quantifying the importance of linkage to selected genes in the context of local adaptation. In practice, however, their application requires accurate estimates of the recombination map, the selective advantage of the beneficial background allele, and the actual migration rate.

\subsection{A non-trivial effect of gene flow}
Our stochastic modelling allows for a more differentiated understanding of the role of gene flow in opposing adaptive divergence. Whereas deterministic theory specifies a critical migration rate beyond which a focal mutation of a given advantage cannot be established \cite[][; see also Figures \ref{fig:stabilityEBPolymMR} and \ref{fig:stabilityEBPolymMqC}]{Buerger:2011uq}, the potential of invasion is far from uniform 
if migration is below this critical value (Figures \ref{fig:invProbFuncM}, \ref{fig:invProbFuncMPolymCont}, and \ref{fig:meanAbsTimeQLEWFrLargeCombM}). For instance, we may define the relative advantage of linkage to a migration--selection polymorphism as the ratio of the quantity of interest \emph{with} a given degree of linkage to that \emph{without} linkage.

A comparison of the two quantities of interest in our case -- invasion probability and mean extinction time -- with respect to migration is instructive \rev{(Figures \ref{fig:meanabstimeQLEOLM} and \ref{fig:ratioTLM2OLMInvProb})}. Starting from zero migration, the relative advantage of linkage in terms of the invasion probability initially increases with the migration rate very slowly, but then much faster as the migration rate approaches the critical value beyond which an unlinked focal mutation cannot invade \rev{(Figure \ref{fig:ratioTLM2OLMInvProb}A)}. Beyond this critical value, the relative advantage is infinite until migration is so high that  even a fully linked mutation cannot be established. In contrast, we have shown that the relative advantage of linkage in terms of the mean extinction time is maximised at an intermediate migration rate (Figure \ref{fig:meanabstimeQLEOLM}).

In conclusion, for very weak migration, the benefit of being linked to a background polymorphism is almost negligible. For intermediate migration rates, the potential of invasion is elevated by linked selection; this is mainly due to a substantially increased mean extinction time of those still rather few mutations that successfully survive the initial phase of stochastic loss. \rev{This argument is based on the increase of the mean extinction time \emph{relative} to unlinked selection. Because, for large populations, \emph{absolute} extinction times become very large as the migration rate decreases (Figures \ref{fig:meanabstimeQLEOLM}C and \ref{fig:meanabstimeQLEOLM}F), the biological relevance of this comparison may be confined to cases in which the mean extinction time of an unlinked mutation is not extremely high.} 
For migration rates close to the critical migration rate, however, any relative advantage of linkage seems to arise via an increased invasion probability, not via an increased mean extinction time. This is because\rev{, in this case, the latter is close to that for no linkage} (compare Figure \ref{fig:ratioTLM2OLMInvProb}A to Figures \ref{fig:meanabstimeQLEOLM}A and \ref{fig:meanabstimeQLEOLM}B). \rev{A final statement about the relative importance of invasion probability versus mean extinction time is not appropriate at this point. This would require extensive numerical work, along with a derivation of a diffusion approximation to the mean extinction time for tight linkage. However, for small populations, our results show that linked selection can increase the mean extinction time to an extent that is biologically relevant, while, at the same time, not affecting the invasion probability much. This suggests that invasion probabilities may not be a sufficient measure for the importance of physical linkage in adaptive divergence.}

\subsection{Standing variation at the background locus}
We have extended some of our analyses to the case where the background locus is polymorphic on the continent and immigrants may therefore carry both the locally beneficial or deleterious allele. This represents a compromise between the extremes of adaptation from standing versus de-novo genetic variation. 
We have shown that the presence of the beneficial background allele on the continent, and hence among immigrants, leads to a lower invasion probability and a shorter extinction time for the focal de-novo mutation. This effect is due to increased competition against a fitter resident population. While this result is of interest as such, it should not be abused to gauge the relative importance of standing versus de-novo variation in the context of local adaptation. For this purpose, invasion probabilities and extinction times of single mutations do matter, but are not sufficient metrics on their own. Factors such as the mutation rate, the mutational target size, and the distribution of selection coefficients must be taken into account \citep{Hermisson:2005uq}. 

\subsection{Footprint of polygenic local adaptation}
A number of previous studies have quantified the effect of divergent selection or genetic conflicts on linked neutral variation in discrete \citep{Bengtsson:1985fk,Charlesworth:1997vn} and continuous space \citep{Barton:1979fk,Petry:1983uq}. They all concluded that a single locus under selection leads to a pronounced reduction in effective gene flow only if selection is strong or linkage to the neutral site is tight. Whereas \citet{Bengtsson:1985fk} found that additional, physically unlinked, loci under selection had no substantial effect on neutral differentiation, \citet{Feder:2010fk} recently suggested that such loci may have an appreciable effect as long as they are not too numerous. When these authors added a large number of unlinked loci under selection, this resulted in a genome-wide reduction of the effective migration rate, such that the baseline level of neutral divergence was elevated and any effect of  linkage to a single selected locus unlikely to be detected. However, for large numbers of selected loci, it is no longer justified to assume that all of them are physically unlinked. This was noted much earlier by \citet{Barton:1986fk}, who therefore considered a linear genome with an arbitrary number of selected loci linked to a focal neutral site. They showed that a large number of linked selected loci is needed to cause a strong reduction in effective migration rate. In such cases, the majority of other genes must be linked to some locus under selection.

The concept of an effective migration rate has played a key role in most of the studies mentioned above (see \citealt{Barton:1986fk}, and \citealt{Charlesworth:1997vn} for a more comprehensive review). However, for models with more than one linked locus under selection, previous studies relied on numerical solutions or simulations to compute the effective migration rate. Recently, \citet{Buerger:2011uq} derived an analytical approximation for a neutral site that is flanked by two selected loci. We have generalised their result to alternative genetic architectures and an arbitrary number of selected loci (Eq.\ \ref{eq:effMigNeutrGeneral}). From this, we predicted the long-term footprint of polygenic local adaptation in terms of the distribution of allele frequencies, population divergence, and coalescent rate at the neutral site. When considered as a function of the position of the neutral site on the chromosome, these quantities reveal patterns that can hopefully be used for inference about the selective process (Figures \ref{fig:statAllDistNeutr} and \ref{fig:coalRateAndNeNeutr}).

We have only considered the case where migration--selection equilibrium has been reached at the selected loci. It would be interesting, though more demanding, to study the transient phase during which locally beneficial mutations (such as $A_1$ in our case) rise in frequency from $p_0 = 1/(2N)$ to the (pseudo-)equilibrium frequency. We expect this to create a temporary footprint similar to that of a partial sweep \citep{Pennings:2006ys,Pennings:2006ly,Pritchard:2010dq,Ralph:2010qf}. Theoretical progress hinges on a description of the trajectory of the linked sweeping alleles, accounting in particular for the stochastic `lag phase' at the beginning. It will then be of interest to study recurrent local sweeps and extend previous theory for panmictic populations \citep{Coop:2012vn,Lessard:2012uq} to include population structure, migration, and spatially heterogeneous selection. The \rev{hitchhiking} effect of a beneficial mutation in a subdivided population has been described in previous studies \cite[e.g.][]{Slatkin:1998kx, Kim:2011uq}, but these did not account for additional linked loci under selection.

One limitation to our prediction of the coalescence rate at linked neutral sites is the assumption of strong migration relative to genetic drift \citep{Nagylaki:1980uq,Notohara:1993kx}. As the effective migration rate decays to zero if the neutral site is very closely linked to a selected site (Figure \ref{fig:mEffSojourn}C), this assumption will be violated. Therefore, our predictions should be interpreted carefully when linkage is tight. Moreover, and even though this seems widely accepted, we are not aware of a rigorous proof showing that an effective migration rate can sufficiently well describe the effect of local selection on linked neutral genealogies \cite[cf.][]{Barton:2004uq}.

Another limitation is that our prediction of linked neutral diversity and divergence (Figure \ref{fig:statAllDistNeutr}) holds only for drift--migration equilibrium. For closely linked neutral sites, which experience very low rates of effective migration, it may take a long time for this equilibrium to be reached. By that time, other evolutionary processes such as background selection and mutation will have interfered with the dynamics at the focal site.

\subsection{Further limitations and future extensions}
We assumed no dominance and no epistasis. Both are known to affect the rate of adaptation and the maintenance of genetic variation \cite[e.g.][]{Charlesworth:1987kx, Bank:2012vn}. Empirical results on dominance effects of beneficial mutations are ambiguous \citep{Vicoso:2006ve}. Some studies showed no evidence for a deviation from additivity, 
whereas others suggested weak recessivity 
(reviewed in \citealt{Orr:2010fk} and \citealt{Presgraves:2008qf}). Empirical evidence for epistasis comes from studies reporting genetic incompatibilities between hybridising populations \citep{Lowry:2008kx,Presgraves:2010vn}. In the classical Dobzhansky-Muller model \citep{Bateson:1909ve,Dobzhansky:1936bh,Muller:1942qf}, such incompatibilities may become expressed during secondary contact after allopatry, even if divergence is neutral. With gene flow, genetic incompatibilities can be maintained only if the involved alles are locally beneficial \citep{Bank:2012vn}. \citet{Bank:2012vn} derived respective conditions using deterministic theory. An extension to a stochastic model focussing on invasion probabilities and extinction times would be desirable.

Our model assumed one-way migration. While this is an important limiting case and applies to a number of natural systems \cite[e.g.][]{King:1995fk}, an extension to two-way migration is of interest, because natural populations or incipient species often exchange migrants mutually \cite[e.g.][]{Janssen:2013kx,Nadeau:2013vn}. Such theory will allow for a direct comparison to recent simulation studies \citep{Yeaman:2011fk, Feder:2010fk,Feder:2012dq,Yeaman:2013uq,Flaxman:2013fk}. \rev{It will also have a bearing on the evolution of suppressed recombination in sex chromosomes \cite[e.g.][]{Rice:1984vn,Rice:1987rt,Fry:2010gf,Jordan:2012fk,Charlesworth:2013zr}}. Deterministic theory suggests that linkage becomes less crucial for the maintenance of locally beneficial alleles the more symmetric gene flow is \citep{Akerman:2013uq}.

When describing the distribution of fitness effects of successful beneficial mutations, we only considered a single mutation. \rev{Future studies should investigate a complete adaptive walk, allowing for mutations at multiple loci to interact via dominance, epistasis, and linkage. Moreover, it would then seem justified to relax the assumption of a constant fitness gradient, especially in the proximity of an optimum, and to account for the fact that the input DFE is not necessarily exponential \citep{Martin:2008fk}.}

In our derivations of sojourn and mean absorption times, we assumed quasi-linkage disequilibrium (QLE). As expected, the approximations break down  if \rev{recombination is weak (e.g.\ Figure \ref{fig:meanabstimeQLEWF}). For tight linkage, when linked selection is most beneficial, an alternative diffusion process needs to be developed.} \rev{However, to determine how weak physical linkage may be such that an invading mutation still has an advantage, an approximation is required that is accurate for moderate and loose linkage. Therefore, the assumption of QLE does not restrict the scope of our results that address the limits to the importance of linked selection.}
%

\subsection{Conclusion}
This study advances our understanding of the effects of physical linkage and maladaptive gene flow on local adaptation. We derived explicit approximations to the invasion probability and extinction time of benefical de-novo mutations that arise in linkage to an established migration--selection polymorphsim. In addition, we obtained an analytical formula for the effective migration rate experienced by a neutral or weakly beneficial site that is linked to an arbitrary number of selected loci. These approximations provide an efficient alternative to simulations \cite[e.g.][]{Feder:2010fk,Feder:2012dq}. Our results strengthen the emerging view that \rev{physically} linked selection \rev{(and hence so-called divergence hitchhiking) is biologically relevant} only if linkage is tight or if selection at the background locus is strong \citep{Petry:1983uq,Barton:1986fk,Feder:2012dq,Flaxman:2013fk}. When these conditions are met, however, the effect of linkage can be substantial. \rev{A definite statement about the importance of `divergence hitchhiking' versus `genome hitchhiking' and complementary processes \cite[cf.][]{Yeaman:2013uq} seems premature, though; it will require further empirical and theoretical work.} We suggest that future theoretical studies (i) obtain analogous approximations for bi- rather than unidirectional gene flow, (ii) account for epistasis \rev{and dominance, (iii) incorporate the distribution of fitness effects of beneficial mutations, and (iv) employ a stochastic modifier-of-recombination model to assess the importance of non-zero optimal recombination rates.} 
Extensions of this kind will further enhance our understanding of polygenic local adaptation and its genetic footprint.

\section*{Acknowledgements}
We thank Ada Akerman for sharing an unpublished manuscript and \emph{Mathematica} code, Josef Hofbauer for help with finding the derivative of the invasion probability as a function of the recombination rate, \rev{the Biomathematics Group at the University of Vienna for stimulating discussions, and Samuel Flaxman and Samuel Yeaman for useful comments on the manuscript}. SA and RB acknowledge financial support by the Austrian Science Fund (FWF projects P21305 and P25188). The computational results presented have been achieved in parts using the Vienna Scientific Cluster (VSC).


\clearpage

\bibliographystyle{/Users/Simon/Documents/LocAdD/pub/paper/genetics}
\begin{small}
\bibliography{/Users/Simon/Documents/Literature/BibDesk/Jshort_dot,/Users/Simon/Documents/Literature/BibDesk/central}
\end{small}
\clearpage




\setcounter{section}{0}
\captionsetup[text]{labelsep=space,size=Large}
\captionsetup[procedure]{labelsep=space}
\captionsetup[figure]{labelsep=space}
\captionsetup[table]{labelsep=space}

\renewcommand{\thefigure}{S\arabic{figure}}
\setcounter{figure}{0}
\section*{Supporting Information: Figures}

\clearpage

\begin{figure}[!ht]
\begin{center}
\includegraphics[width=1\textwidth]{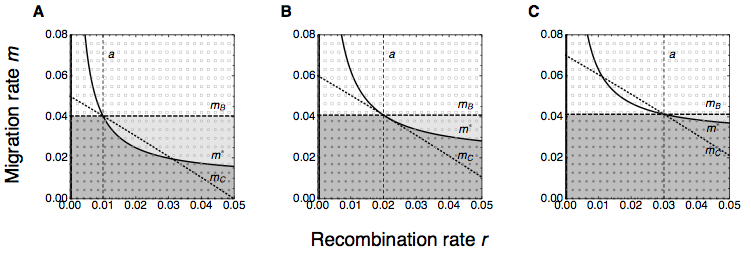}
\end{center}
\caption{
{\textrm{Critical migration rates and invasion of $A_1$ for a monomorphic continent.}} Dark grey: invasion of $A_1$ via the unstable marginal equilibrium $E_\bloc$; light grey: no invasion of $A_1$, stable marginal equilibrium $E_\bloc$; white: no invasion, fixation of continental haplotype $A_2B_2$ and convergence to the monomorphic equilibrium $E_{C}$, at which the island population is fixed for the continental haplotype $A_2B_2$. Numerical iterations of invasion dynamics where performed at coordinates indicated by grey symbols (File \ref{prc:determDiscrNB}). Different symbols show which equilibrium is reached:  $\bullet$ $E_+$; $\circ$ $E_\bloc$;  $\smallsquare$ $E_{C}$. Initial values for iterations were $\left\{p_0, q_0, D_0\right\} = \left\{0, \qeqb, 0\right\}$, where $\qeqb$ is the frequency of $B_1$ at $E_\bloc$. Iterations were stopped when successive changes in each coordinate became smaller than the numerical machine precision.  The thick, almost-vertical line close to $r = 0$ is for the critical migration rate $m^{\ast}$. This curve crosses the $r$ axis at $r = a(b-a)/(1-2a+b)$, which is denoted by a vertical dashed line that can hardly be seen. The second vertical dashed line corresponds to $r = a$. \textbf{(A)} $a=0.01, b=0.04$. \textbf{(B)} $a=0.02, b=0.04$. \textbf{(C)} $a=0.03, b=0.04$.\\
}
\label{fig:stabilityEBMonomCont}
\end{figure}

\clearpage

\begin{figure}[!ht]
\begin{center}
\includegraphics[width=0.8\textwidth]{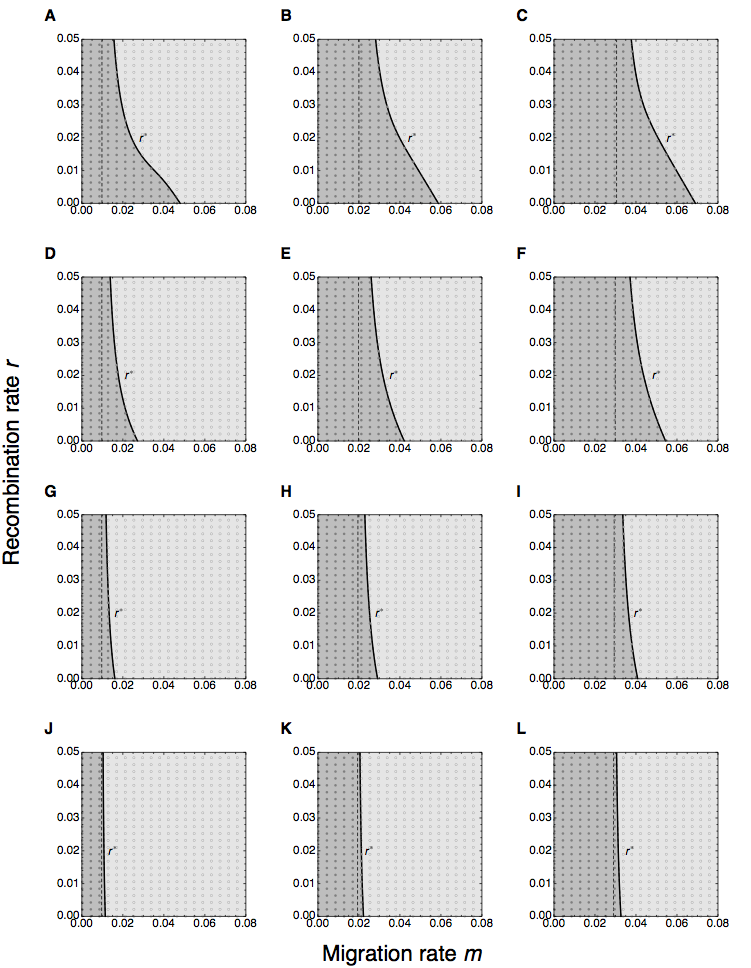}
\end{center}
\caption{
{\textrm{Critical recombination rate and invasion of $A_1$ for a polymorphic continent.}} Dark grey: invasion of $A_1$ via the unstable marginal equilibrium $E_\bloc$; light grey: no invasion of $A_1$, stable marginal equilibrium $E_\bloc$. Numerical iterations of invasion dynamics where performed at coordinates indicated by grey symbols (File \ref{prc:determDiscrNB}). Different symbols show which equilibrium is reached:  $\bullet$ $E_+$; $\circ$ $E_\bloc$. Initial values for iterations were $\left\{p_0, q_0, D_0\right\} = \left\{0, \qeqb, 0\right\}$, where $\qeqb$ is the frequency of $B_1$ at $E_\bloc$. The vertical dashed line indicates the pole of the function $r^{\ast}(m)$ from Eq.\ \eqref{eq:rCritPolymCont}. In the left column (\textbf{A}, \textbf{D}, \textbf{G}, and \textbf{J}), the selection coefficients are $a=0.01$, $b=0.04$ ($a<b/2$), in the middle column (\textbf{B}, \textbf{E}, \textbf{H}, and \textbf{K}) they are $a=0.02$, $b=0.04$ ($a=b/2$), and in the right column (\textbf{C}, \textbf{F}, \textbf{I}, and \textbf{L}) they are $a=0.03$, $b=0.04$ ($a>b/2$). From top to bottom, the continental frequency of $B_1$ increases and takes values of $q_c = 0.01$ in \textbf{(A)}--\textbf{(C)}, $ q_c = 0.2$ in \textbf{(D)}--\textbf{(F)}, $q_c = 0.5$ in \textbf{(G)}--\textbf{(I)}, and $q_c = 0.8$ in \textbf{(J)}--\textbf{(L)}.\\
}
\label{fig:stabilityEBPolymCont}
\end{figure}

\clearpage

\begin{figure}[!ht]
\begin{center}
\includegraphics[width=1\textwidth]{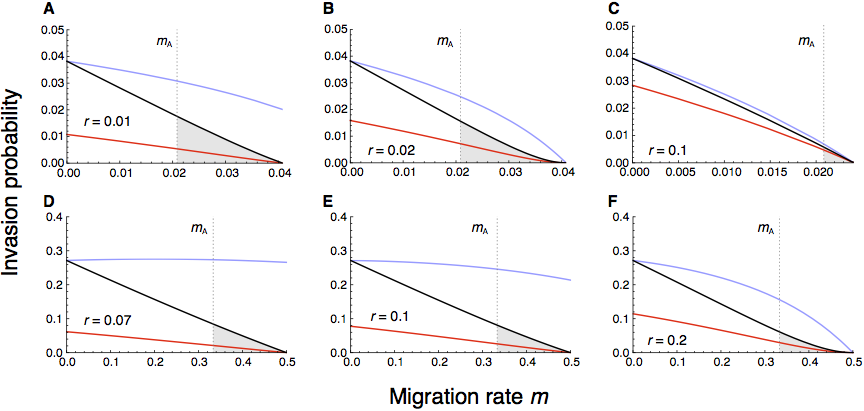}
\end{center}
\caption{
{\textrm{Invasion probability of $A_1$ as a function of the migration rate for a monomorphic continent.}} Shown are numerical solutions to the branching process, conditional on initial occurrence of $A_1$ on background $B_1$ (blue), on background $B_2$ (red), and when averaged across backgrounds with weights determined by the equilibrium frequency $\qeqb$ of $B_1$ (black). The vertical dashed line shows $m_{\aloc} = a/(1-b)$, the critical migration rate below which $A_1$ can invade without linkage to the background locus. The shaded area thus indicates where $A_1$ has a non-zero average invasion probability exclusively due to linkage to locus $\bloc$. \textbf{(A)}--\textbf{(C)} Weak selection: $a = 0.02$, $b = 0.04$ and $r$ as given in the panels. \textbf{(D)}--\textbf{(F)} Strong selection: $a = 0.2$, $b = 0.4$ and $r$ as given in the panels. In this case, if linkage is tight ($r$ small), the invasion probability conditional on the beneficial background increases with $m$ as long as $m$ is sufficiently small, and only starts decreasing if $m$ is much larger (blue curve in panel \textbf{D}). This is because migration reduces the fitness of the resident population (consisting of $A_2B_1$ and $A_2B_2$) more strongly than it reduces the marginal fitness of type $A_1B_1$, which is favourable to type $A_1B_1$. As migration becomes stronger, though, the reduction in marginal fitness of $A_1B_1$ becomes dominant. The parameter combination in \textbf{(D)} was arbitrarily chosen to illustrate this effect \rev{(for a detailed explanation, see section 5 of File \ref{txt:SI})}. For $r < 0.07$, the maximum of the blue curve is shifted further to the right.\\
}
\label{fig:invProbFuncM}
\end{figure}

\clearpage

\begin{figure}[!ht]
\begin{center}
\includegraphics[width=1\textwidth]{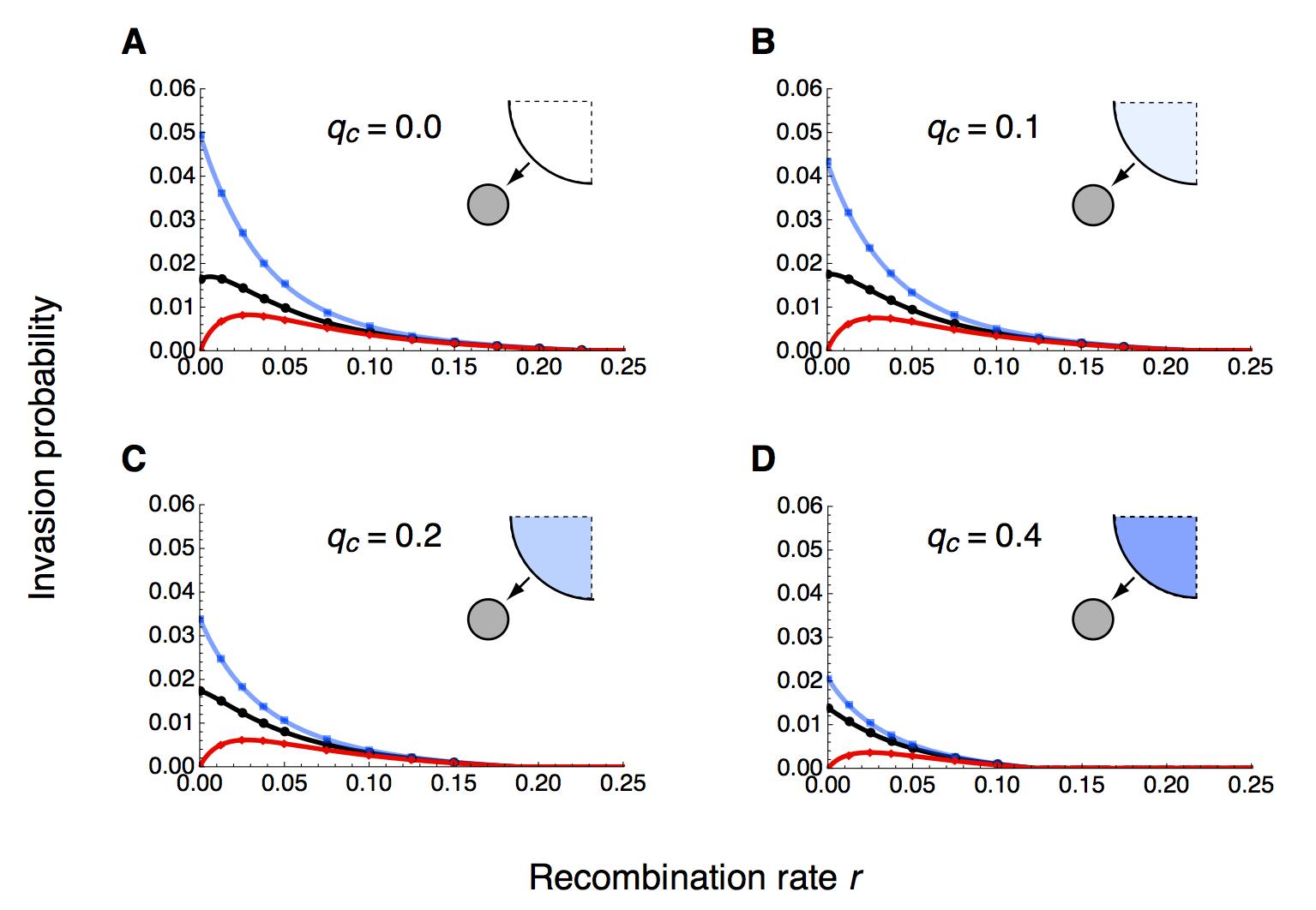}
\end{center}
\caption{
{\textrm{Invasion probability of $A_1$ as a function of the recombination rate and the continental frequency of $B_1$.}} Panels are for different values of the continental frequency $q_c$ of the beneficial background allele ($B_1$). \rev{Curves} show numerical solutions to the branching process \rev{(Eq.\ \ref{eq:pgfExplAddPolymCont})}, conditional on initial occurrence of $A_1$ on background $B_1$ (blue), on background $B_2$ (red), and when averaged across backgrounds with weights determined by the equilibrium frequency $\qeqb$ of $B_1$ (black). \rev{Dots represent the point estimates across $10^{6}$ simulations under the branching-process assumptions (see Methods). Error bars span twice the standard error on each side of the point estimates, but are too short to be visible}. Parameters other than $q_c$ are the same in all panels: $a = 0.03$, $b = 0.04$ and $m = 0.032$.\\
}
\label{fig:invProbFuncRPolymCont}
\end{figure}

\clearpage

\begin{figure}[!ht]
\begin{center}
\includegraphics[width=1\textwidth]{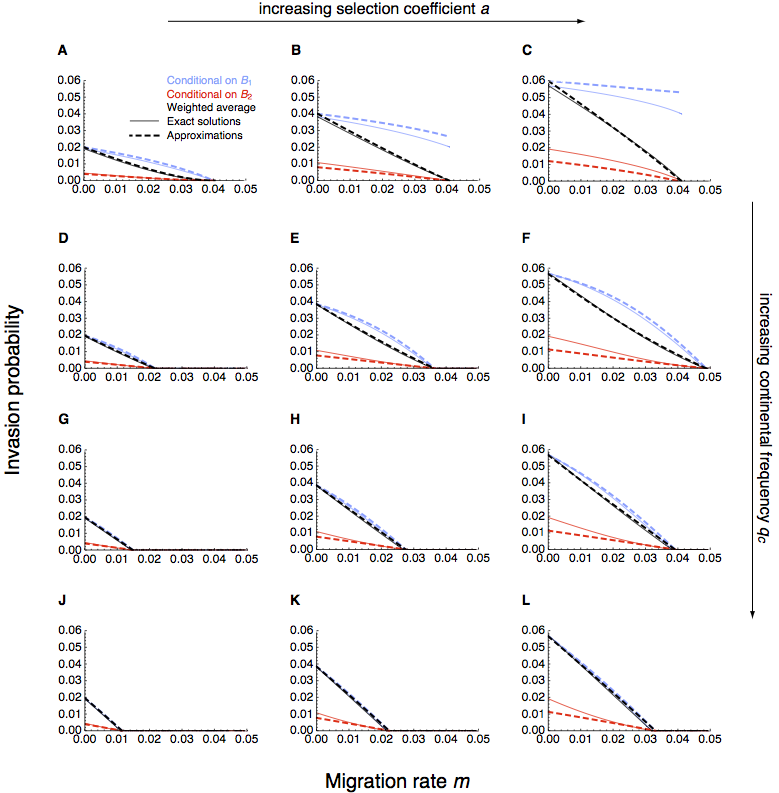}
\end{center}
\caption{
{\textrm{Invasion probability of $A_1$ as a function of the migration rate.}} Panels are for varying selective advantage $a$ and continental frequency $q_c$ of the beneficial background allele $B_1$. Invasion probabilities are shown conditional on initial occurrence of $A_1$ on background $B_1$ (blue), on background $B_2$ (red), and as a weighted average across the two backgrounds (black). Solid curves show exact numerical solutions to the branching process, whereas thick dashed curves show the analytical approximations valid for weak evolutionary forces and a slightly supercritical branching process (see section 4 of File \ref{txt:SI}, and Eqs.\ 7--9 in File \ref{prc:stochDiscrSlightlySupercritBPNB}). In all panels, $b=0.04$ and $r=0.01$. The selective advantage $a$ of $A_1$ increases from left to right, taking values of $a=0.01$ in panels \textbf{(A)}, \textbf{(D)}, \textbf{(G)}, \textbf{(J)}, $a=0.02$ in panels \textbf{(B)}, \textbf{(E)}, \textbf{(H)}, \textbf{(K)}, and $a=0.03$ in panels \textbf{(C)}, \textbf{(F)}, \textbf{(I)} and \textbf{(L)}. The continental frequency $q_c$ of $B_1$ increases from top to bottom, taking values of $q_c=0$ in panels \textbf{(A)}--\textbf{(C)}, $q_c= 0.2$ in panels \textbf{(D)}--\textbf{(F)}, $q_c=0.5$ in panels \textbf{(G)}--\textbf{(I)}, and $q_c = 0.8$ in panels \textbf{(J)}--\textbf{(L)}.\\
}
\label{fig:invProbFuncMPolymCont}
\end{figure}

\clearpage

\begin{figure}[!ht]
\begin{center}
\includegraphics[width=1\textwidth]{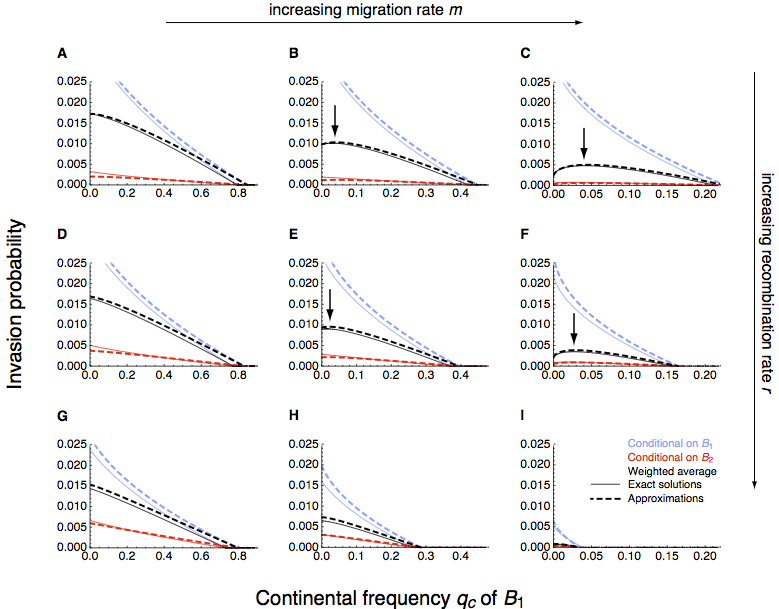}
\end{center}
\caption{
{\textrm{Invasion probability of $A_1$ as a function of the continental frequency of $B_1$.}} Panels are for varying migration and recombination rates. Invasion probabilities are shown conditional on initial occurrence of $A_1$ on background $B_1$ (blue), on background $B_2$ (red), and as a weighted average across the two backgrounds (black). Solid curves show exact numerical solutions to the branching process, whereas thick dashed curves show the analytical approximations valid for weak evolutionary forces and a slightly supercritical branching process (see section 3 of File \ref{txt:SI}, and Eqs.\ 7--9 in File \ref{prc:stochDiscrSlightlySupercritBPNB}). In all panels, $a=0.02$ and $b=0.04$. The migration rate $m$ increases from left to right, taking values of $m=0.022$ in panels \textbf{(A)}, \textbf{(D)}, \textbf{(G)}, $m=0.03$ in panels \textbf{(B)}, \textbf{(E)}, \textbf{(H)}, and $m=0.038$ in panels \textbf{(C)}, \textbf{(F)}, and \textbf{(I)}. The recombination rate increases from top to bottom, taking values of $r=0.005$ in panels \textbf{(A)}--\textbf{(C)}, $r=0.01$ in panels \textbf{(D)}--\textbf{(F)}, and $r=0.02$ in panels \textbf{(G)}--\textbf{(I)}. Arrows indicate where the optimal $q_c$ is non-zero.\\
}
\label{fig:invProbFuncqCPolymCont}
\end{figure}

\clearpage

\begin{figure}[!ht]
\begin{center}
\includegraphics[width=1\textwidth]{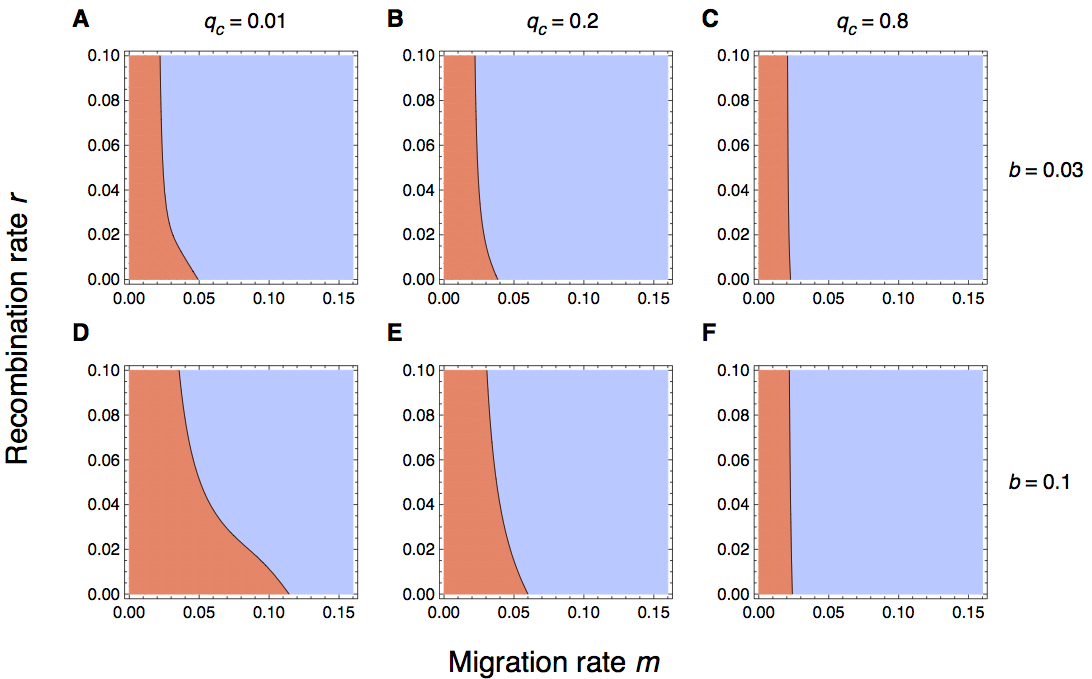}
\end{center}
\caption{
{\textrm{Asymptotic stability of the marginal one-locus migration--selection equilibrium $\tilde{E}_\bloc$ in continuous time I.}} Light blue areas indicate where $\tilde{E}_\bloc$ is asymptotically stable and $A_1$ cannot invade ($\tilde{\nu} < 0$; $\tilde{\nu}$ as in Eq.\ \ref{eq:leadEvalPolymContCont}, File \ref{txt:SI}). Orange areas indicate where $\tilde{E}_\bloc$ is unstable and $A_1$ may potentially invade ($\tilde{\nu} > 0$). The black curve represents the critical recombination rate $\tilde{r}_{B}$ given in Eq.\ \eqref{eq:rBPolymContCont}, as a function of the migration rate. The selection coefficient $a$ in favour of $A_1$ is 0.02 throughout, the selection doefficient $b$ in favour of $B_1$ is 0.03 in the first row (\textbf{A}--\textbf{C}) and 0.1 in the second (\textbf{D}--\textbf{F}). In each row, the continental frequency of $B_1$ increases from left to right, taking values of $q_c=0.01$ in \textbf{(A)} and \textbf{(D)}, $q_c=0.2$ in \textbf{(B)} and \textbf{(E)}, and $q_c=0.8$ in \textbf{(C)} and \textbf{(F)}.\\
}
\label{fig:stabilityEBPolymMR}
\end{figure}

\clearpage

\begin{figure}[!ht]
\begin{center}
\includegraphics[width=1\textwidth]{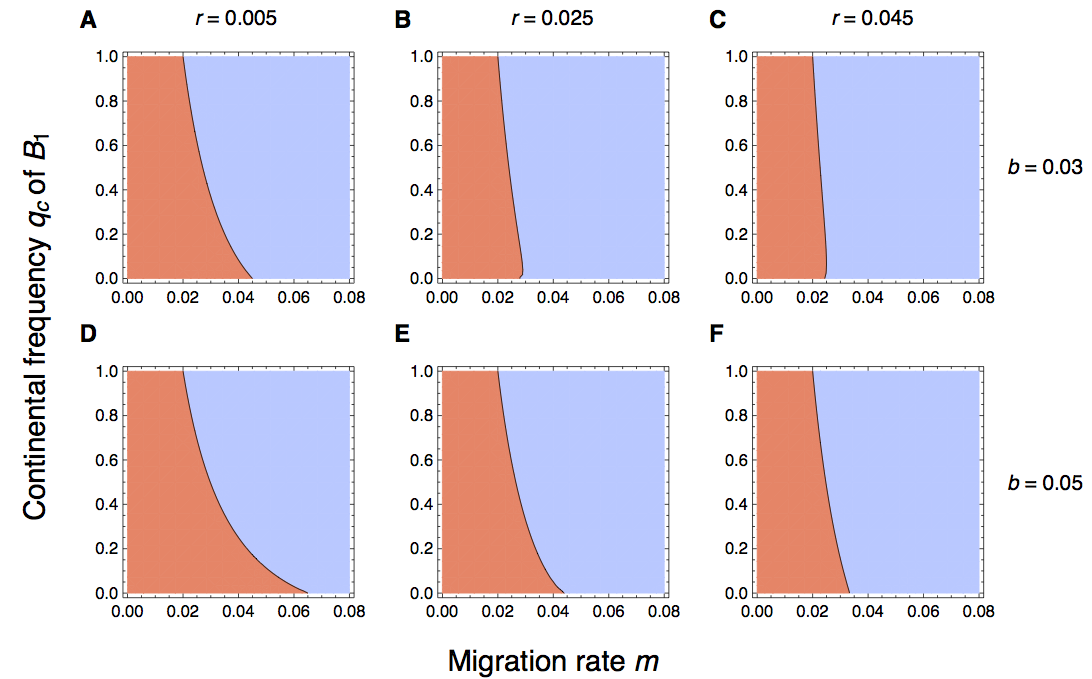}
\end{center}
\caption{
{\textrm{Asymptotic stability of the marginal one-locus migration--selection equilibrium $\tilde{E}_\bloc$ in continuous time II.}} Light blue areas indicate where $\tilde{E}_\bloc$ is asymptotically stable and $A_1$ cannot invade ($\tilde{\nu} < 0$; $\tilde{\nu}$ as in Eq.\ \ref{eq:leadEvalPolymContCont}, File \ref{txt:SI}). Orange areas indicate where $\tilde{E}_\bloc$ is unstable and $A_1$ may potentially invade ($\tilde{\nu} > 0$). The black curve corresponds to a combination of $\tilde{q}_{c-}^{\ast\ast}$ and $\tilde{q}_{c+}^{\ast\ast}$ as described in section 6 of File \ref{txt:SI}, as a function of the migration rate. The selection coefficient $a$ in favour of $A_1$ is 0.02 throughout, the selection doefficient $b$ in favour of $B_1$ is 0.03 in the first row (\textbf{A}--\textbf{C}) and 0.05 in the second (\textbf{D}--\textbf{F}). In each row, the recombination rate increases from left to right, taking values of $r=0.005$ in \textbf{(A)} and \textbf{(D)}, $r=0.025$ in \textbf{(B)} and \textbf{(E)}, and $r=0.045$ in \textbf{(C)} and \textbf{(F)}.\\
}
\label{fig:stabilityEBPolymMqC}
\end{figure}

\clearpage

\begin{figure}[!ht]
\begin{center}
\includegraphics[width=1\textwidth]{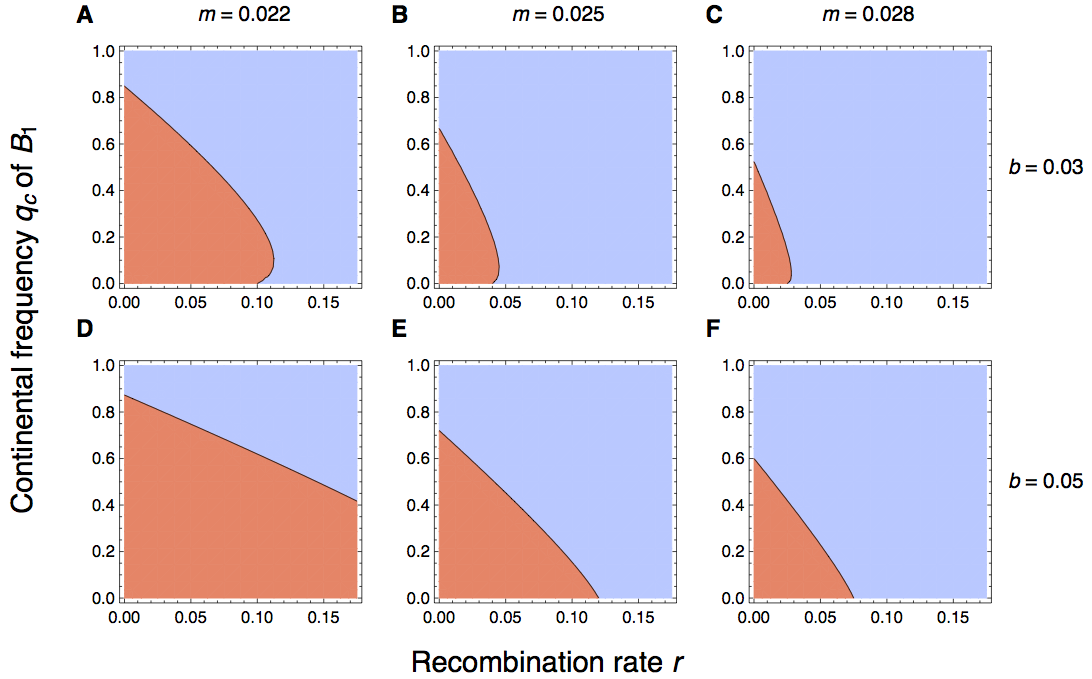}
\end{center}
\caption{
{\textrm{Asymptotic stability of the marginal one-locus migration--selection equilibrium $\tilde{E}_\bloc$ in continuous time III.}}
Light blue areas indicate where $\tilde{E}_\bloc$ is asymptotically stable and $A_1$ cannot invade ($\tilde{\nu} < 0$; $\tilde{\nu}$ as in Eq.\ \ref{eq:leadEvalPolymContCont}, File \ref{txt:SI}). Orange areas indicate where $\tilde{E}_\bloc$ is unstable and $A_1$ may potentially invade ($\tilde{\nu} > 0$). The black curve corresponds to a combination of $\tilde{q}_{c-}^{\ast\ast}$ and $\tilde{q}_{c+}^{\ast\ast}$ as described in section 6 of File \ref{txt:SI}, as a function of the recombination rate. 
The selection coefficient $a$ in favour of $A_1$ is 0.02 throughout, the selection doefficient $b$ in favour of $B_1$ is 0.03 in the first row (\textbf{A}--\textbf{C}) and 0.05 in the second (\textbf{D}--\textbf{F}). In each row, the migration rate increases from left to right, taking values of $m=0.022$ in \textbf{(A)} and \textbf{(D)}, $m=0.025$ in \textbf{(B)} and \textbf{(E)}, and $m=0.028$ in \textbf{(C)} and \textbf{(F)}.\\
}
\label{fig:stabilityEBPolymRqC}
\end{figure}

\clearpage

\begin{figure}[!ht]
\begin{center}
\includegraphics[width=1\textwidth]{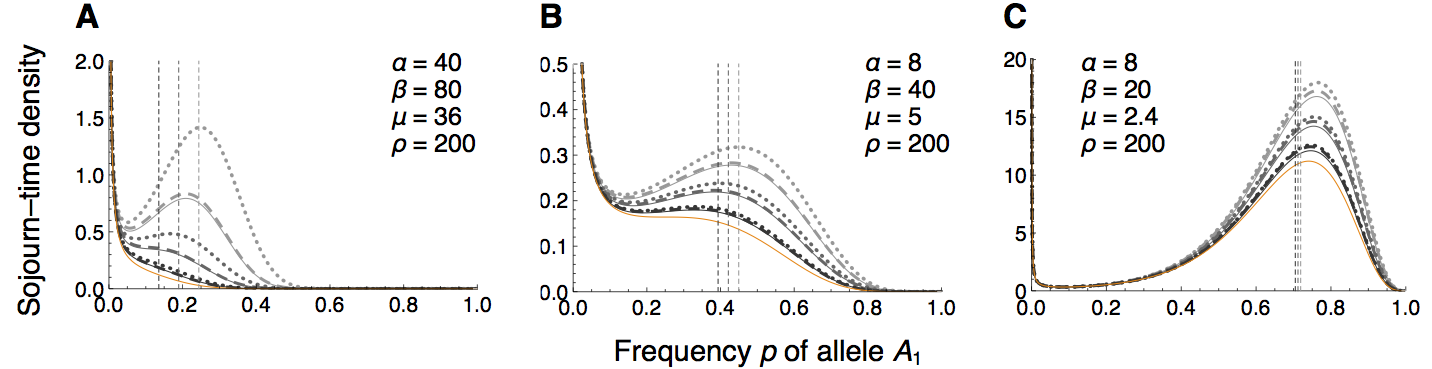}
\end{center}
\caption{
{\textrm{Diffusion approximation to the sojourn-time density of $A_1$ under quasi-linkage equilibrium for a polymorphic continent.}} Comparison of the sojourn-time density (STD) $\ttwoQLE$ (thin curves, Eq.\ \ref{eq:STD2QLE}) to the approximation valid for small $p_0$, $\ttwotilQLE$ (dashed curves, analogous to Eq.\ \ref{eq:STD2QLEApprox} in File \ref{txt:SI}) 
and the one based on the additional assumption of $\rho \gg 0$, $\ttwotilQLErho$ (dotted curves, Eq.\ \ref{eq:STD2QLEApproxRhoLargePolymCont}) assuming a polymorphic continent. The continental frequency $q_c$ of $B_1$ increases from light to dark grey, taking values of 0.2, 0.5, and 0.8. The STD for the one-locus model, $\ttwotilOLM$, is shown in orange as a reference. Vertical lines give the deterministic frequency $\hat{p}_{+}$ of $A_1$ at the respective fully-polymorphic equilibrium (computed in File \ref{prc:stochDiffApproxQLENB}). \textbf{(A)} Strong evolutionary forces relative to genetic drift. 
\textbf{(B)} Strong asymmetry in selection coefficients, and moderate migration. 
\textbf{(C)} Recombination ten times stronger than selection at locus $\bloc$. 
In all panels, $p_0 = 0.005$, which corresponds to an island population of size $N=100$ and a single initial copy of $A_1$. 
Panels \textbf{(A)}, \textbf{(B)} and \textbf{(C)} correspond to Figures \ref{fig:stdQLE}C, \ref{fig:stdQLE}D and \ref{fig:stdQLE}E for a monomorphic continent ($q_c = 0$), respectively.\\
}
\label{fig:stdQLEPolym}
\end{figure}

\clearpage

\begin{figure}[!ht]
\begin{center}
\includegraphics[width=1\textwidth]{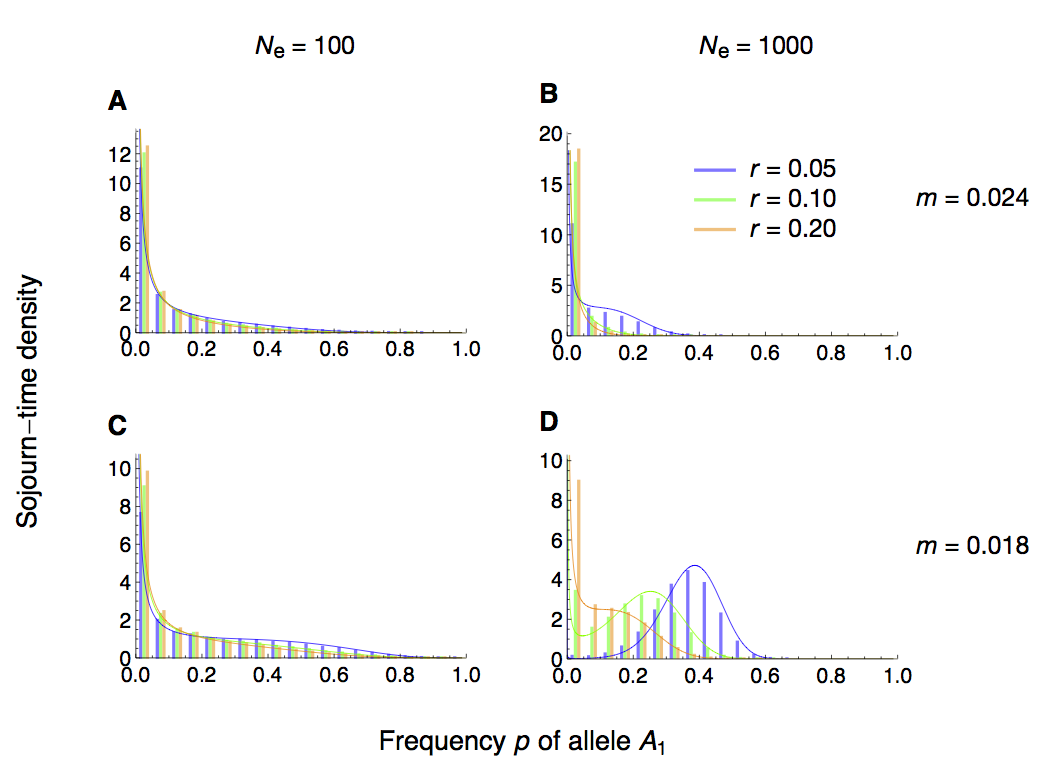}
\end{center}
\caption{
{\textrm{Comparison of analytical and simulated sojourn-time densites of $A_1$ for a monomorphic continent.} Results are shown for various recombination rates $r$. Histograms were obtained from $10^6$ simulations (see Methods) and curves give the diffusion approximation $\ttwotilQLE$ from Eq.\ \eqref{eq:STD2QLEApprox}. Throughout,  $a=0.02$, $b=0.04$ and $p_0 = 1/(2N)$ (we assumed $N_e = N$). In the first row, migration is relatively strong compared to selection in favour of $A_1$ ($m=0.024 > a$), in the second row it is relatively weak ($m = 0.018 < a$). In the left column, the effective population size is small ($N_e = 100$) and drift dominates, whereas in the right column, $N_e = 1000$ and deterministic forces become more important.}\\
}
\label{fig:STDCompSimMonomCont}
\end{figure}

\clearpage

\begin{figure}[!ht]
\begin{center}
\includegraphics[width=1\textwidth]{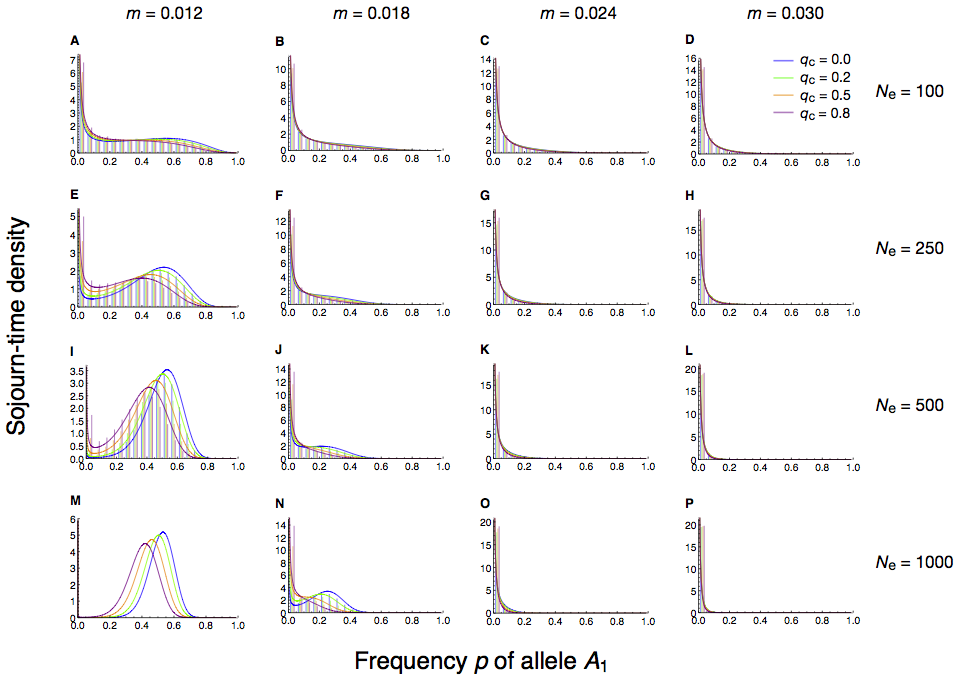}
\end{center}
\caption{
{\textrm{Comparison of analytical and simulated sojourn-time densites of $A_1$ for a polymorphic continent.} Results are shown for various migration rates $m$ and continental frequencies $q_c$ of $B_1$. Histograms were obtained from $10^6$ simulations (see Methods) and curves give the diffusion approximation under the assumption of quasi-linkage equilibrium, $\ttwotilQLE$, from Eq.\ \eqref{eq:STD2QLEApprox}. Throughout,  $a=0.02$, $b=0.04$, $r=0.1$ and $p_0 = 1/(2N)$ (we assumed $N_e = N$). From the top to the bottom row, the effective population size $N_e$ increases and therefore genetic drift becomes less important. From the left to the right column, the migration rate $m$ increases, making it more difficult for $A_1$ to survive. No simulations were completed for the parameter combination in panel \textbf{(M)}, as they were too time-consuming.}\\
}
\label{fig:STDCompSimPolymCont}
\end{figure}

\clearpage

\begin{figure}[!ht]
\begin{center}
\includegraphics[width=1\textwidth]{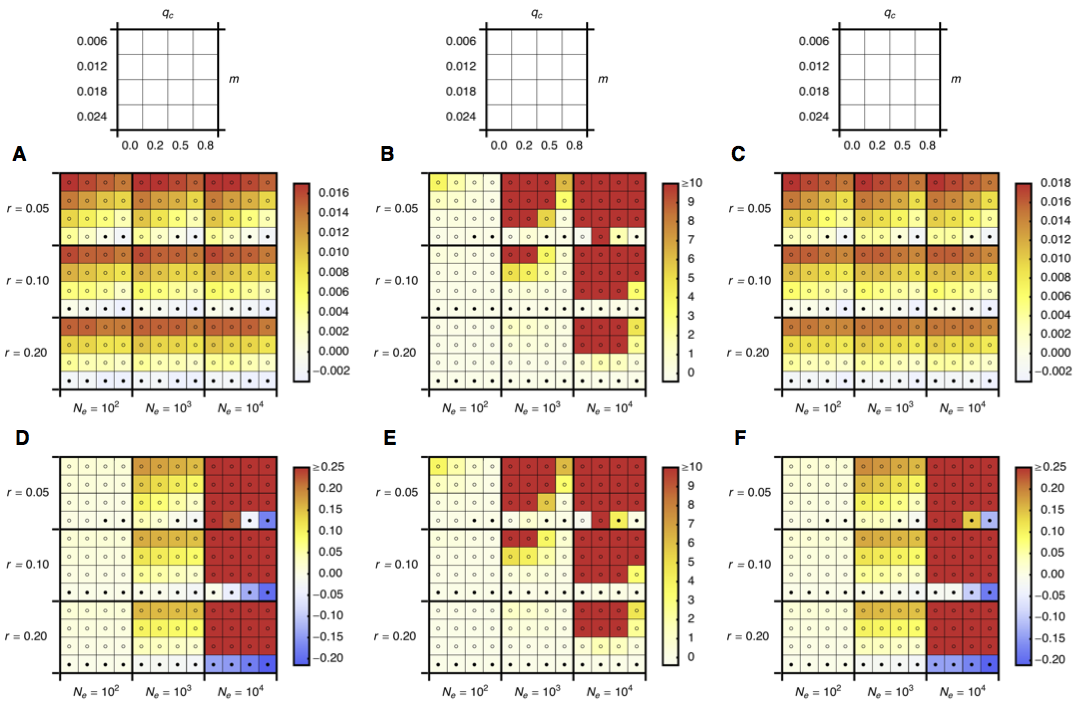}
\end{center}
\caption{
{\textrm{Comparison of various diffusion approximations of the mean absorption time of $A_1$.}} \textbf{(A)} The error of $\tbartilQLE$ (Eq.\ \ref{eq:meanAbsTimeQLEApprox} in File \ref{txt:SI}) 
relative to $\tbarQLE$ (Eq.\ \ref{eq:meanAbsTimeQLE}) for various parameter combinations and an initial frequency of $A_1$ equal to $p_0 = 1/(2N)$ (we assumed $N_e = N$). Squares bounded by thick lines delimit combinations of values of the recombination rate $r$ and the effective population size $N_e$. Within each of them, squares bounded by thin lines correspond to combinations of values of the migration rate $m$ and the continental frequency $q_c$ of $A_1$, as shown in the small panels on top. The colour code assigns deeper blue to more negative, and deeper red to more positive values. Empty (filled) circles indicate that the marginal one-locus equilibrium $\tilde{E}_{\bloc}$ is unstable (stable) and $A_1$ can (not) be established under deterministic dynamics. Selection coefficients are $a=0.02$ and $b=0.04$. \textbf{(B)} The error of $\tbarQLErho$ (Eq.\ \ref{eq:meanAbsTimeQLERhoLarge} in File \ref{txt:SI}) 
relative to $\tbarQLE$ for $p_0 = 1/(2N)$. Other details as for panel \textbf{(A)}. \textbf{(C)} The error of $\tbartilQLErho$ (Eq.\ \ref{eq:meanAbsTimeQLEApproxRhoLarge} in File \ref{txt:SI}) 
relative to $\bar{t}_{\mathrm{QLE}, \rho \gg 0}$ for $p_0 = 1/(2N)$. Other details as for panel \textbf{(A)}. \textbf{(D)} As in panel \textbf{(A)}, but for an initial frequency of $A_1$ equal to $p_0 = 0.005$, independently of $N$. \textbf{(E)} As in panel \textbf{(B)}, but for $p_0 = 0.005$ fixed. \textbf{(F)} As in panel \textbf{(C)}, but for $p_0 = 0.005$ fixed. Simulations were as described in Methods. Numerical values for errors represented in panels \textbf{(A)} to \textbf{(C)} and \textbf{(D)} to \textbf{(F)} are shown in Tables \ref{tab:relErrorp0Small} to \ref{tab:relErrorp0SmallGivenrhoLarge} and \ref{tab:relErrorp0Smallp0Fixed} to \ref{tab:relErrorp0SmallGivenrhoLargep0Fixed}, respectively.\\
}
\label{fig:relErrDiffApproxComb}
\end{figure}

\clearpage

\begin{figure}[!ht]
\begin{center}
\includegraphics[width=1\textwidth]{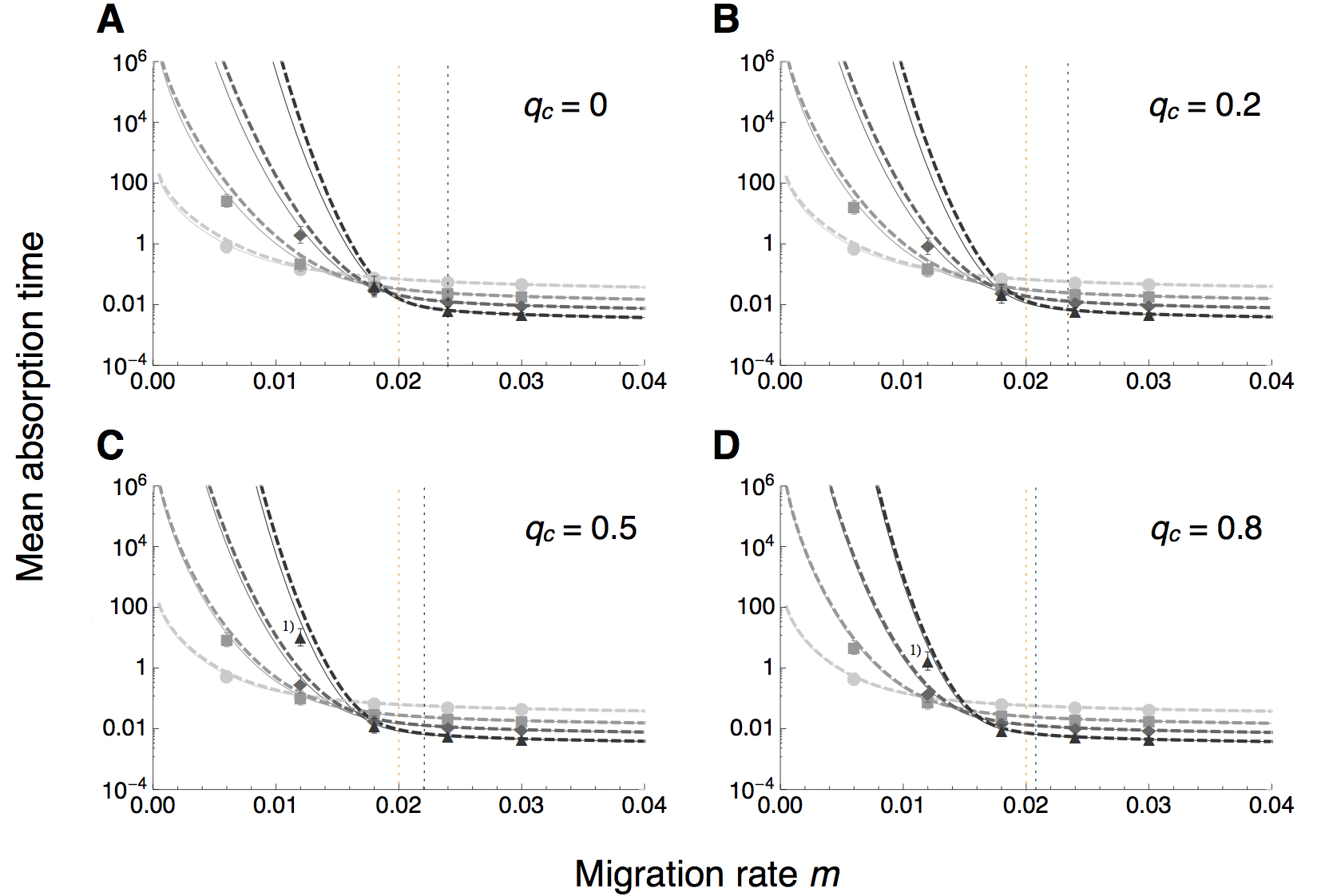}
\end{center}
\caption{
\textrm{Mean absorption time of $A_1$ as a function of the migration rate.} Two approximations derived under the assumption of quasi-linkage eqilibrium (QLE) are compared. Solid curves show $\tbarQLE$ (Eq.\ \ref{eq:meanAbsTimeQLE}) and thick dashed curves $\bar{t}_{\mathrm{QLE}, \rho \gg 0}$ (Eq.\ \ref{eq:meanAbsTimeQLERhoLarge} in File \ref{txt:SI}). 
The effective population size $N_e$ increases from light to dark grey, taking values of 100, 250, 500, and 1000. The vertical dotted lines denote the critical values of $m$ below which $A_1$ can invade in the deterministic one-locus (orange) and two-locus (black) model. Dots and whiskers show the mean and 95\% empirical interquantile range across 1000 runs of the mean absorption time in 1000 replicates per run. Where points and whiskers are missing, simulations could not be completed within the time limit of 72 hours per replicate on the computer cluster. Data points labelled by 1) are from parameter combinations for which fewer than 1000 replicates per run could be realised, because some took longer than the limit of 72 hours. \textbf{(A)} Monomorphic continent: $q_c = 0$. \textbf{(B)}--\textbf{(D)} Polymorphic continent with continental frequency of $B_1$ equal to $q_c = 0.2$, $q_c = 0.5$, and $q_c = 0.8$, respectively. Other parameters are $a=0.02$, $b=0.04$, $r=0.1$, and  $p_0=1/(2 N)$ (we assumed $N_e=N$). Time is in multiples of $2N_e$ generations and plotted on the log scale.\\
}
\label{fig:meanAbsTimeQLEWFrLargeCombM}
\end{figure}

\clearpage

\begin{figure}[!ht]
\begin{center}
\includegraphics[width=1\textwidth]{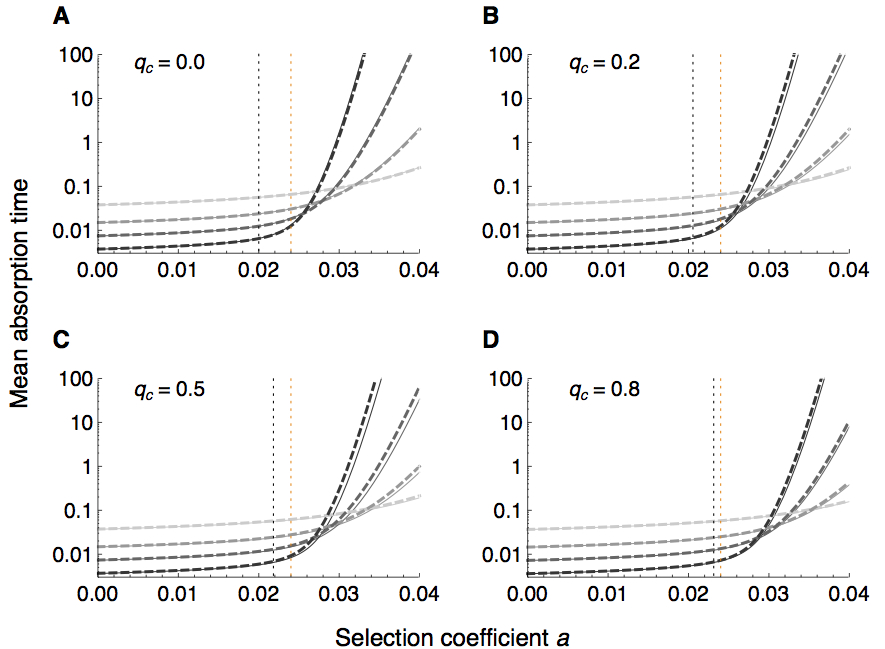}
\end{center}
\caption{
\textrm{Mean absorption time of $A_1$ as a function of its selective advantage.} Two approximations derived under the assumption of quasi-linkage eqilibrium (QLE) are compared. Solid curves show $\tbarQLE$ (Eq.\ \ref{eq:meanAbsTimeQLE}) and thick dashed curves $\bar{t}_{\mathrm{QLE}, \rho \gg 0}$ (Eq.\ \ref{eq:meanAbsTimeQLERhoLarge} in File \ref{txt:SI}). 
The effective population size $N_e$ increases from light to dark grey, taking values of 100, 250, 500, and 1000. The vertical dotted lines denote the critical values of $a$ above which $A_1$ can invade in the deterministic one-locus (orange) and two-locus (black) model. \textbf{(A)} Monomorphic continent ($q_c = 0$). \textbf{(B)}--\textbf{(D)} Polymorphic continent with $q_c$ equal to 0.2 in \textbf{(B)}, 0.5 in \textbf{(C)} and 0.8 in \textbf{(D)}. Other parameters are $b=0.04$, $m= 0.024$, $r=0.1$, and  $p_0 = 1/(2N)$ (we assumed $N_e = N$). Time is in multiples of $2N_e$ generations and plotted on the log scale.\\
}
\label{fig:meanAbsTime_effectOfA_variousqC}
\end{figure}

\clearpage

\begin{figure}[!ht]
\begin{center}
\includegraphics[width=1\textwidth]{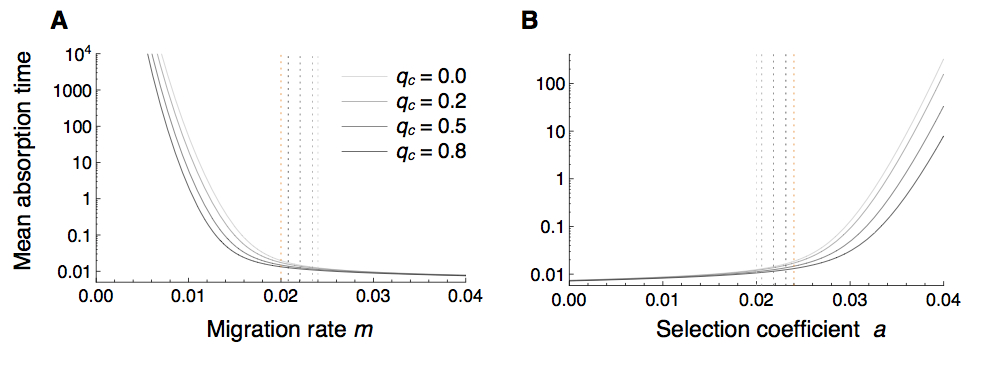}
\end{center}
\caption{
\textrm{Effect of the continental frequency $q_c$ of $B_1$ on the mean absorption time of $A_1$.} Curves show the diffusion approximation $\bar{t}_{\textbf{QLE}}$ (Eq.\ \ref{eq:meanAbsTimeQLE}), derived under the assumption of quasi-linkage equilibrium. The continental frequency $q_c$ of $B_1$ increases from light to dark grey, taking values of 0, 0.2, 0.5, and 0.8. \textbf{(A)} The mean absorption time is given in multiples of $2 N_e$ generations as a function of the migration rate. Vertical dotted lines denote the critical values of $m$ below which $A_1$ can invade in the respective deterministic two-locus model (grey) and, as a reference, in the one-locus model (orange). The selection coefficient in favour of $A_1$ is $a=0.02$. \textbf{(B)} As in \textbf{(A)}, but as a function of the selective advantage of allele $A_1$. The migration rate is $m = 0.024$. In both panels, $b = 0.04$, $r = 0.1$, $N_e = 500$, and $p_0 = 1/(2N)$ (we assumed $N_e = N$).\\
}
\label{fig:meanAbsTime_effectOfMandA_variousqC}
\end{figure}

\clearpage

\begin{figure}[!ht]
\begin{center}
\includegraphics[width=1\textwidth]{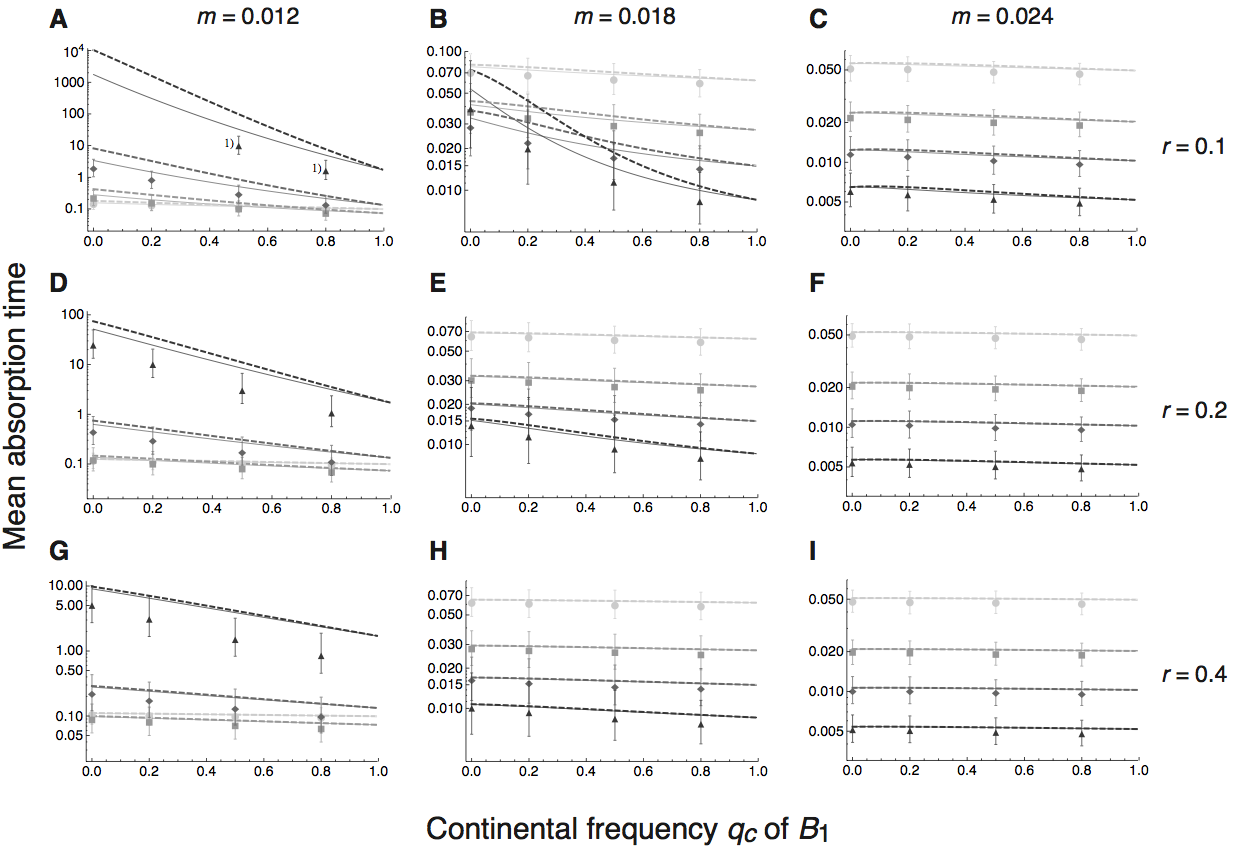}
\end{center}
\caption{
\textrm{Mean absorption time of $A_1$ as a function of the continental frequency $B_1$.} Two approximations derived under the assumption of quasi-linkage eqilibrium (QLE) are compared. Solid curves show $\tbarQLE$ (Eq.\ \ref{eq:meanAbsTimeQLE}) and thick dashed curves $\bar{t}_{\mathrm{QLE}, \rho \gg 0}$ (Eq.\ \ref{eq:meanAbsTimeQLERhoLarge} in File \ref{txt:SI}). 
The effective population size $N_e$ increases from light to dark grey, taking values of 100, 250, 500, and 1000. Dots and whiskers show the mean and 95\% empirical interquantile range across 1000 runs of the mean absorption time in 1000 replicates per run. Where points and whiskers are missing, simulations could not be completed within the time limit of 72 hours per replicate on the computer cluster. Data points labelled by 1) represent parameter combinations for which fewer than 1000 replicates per run could be realised, because some took longer than the limit of 72 hours. The migration rate $m$ increases from the left to the right column, taking values of 0.012, 0.018, and 0.024. The recombination rate $r$ increases from the top to the bottom row, taking values of 0.1, r=0.2, and r=0.4. Other parameters are $a=0.02$, $b=0.04$, and $p_0 = 1/(2N)$ (we assumed $N_e = N$). Time is in multiples of $2N_e$ generations and plotted on the log scale.\\
}
\label{fig:meanAbsTimeQLEWFrLargeqCComb}
\end{figure}

\clearpage

\begin{figure}[!ht]
\begin{center}
\includegraphics[width=1\textwidth]{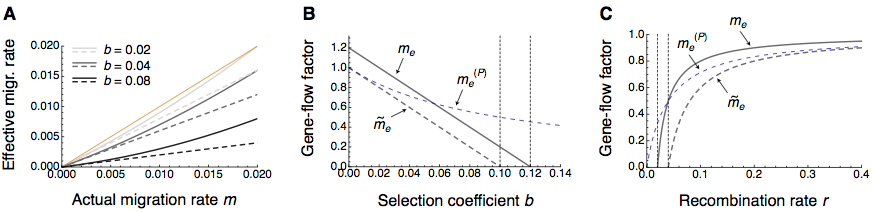}
\end{center}
\caption{
{\textrm{Effective migration rate at a weakly beneficial mutation arising in linkage to a migration--selection polymorphism.}} \textbf{(A)} The effective migration rate under the QLE approximation up to second ($m_e$, Eq.\ \ref{eq:mEffQLE}, solid) and first ($\tilde{m}_e$, Eq.\ \ref{eq:mEffQLEweakMig}, dashed) order of the actual migration rate $m$. The orange curve has a slope of 1 and represents the marginal case of linkage to a neutral background ($b=0$). Parameter values are $b=0.02$ (light grey), $b=0.04$ (medium grey), $b=0.08$ (black), and $r=0.1$. \textbf{(B)} The gene-flow factor (ratio of effective to actual migration rate, \citealt{Bengtsson:1985fk}) as a function of the selective advantage $b$ of the beneficial background allele $B_1$. Grey solid and dashed curves show the gene-flow factor computed using $m_e$ and $\tilde{m}_e$, respectively. The curves cross the horizontal axis at $b = m+r$ and $b = r$, respectively (vertical lines). The blue dashed curve gives the gene-flow factor for \citeauthor{Petry:1983uq}'s \citeyearpar{Petry:1983uq} $m_e^{(P)}$ in Eq.\ \eqref{eq:mEffPetry}. Parameters are $m=0.02$ and $r=0.1$. \textbf{(C)} As in panel $\textbf{(B)}$, but as a function of the recombination rate. Vertical dotted lines indicate $r = b-m$ and $r=b$ for $m_e/m$ and $\tilde{m}_e/m$, respectively. Parameters are $b = 0.04$ and $m=0.02$.\\
}
\label{fig:mEffSojourn}
\end{figure}

\clearpage

\begin{figure}[!ht]
\begin{center}
\includegraphics[width=1\textwidth]{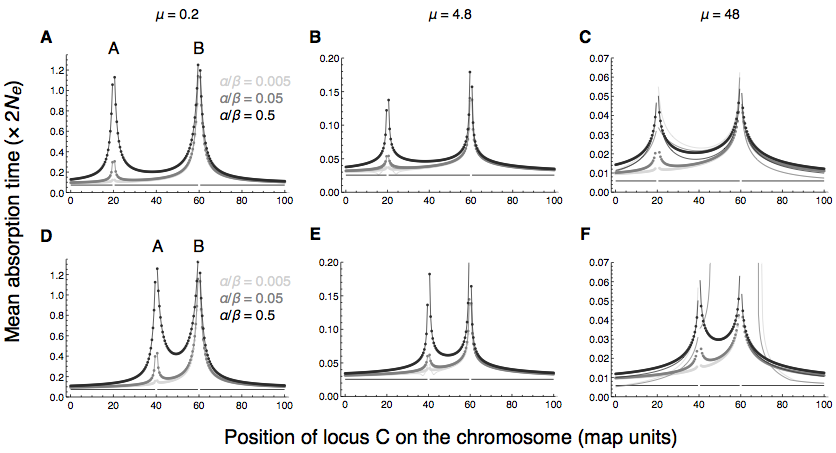}
\end{center}
\caption{
{\textrm{Effect of linkage to two selected sites on the absorption time of a neutral mutation.}} The mutation occurs at the neutral locus $\cloc$. The loci \textsf{A} and \textsf{B} under selection are located 20 and 60 map units from the left end of the chromosome in panels \textbf{(A)}--\textbf{(C)}, whereas locus \textsf{A} is located 40 map units from the left end of the chromosome in panels \textbf{(D)}--\textbf{(F)}. One map unit (centimorgan) corresponds to a recombination rate of $r=0.01$ and the effective population size is $N_e = 100$. The scaled selection coefficient in favour of $B_1$ is $\beta = 80$ and the scaled migration rate increases from left to right from $\mu = 0.2$ in \textbf{(A)} and \textbf{(D)} to $\mu = 4.8$ in \textbf{(B)} and \textbf{(E)} and $\mu = 48$ in \textbf{(C)} and \textbf{(F)}. From light to dark, $\alpha/\beta$ takes values of 0.005, 0.05, and 0.5, where $\alpha$ is the scaled selection coefficient in favour of $A_1$. Points show values computed using the approximate effective migration rates in Eq.\ \eqref{eq:effMigNeutr} and curves are based on numerically computed exact effective migration rates (Procecure \ref{prc:stochDiffNeutrVar}). For $\mu$ large and $\alpha$ small (light grey curves in \textbf{F}), the latter were affected by numerical errors causing strong deviation. 
The horizontal black line denotes the baseline for free recombination between locus \textsf{C} and the selected sites.\\
}
\label{fig:meanabstimeNeutr}
\end{figure}

\clearpage

\begin{figure}[!ht]
\begin{center}
\includegraphics[width=1\textwidth]{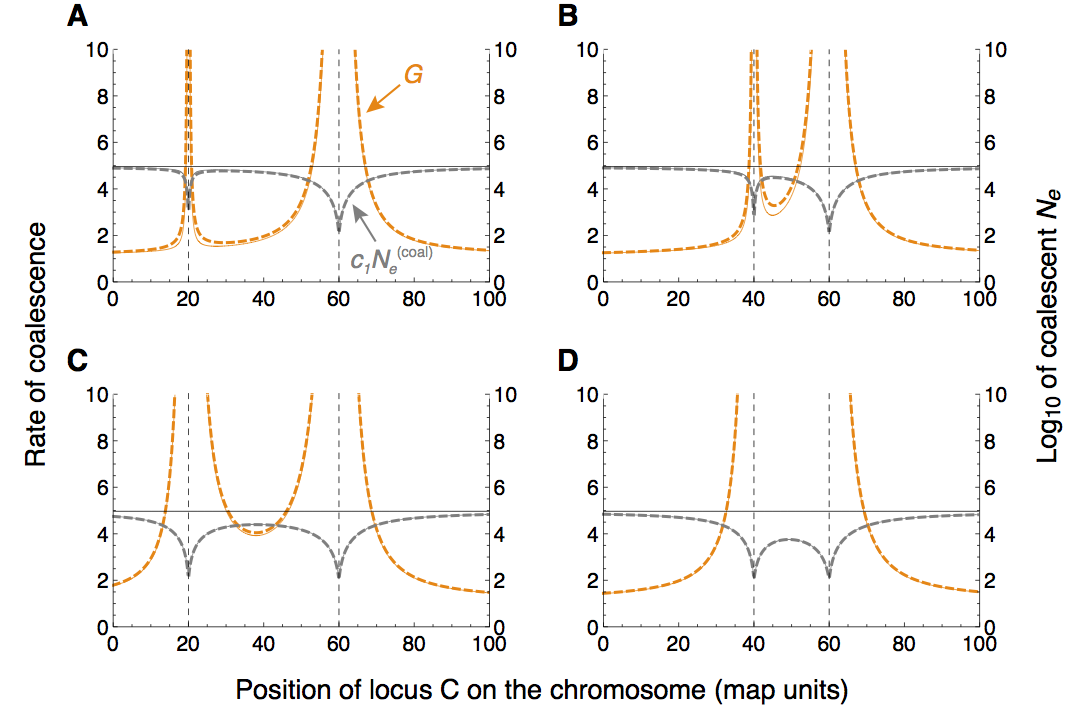}
\end{center}
\caption{
{\textrm{Effect on neutral coalescence of linkage to two sites at migration--selection balance.}} The rate of coalescence $G$ (orange, see Eq.\ \ref{eq:coalRateNeutr}) and the coalescent effective size of the island population, $c_1\bar{N}/G$, are given as a function of the position (in map units) of the neutral locus \textsf{C}. Solid and thick dashed curves are for values computed using the exact and approximate (Eq.\ \ref{eq:effMigNeutr}) effective migration rate, respectively (they overlap almost completely). One map unit (centimorgan) corresponds to a recombination rate of $r=0.01$ and the position of the sites under selection is indicated by vertical dashed lines. The total population size is $\bar{N} = 10^8$, the fraction of the island is $N_1/\bar{N} = c_1 = 0.01$ and the selection coefficient at locus \textsf{B} (position 60) is $b=0.4$. \textbf{(A)} and \textbf{(B)} The migration rate to the island is of the same order of magnitude as selection at locus \textsf{A}: $a=0.02$, $m_1 = 0.024$. \textbf{(C)} and \textbf{(D)} Immigration is weak compared to selection at locus \textsf{A}: $a=0.2$, $m_1 = 0.024$. Throughout, $m_1/m_2 = c_2/c_1 = (1 - c_1)/c_1$, so actual migration is conservative \cite[p.\ 194]{Wakeley:2009kx}. The horizontal black line gives the baseline-effective population size at the neutral locus in the absence of linked selection. For alternative parameter combinations, see File \ref{prc:stochDiffNeutrVar}.\\
}
\label{fig:coalRateAndNeNeutr}
\end{figure}

\clearpage

\begin{figure}[!ht]
\begin{center}
\includegraphics[width=1\textwidth]{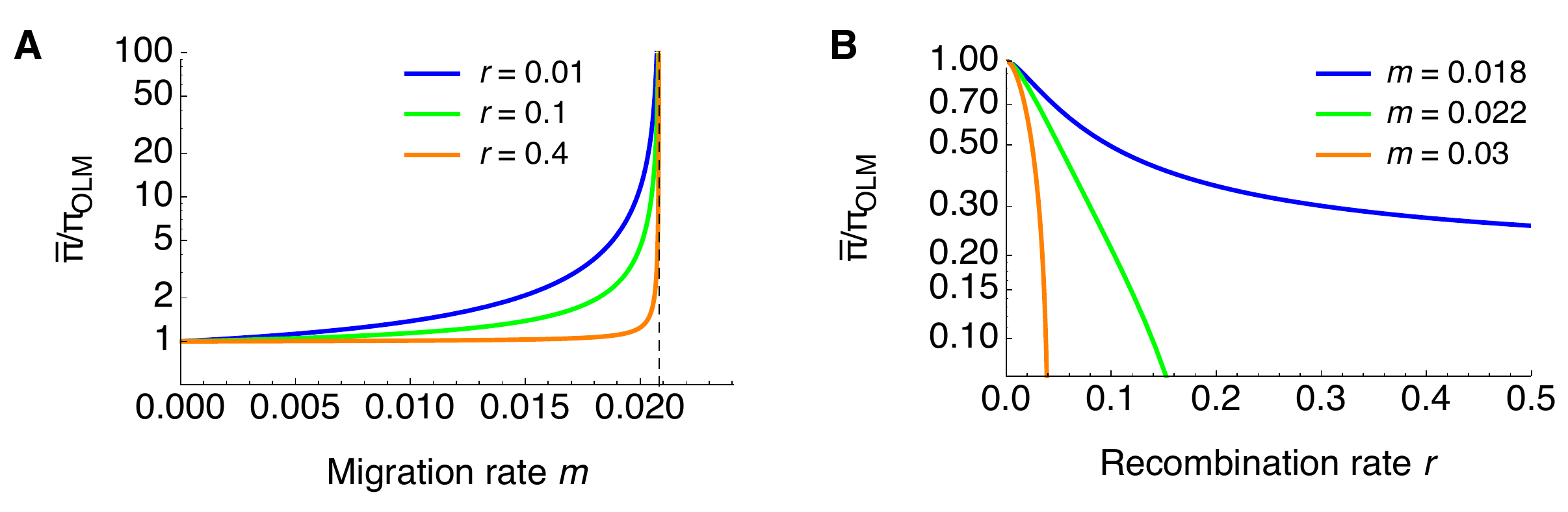}
\end{center}
\caption{
{\textrm{Mean invasion probability of $A_1$ with linkage to a background polymorphism compared to no linkage.}} Curves show the ratio of the weighted mean invasion probability, $\bar{\pi}$, divided by that of the one-locus model, $\pi_{\mathrm{OLM}}$ ($r = 0.5$). The ratio was computed from numerical solutions to the branching process (Eq.\ \ref{eq:bpSol}) and is shown as a function of the migration ($m$) and recombination ($r$) rate in panels \textbf{(A)} and \textbf{(B)}, respectively. The vertical dashed line in panel \textbf{(A)} shows the critical migration rate $a/(1-b)$, beyond which allele $A_1$ cannot be established under the deterministic one-locus model. In panel $\textbf{(B)}$, for $m = 0.018$ (blue curve), allele $A_1$ can be established independently of $r$. For stronger migration (green and orange curves), $A_1$ can be established only if $r$ is below a critical value (where the green and orange curves cross the x-axis, respectively). Other parameter values are $a = 0.02$, $b = 0.04$, and $q_c = 0$. Compare to Figure \ref{fig:meanabstimeQLEOLM} for the relative effect of $m$ on mean extinction time.\\
}
\label{fig:ratioTLM2OLMInvProb}
\end{figure}

\clearpage

\begin{figure}[!ht]
	\begin{center}
	\includegraphics[width = 1\textwidth]{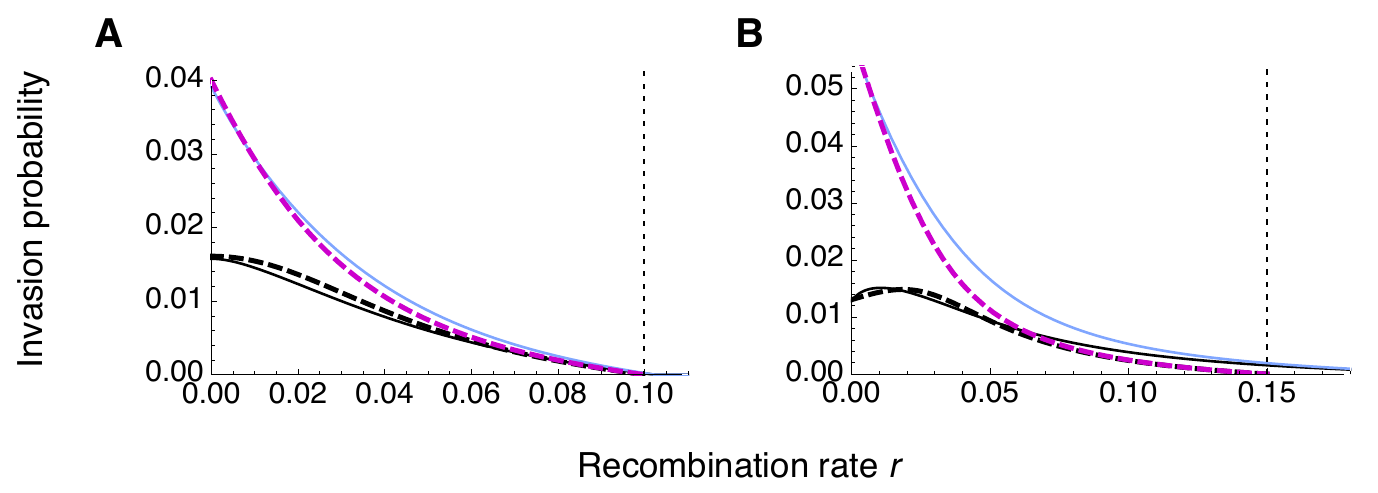}
	\end{center}
	\caption{
	\textrm{Comparison of branching-proccess and `splicing' approximations to the invasion probability of $A_1$ as a function of the recombination rate}. Black curves represent the branching-process solution averaged across the two backgrounds ($B_1$ and $B_2$). The solid curve gives the exact numerical solution and the dashed curve the analytical approximation for a slighty-supercritical process (based on Eq.\ \ref{eq:approxInvProbAdd}). The dashed purple curve represents the approximation based on the `splicing approach' as proposed by \citet{Yeaman:2013uq}. As a reference, the thin blue curve gives the numerical branching-process solution conditional on $A_1$ arising on the beneficial background $B_1$. \textbf{(A)} A case where $r_{\mathrm{opt}} = 0$; $a = 0.02$, $m = 0.024$. \textbf{(B)} A case where $r_{\mathrm{opt}} > 0$; $a = 0.03$, $m = 0.032$. In both panels, $b = 0.04$, and the vertical dotted line gives the critical recombination rate below which $A_1$ can invade according to deterministic continuous-time theory.
	}
	\label{fig:invProbCompAnalytApprox}
\end{figure}

\clearpage
\renewcommand{\thetable}{S\arabic{table}}
\setcounter{table}{0}
\section*{Supporting Information: Tables}

\begin{table}[!ht]
\centering 
\topcaption{
\bf{Simulated sojourn and absorption times and comparison to diffusion approximation. (Electronically only)
}
}
\label{tab:summaryTableSTDAndMAT}
\end{table}

\begin{sidewaystable}[!ht]
\centering 
\begin{threeparttable}
\topcaption{
\bf{The effect of assuming $p_0$ small in the diffusion approximation of the mean absorption time.}
\label{tab:relErrorp0Small}}
\small
\begin{tabular}{@{\extracolsep{\fill}} r@{.}l r@{.}l  r@{.}l r@{.}l r@{.}l r@{.}l r@{.}l r@{.}l r@{.}l r@{.}l r@{.}l r@{.}l r@{.}l r@{.}l @{}}
\toprule 
\multicolumn{4}{c}{} & \multicolumn{8}{c}{$N_e = 100$} & \multicolumn{8}{c}{$N_e = 10^3$} & \multicolumn{8}{c}{$N_e = 10^4$} \\
\cmidrule(lr){5-12}
\cmidrule(lr){13-20}
\cmidrule(l){21-28}
\multicolumn{2}{c}{$r$} & \multicolumn{2}{c}{$m$} & \multicolumn{2}{c}{$q_c = 0$} & \multicolumn{2}{c}{$q_c = 0.2$} & \multicolumn{2}{c}{$q_c = 0.5$} & \multicolumn{2}{c}{$q_c = 0.8$} &  \multicolumn{2}{c}{$q_c = 0$} & \multicolumn{2}{c}{$q_c = 0.2$} & \multicolumn{2}{c}{$q_c = 0.5$} & \multicolumn{2}{c}{$q_c = 0.8$} & \multicolumn{2}{c}{$q_c = 0$} & \multicolumn{2}{c}{$q_c = 0.2$} & \multicolumn{2}{c}{$q_c = 0.5$} & \multicolumn{2}{c}{$q_c = 0.8$}\\
\midrule
0&05 & 0&006 & 0&017 & 0&016 & 0&015 & 0&014 & 0&017 & 0&017 & 0&016 & 0&015 & 0&017 & 0&017 & 0&016 & 0&015 \\
0&05 & 0&012 & 0&014 & 0&012 & 0&010 & 0&009 & 0&014 & 0&012 & 0&011 & 0&009 & 0&014 & 0&012 & 0&011 & 0&009 \\
0&05 & 0&018 & 0&009 & 0&007 & 0&005 & 0&003 & 0&010 & 0&008 & 0&005 & 0&003 & 0&010 & 0&008 & 0&005 & 0&003 \\
0&05 & 0&024 & 0&004 & 0&002 & 0&000 & $-$0&002 & 0&004 & 0&002 & 0&000 & $-$0&002 & 0&004 & 0&002 & 0&000 & $-$0&002 \\ [0.8ex]
0&10 & 0&006 & 0&016 & 0&015 & 0&015 & 0&014 & 0&016 & 0&015 & 0&015 & 0&014 & 0&016 & 0&015 & 0&015 & 0&014 \\
0&10 & 0&012 & 0&011 & 0&010 & 0&009 & 0&008 & 0&011 & 0&010 & 0&009 & 0&009 & 0&011 & 0&010 & 0&009 & 0&009 \\
0&10 & 0&018 & 0&005 & 0&005 & 0&004 & 0&002 & 0&006 & 0&005 & 0&004 & 0&003 & 0&006 & 0&005 & 0&004 & 0&003 \\
0&10 & 0&024 & 0&000 & $-$0&001 & $-$0&002 & $-$0&003 & 0&000 & $-$0&001 & $-$0&002 & $-$0&003 & 0&000 & $-$0&001 & $-$0&002 & $-$0&003 \\  [0.8ex]
0&20 & 0&006 & 0&015 & 0&015 & 0&014 & 0&014 & 0&015 & 0&015 & 0&015 & 0&014 & 0&015 & 0&015 & 0&015 & 0&014 \\
0&20 & 0&012 & 0&009 & 0&009 & 0&008 & 0&008 & 0&010 & 0&009 & 0&009 & 0&008 & 0&010 & 0&009 & 0&009 & 0&008 \\
0&20 & 0&018 & 0&004 & 0&003 & 0&003 & 0&002 & 0&004 & 0&003 & 0&003 & 0&002 & 0&004 & 0&004 & 0&003 & 0&002 \\
0&20 & 0&024 & $-$0&002 & $-$0&002 & $-$0&003 & $-$0&003 & $-$0&002 & $-$0&002 & $-$0&003 & $-$0&003 & $-$0&002 & $-$0&002 & $-$0&003 & $-$0&003 \\
\bottomrule  
\end{tabular}
\begin{tablenotes}[para, flushleft]
The relative error $\tbartilQLE/\tbarQLE - 1$ is tablulated. The initial frequency of the focal mutant $A_1$ is $p_0 = 1/(2N)$ (we assumed $N_e = N$). Other parameters are $a = 0.02$ and $b = 0.04$. For a graphical representation, see Figure \ref{fig:relErrDiffApproxComb}A.
\end{tablenotes}
\end{threeparttable}
\end{sidewaystable}

\begin{sidewaystable}[!ht]
\centering 
\begin{threeparttable}
\topcaption{
\bf{The effect of assuming $\rho$ large in $M(p)$ when deriving the diffusion approximation to the mean absorption time.}
\label{tab:relErrorrhoLarge}}
\scriptsize
\begin{tabular}{@{\extracolsep{\fill}} r@{.}l r@{.}l  r@{.}l r@{.}l r@{.}l r@{.}l r@{.}l r@{.}l r@{.}l r@{.}l r@{.}l r@{.}l r@{.}l r@{.}l @{}}
\toprule 
\multicolumn{4}{c}{} & \multicolumn{8}{c}{$N_e = 100$} & \multicolumn{8}{c}{$N_e = 10^3$} & \multicolumn{8}{c}{$N_e = 10^4$} \\
\cmidrule(lr){5-12}
\cmidrule(lr){13-20}
\cmidrule(l){21-28}
\multicolumn{2}{c}{$r$} & \multicolumn{2}{c}{$m$} & \multicolumn{2}{c}{$q_c = 0$} & \multicolumn{2}{c}{$q_c = 0.2$} & \multicolumn{2}{c}{$q_c = 0.5$} & \multicolumn{2}{c}{$q_c = 0.8$} &  \multicolumn{2}{c}{$q_c = 0$} & \multicolumn{2}{c}{$q_c = 0.2$} & \multicolumn{2}{c}{$q_c = 0.5$} & \multicolumn{2}{c}{$q_c = 0.8$} & \multicolumn{2}{c}{$q_c = 0$} & \multicolumn{2}{c}{$q_c = 0.2$} & \multicolumn{2}{c}{$q_c = 0.5$} & \multicolumn{2}{c}{$q_c = 0.8$}\\
\midrule
0&05 & 0&006 & 3&818 & 1&995 & 0&770 & 0&211 & 6&558$\times$10\textuperscript{5} & 1&892$\times$10\textsuperscript{4} & 242&595 & 6&217 & 1&397$\times$10\textsuperscript{57} & 1&503$\times$10\textsuperscript{42} & 4&312$\times$10\textsuperscript{23} & 3&395$\times$10\textsuperscript{8}\\
0&05 & 0&012 & 1&391 & 0&901 & 0&382 & 0&106 & 1&288$\times$10\textsuperscript{4} & 2&086$\times$10\textsuperscript{3} & 77&967 & 3&411 & 8&300$\times$10\textsuperscript{40} & 1&700$\times$10\textsuperscript{33} & 1&534$\times$10\textsuperscript{19} & 4&275$\times$10\textsuperscript{6}\\
0&05 & 0&018 & 0&280 & 0&299 & 0&167 & 0&054 & 27&006 & 41&590 & 5&361 & 0&371 & 6&349$\times$10\textsuperscript{14} & 1&583$\times$10\textsuperscript{17} & 1&140$\times$10\textsuperscript{10} & 448&323\\
0&05 & 0&024 & 0&025 & 0&118 & 0&092 & 0&035 & $-$0&014 & 0&620 & 0&235 & 0&056 & $-$0&427 & 1&367$\times$10\textsuperscript{4} & 1&808 & 0&075\\ [0.8ex]
0&10 & 0&006 & 0&404 & 0&301 & 0&169 & 0&060 & 22&759 & 11&577 & 3&772 & 0&843 & 3&891$\times$10\textsuperscript{13} & 7&582$\times$10\textsuperscript{10} & 5&379$\times$10\textsuperscript{6} & 435&536\\
0&10 & 0&012 & 0&149 & 0&129 & 0&080 & 0&030 & 4&922 & 4&152 & 1&995 & 0&539 & 5&884$\times$10\textsuperscript{7} & 1&595$\times$10\textsuperscript{7} & 7&357$\times$10\textsuperscript{4} & 86&049\\
0&10 & 0&018 & 0&033 & 0&048 & 0&037 & 0&015 & 0&372 & 0&565 & 0&292 & 0&073 & 48&506 & 487&778 & 88&493 & 3&323\\
0&10 & 0&024 & 0&003 & 0&022 & 0&022 & 0&010 & $-$0&008 & 0&038 & 0&037 & 0&015 & $-$0&029 & 0&077 & 0&053 & 0&018\\ [0.8ex]
0&20 & 0&006 & 0&083 & 0&067 & 0&042 & 0&017 & 1&181 & 0&915 & 0&528 & 0&191 & 2&241$\times$10\textsuperscript{3} & 623&130 & 66&327 & 4&677\\
0&20 & 0&012 & 0&027 & 0&027 & 0&019 & 0&008 & 0&439 & 0&442 & 0&313 & 0&125 & 39&447 & 41&132 & 15&525 & 2&415\\
0&20 & 0&018 & 0&006 & 0&010 & 0&009 & 0&004 & 0&029 & 0&056 & 0&044 & 0&017 & 0&643 & 1&783 & 1&417 & 0&405\\
0&20 & 0&024 & 0&000 & 0&005 & 0&006 & 0&003 & $-$0&001 & 0&006 & 0&008 & 0&004 & $-$0&003 & 0&008 & 0&010 & 0&005\\
\bottomrule  
\end{tabular}
\begin{tablenotes}[para, flushleft]
The relative error $\tbarQLErho/\tbarQLE - 1$ is tabulated. The initial frequency of the focal mutant $A_1$ is $p_0 = 1/(2N)$ (we assumed $N_e = N$). Other parameters are $a = 0.02$ and $b = 0.04$. For a graphical representation, see Figure \ref{fig:relErrDiffApproxComb}B.
\end{tablenotes}
\end{threeparttable}
\end{sidewaystable}

\begin{sidewaystable}[!ht]
\centering 
\begin{threeparttable}
\topcaption{
\bf{The effect of assuming $p_0$ small in the diffusion approximation to the mean absorption time, given the assumption of $\rho$ large in $M(p)$.}
\label{tab:relErrorp0SmallGivenrhoLarge}}
\footnotesize
\begin{tabular}{@{\extracolsep{\fill}} r@{.}l r@{.}l  r@{.}l r@{.}l r@{.}l r@{.}l r@{.}l r@{.}l r@{.}l r@{.}l r@{.}l r@{.}l r@{.}l r@{.}l @{}}
\toprule 
\multicolumn{4}{c}{} & \multicolumn{8}{c}{$N_e = 100$} & \multicolumn{8}{c}{$N_e = 10^3$} & \multicolumn{8}{c}{$N_e = 10^4$} \\
\cmidrule(lr){5-12}
\cmidrule(lr){13-20}
\cmidrule(l){21-28}
\multicolumn{2}{c}{$r$} & \multicolumn{2}{c}{$m$} & \multicolumn{2}{c}{$q_c = 0$} & \multicolumn{2}{c}{$q_c = 0.2$} & \multicolumn{2}{c}{$q_c = 0.5$} & \multicolumn{2}{c}{$q_c = 0.8$} &  \multicolumn{2}{c}{$q_c = 0$} & \multicolumn{2}{c}{$q_c = 0.2$} & \multicolumn{2}{c}{$q_c = 0.5$} & \multicolumn{2}{c}{$q_c = 0.8$} & \multicolumn{2}{c}{$q_c = 0$} & \multicolumn{2}{c}{$q_c = 0.2$} & \multicolumn{2}{c}{$q_c = 0.5$} & \multicolumn{2}{c}{$q_c = 0.8$}\\
\midrule
0&05 & 0&006 & 0&018 & 0&017 & 0&016 & 0&015 & 0&018 & 0&017 & 0&016 & 0&015 & 0&018 & 0&017 & 0&016 & 0&015 \\
0&05 & 0&012 & 0&015 & 0&013 & 0&011 & 0&009 & 0&015 & 0&014 & 0&012 & 0&009 & 0&015 & 0&014 & 0&012 & 0&009 \\
0&05 & 0&018 & 0&010 & 0&009 & 0&006 & 0&004 & 0&010 & 0&009 & 0&007 & 0&004 & 0&010 & 0&009 & 0&007 & 0&004 \\
0&05 & 0&024 & 0&003 & 0&003 & 0&001 & $-$0&002 & 0&004 & 0&004 & 0&001 & $-$0&002 & 0&004 & 0&004 & 0&001 & $-$0&002 \\[0.8ex]
0&10 & 0&006 & 0&016 & 0&016 & 0&015 & 0&014 & 0&016 & 0&016 & 0&015 & 0&014 & 0&016 & 0&016 & 0&015 & 0&014 \\
0&10 & 0&012 & 0&011 & 0&010 & 0&009 & 0&008 & 0&011 & 0&011 & 0&010 & 0&009 & 0&011 & 0&011 & 0&010 & 0&009 \\
0&10 & 0&018 & 0&006 & 0&005 & 0&004 & 0&003 & 0&006 & 0&005 & 0&004 & 0&003 & 0&006 & 0&005 & 0&004 & 0&003 \\
0&10 & 0&024 & 0&000 & 0&000 & $-$0&001 & $-$0&003 & 0&000 & 0&000 & $-$0&001 & $-$0&003 & 0&000 & 0&000 & $-$0&001 & $-$0&003 \\[0.8ex]
0&20 & 0&006 & 0&015 & 0&015 & 0&014 & 0&014 & 0&015 & 0&015 & 0&015 & 0&014 & 0&015 & 0&015 & 0&015 & 0&014 \\
0&20 & 0&012 & 0&009 & 0&009 & 0&009 & 0&008 & 0&010 & 0&009 & 0&009 & 0&008 & 0&010 & 0&009 & 0&009 & 0&008 \\
0&20 & 0&018 & 0&004 & 0&003 & 0&003 & 0&002 & 0&004 & 0&004 & 0&003 & 0&002 & 0&004 & 0&004 & 0&003 & 0&002 \\
0&20 & 0&024 & $-$0&002 & $-$0&002 & $-$0&002 & $-$0&003 & $-$0&002 & $-$0&002 & $-$0&002 & $-$0&003 & $-$0&002 & $-$0&002 & $-$0&002 & $-$0&003\\
\bottomrule  
\end{tabular}
\begin{tablenotes}[para, flushleft]
The relative error $\tbartilQLErho/\bar{t}_{\mathrm{QLE}, \rho \gg 0} - 1$ is tabulated. The initial frequency of the focal mutant $A_1$ is $p_0 = 1/(2N)$ (we assumed $N_e = N$). Other parameters are $a = 0.02$ and $b = 0.04$. For a graphical representation, see Figure \ref{fig:relErrDiffApproxComb}C.
\end{tablenotes}
\end{threeparttable}
\end{sidewaystable}

\begin{sidewaystable}[!ht]
\centering 
\begin{threeparttable}
\topcaption{
\bf{The mean absorption time under the QLE approximation relative to the one without linkage.}
\label{tab:ratioTAbsQLEToTAbsOLM}}
\scriptsize
\begin{tabular}{@{\extracolsep{\fill}} r@{.}l r@{.}l  r@{.}l r@{.}l r@{.}l r@{.}l r@{.}l r@{.}l r@{.}l r@{.}l r@{.}l r@{.}l r@{.}l r@{.}l @{}}
\toprule 
\multicolumn{4}{c}{} & \multicolumn{8}{c}{$N_e = 100$} & \multicolumn{8}{c}{$N_e = 10^3$} & \multicolumn{8}{c}{$N_e = 10^4$} \\
\cmidrule(lr){5-12}
\cmidrule(lr){13-20}
\cmidrule(l){21-28}
\multicolumn{2}{c}{$r$} & \multicolumn{2}{c}{$m$} & \multicolumn{2}{c}{$q_c = 0$} & \multicolumn{2}{c}{$q_c = 0.2$} & \multicolumn{2}{c}{$q_c = 0.5$} & \multicolumn{2}{c}{$q_c = 0.8$} &  \multicolumn{2}{c}{$q_c = 0$} & \multicolumn{2}{c}{$q_c = 0.2$} & \multicolumn{2}{c}{$q_c = 0.5$} & \multicolumn{2}{c}{$q_c = 0.8$} & \multicolumn{2}{c}{$q_c = 0$} & \multicolumn{2}{c}{$q_c = 0.2$} & \multicolumn{2}{c}{$q_c = 0.5$} & \multicolumn{2}{c}{$q_c = 0.8$}\\
\midrule
0&05 & 0&006 & 3&887 & 2&738 & 1&763 & 1&228 & 1&377$\times$10\textsuperscript{6} & 4&280$\times$10\textsuperscript{4} & 468&965 & 9&837 & 1&037$\times$10\textsuperscript{181} & 1&053$\times$10\textsuperscript{166} & 3&122$\times$10\textsuperscript{146} & 5&560$\times$10\textsuperscript{129}\\
0&05 & 0&012 & 2&563 & 1&897 & 1&386 & 1&116 & 3&898$\times$10\textsuperscript{10} & 7&261$\times$10\textsuperscript{8} & 8&047$\times$10\textsuperscript{6} & 2&895$\times$10\textsuperscript{5} & 3&186$\times$10\textsuperscript{94} & 1&105$\times$10\textsuperscript{77} & 1&511$\times$10\textsuperscript{57} & 2&154$\times$10\textsuperscript{42}\\
0&05 & 0&018 & 1&679 & 1&407 & 1&187 & 1&059 & 2&092$\times$10\textsuperscript{4} & 1&212$\times$10\textsuperscript{3} & 118&033 & 46&914 & 5&463$\times$10\textsuperscript{38} & 3&218$\times$10\textsuperscript{25} & 1&172$\times$10\textsuperscript{13} & 1&268$\times$10\textsuperscript{6}\\
0&05 & 0&024 & 1&335 & 1&224 & 1&111 & 1&037 & 246&649 & 150&044 & 111&839 & 97&857 & \multicolumn{2}{c}{NA} & \multicolumn{2}{c}{NA} & \multicolumn{2}{c}{NA} & \multicolumn{2}{c}{NA}\\ [0.8ex]
0&10 & 0&006 & 2&183 & 1&815 & 1&418 & 1&140 & 4&032$\times$10\textsuperscript{3} & 608&988 & 45&294 & 4&288 & 6&045$\times$10\textsuperscript{155} & 4&043$\times$10\textsuperscript{147} & 2&273$\times$10\textsuperscript{136} & 1&374$\times$10\textsuperscript{126}\\
0&10 & 0&012 & 1&580 & 1&403 & 1&209 & 1&071 & 5&262$\times$10\textsuperscript{7} & 9&268$\times$10\textsuperscript{6} & 9&804$\times$10\textsuperscript{5} & 1&497$\times$10\textsuperscript{5} & 2&960$\times$10\textsuperscript{65} & 6&234$\times$10\textsuperscript{57} & 6&338$\times$10\textsuperscript{47} & 2&248$\times$10\textsuperscript{39}\\
0&10 & 0&018 & 1&256 & 1&189 & 1&104 & 1&037 & 224&875 & 118&051 & 60&712 & 41&759 & 7&771$\times$10\textsuperscript{16} & 1&157$\times$10\textsuperscript{13} & 2&039$\times$10\textsuperscript{8} & 1&085$\times$10\textsuperscript{5}\\
0&10 & 0&024 & 1&131 & 1&109 & 1&064 & 1&024 & 116&181 & 111&037 & 102&175 & 95&712 & \multicolumn{2}{c}{NA} & \multicolumn{2}{c}{NA} & \multicolumn{2}{c}{NA} & \multicolumn{2}{c}{NA}\\ [0.8ex]
0&20 & 0&006 & 1&519 & 1&386 & 1&218 & 1&079 & 91&941 & 34&912 & 8&699 & 2&325 & 2&582$\times$10\textsuperscript{139} & 1&660$\times$10\textsuperscript{135} & 1&588$\times$10\textsuperscript{129} & 3&019$\times$10\textsuperscript{123}\\
0&20 & 0&012 & 1&247 & 1&188 & 1&108 & 1&040 & 1&550$\times$10\textsuperscript{6} & 7&361$\times$10\textsuperscript{5} & 2&525$\times$10\textsuperscript{5} & 9&391$\times$10\textsuperscript{4} & 6&923$\times$10\textsuperscript{49} & 3&274$\times$10\textsuperscript{46} & 5&148$\times$10\textsuperscript{41} & 1&717$\times$10\textsuperscript{37}\\
0&20 & 0&018 & 1&112 & 1&091 & 1&055 & 1&021 & 63&566 & 55&483 & 45&480 & 38&849 & 4&712$\times$10\textsuperscript{8} & 3&624$\times$10\textsuperscript{7} & 6&508$\times$10\textsuperscript{5} & 2&380$\times$10\textsuperscript{4}\\
0&20 & 0&024 & 1&059 & 1&054 & 1&035 & 1&014 & 101&157 & 100&356 & 97&307 & 94&249 & \multicolumn{2}{c}{NA} & \multicolumn{2}{c}{NA} & \multicolumn{2}{c}{NA} & \multicolumn{2}{c}{NA}\\
\bottomrule  
\end{tabular}
\begin{tablenotes}[para, flushleft]
Tabulated is the ratio $\tbarQLE / \bar{t}_{\mathrm{OLM}}$. The initial frequency of the focal mutant $A_1$ is $p_0 = 1/(2N)$ (we assumed $N_e = N$). Other parameters are $a = 0.02$ and $b = 0.04$. NA denotes cases where $\bar{t}_{\mathrm{OLM}}$ is numerically indistinguishable from 0 and hence the ratio $\tbarQLE/\bar{t}_{\mathrm{OLM}}$ not defined.
\end{tablenotes}
\end{threeparttable}
\end{sidewaystable}

\begin{sidewaystable}[!ht]
\centering 
\begin{threeparttable}
\topcaption{
\bf{The error of $\tbartilQLE$ relative to $\tbarQLE$ as in Table \ref{tab:relErrorp0Small}, but for $p_0 = 0.005$ fixed instead of $p_0 = 1/(2N)$.}
\label{tab:relErrorp0Smallp0Fixed}}
\footnotesize
\begin{tabular}{@{\extracolsep{\fill}} r@{.}l r@{.}l  r@{.}l r@{.}l r@{.}l r@{.}l r@{.}l r@{.}l r@{.}l r@{.}l r@{.}l r@{.}l r@{.}l r@{.}l @{}}
\toprule 
\multicolumn{4}{c}{} & \multicolumn{8}{c}{$N_e = 100$} & \multicolumn{8}{c}{$N_e = 10^3$} & \multicolumn{8}{c}{$N_e = 10^4$} \\
\cmidrule(lr){5-12}
\cmidrule(lr){13-20}
\cmidrule(l){21-28}
\multicolumn{2}{c}{$r$} & \multicolumn{2}{c}{$m$} & \multicolumn{2}{c}{$q_c = 0$} & \multicolumn{2}{c}{$q_c = 0.2$} & \multicolumn{2}{c}{$q_c = 0.5$} & \multicolumn{2}{c}{$q_c = 0.8$} &  \multicolumn{2}{c}{$q_c = 0$} & \multicolumn{2}{c}{$q_c = 0.2$} & \multicolumn{2}{c}{$q_c = 0.5$} & \multicolumn{2}{c}{$q_c = 0.8$} & \multicolumn{2}{c}{$q_c = 0$} & \multicolumn{2}{c}{$q_c = 0.2$} & \multicolumn{2}{c}{$q_c = 0.5$} & \multicolumn{2}{c}{$q_c = 0.8$}\\
\midrule
0&05 & 0&006 & 0&017 & 0&016 & 0&015 & 0&014 & 0&182 & 0&174 & 0&163 & 0&153 & 2&551 & 2&418 & 2&239 & 2&079 \\
0&05 & 0&012 & 0&014 & 0&012 & 0&010 & 0&009 & 0&144 & 0&128 & 0&108 & 0&092 & 1&943 & 1&690 & 1&389 & 1&144 \\
0&05 & 0&018 & 0&009 & 0&007 & 0&005 & 0&003 & 0&099 & 0&077 & 0&051 & 0&030 & 1&249 & 0&934 & 0&599 & 0&347 \\
0&05 & 0&024 &0&004 & 0&002 & 0&000 & $-$0&002 & 0&042 & 0&020 & $-$0&004 & $-$0&023 & 0&492 & 0&229 & $-$0&039 & $-$0&175 \\ [0.8ex]
0&10 & 0&006 & 0&016 & 0&015 & 0&015 & 0&014 & 0&166 & 0&162 & 0&156 & 0&150 & 2&297 & 2&228 & 2&131 & 2&039 \\
0&10 & 0&012 & 0&011 & 0&010 & 0&009 & 0&008 & 0&115 & 0&108 & 0&097 & 0&088 & 1&490 & 1&379 & 1&227 & 1&088 \\
0&10 & 0&018 & 0&005 & 0&005 & 0&004 & 0&002 & 0&059 & 0&050 & 0&037 & 0&025 & 0&698 & 0&590 & 0&434 & 0&295 \\
0&10 & 0&024 & 0&000 & $-$0&001 & $-$0&002 & $-$0&003 & 0&000 & $-$0&005 & $-$0&016 & $-$0&027 & $-$0&003 & $-$0&047 & $-$0&131 & $-$0&199 \\  [0.8ex]
0&20 & 0&006 & 0&015 & 0&015 & 0&014 & 0&014 & 0&157 & 0&155 & 0&152 & 0&148 & 2&148 & 2&113 & 2&062 & 2&012 \\
0&20 & 0&012 & 0&009 & 0&009 & 0&008 & 0&008 & 0&099 & 0&096 & 0&090 & 0&085 & 1&250 & 1&200 & 1&124 & 1&050 \\
0&20 & 0&018 & 0&004 & 0&003 & 0&003 & 0&002 & 0&038 & 0&035 & 0&029 & 0&022 & 0&445 & 0&404 & 0&331 & 0&257 \\
0&20 & 0&024 & $-$0&002 & $-$0&002 & $-$0&003 & $-$0&003 & $-$0&018 & $-$0&019 & $-$0&024 & $-$0&029 & $-$0&141 & $-$0&149 & $-$0&181 & $-$0&215 \\
\bottomrule  
\end{tabular}
\begin{tablenotes}[para, flushleft]
The relative error is computed as $\tbartilQLE/\tbarQLE - 1$. It quantifies the effect of assuming $p_0$ small in the derivation of the diffusion approximation of the mean absorption time. Other parameters are $a = 0.02$ and $b = 0.04$. For a graphical representation, see Figure \ref{fig:relErrDiffApproxComb}D.
\end{tablenotes}
\end{threeparttable}
\end{sidewaystable}

\begin{sidewaystable}[!ht]
\centering 
\begin{threeparttable}
\topcaption{
\bf{The error of $\bar{t}_{\mathrm{QLE}\rho \gg 0}$ relative to $\tbarQLE$ as in Table \ref{tab:relErrorrhoLarge}, but for $p_0 = 0.005$ fixed instead of $p_0 = 1/(2N)$.}
\label{tab:relErrorrhoLargep0Fixed}}
\scriptsize
\begin{tabular}{@{\extracolsep{\fill}} r@{.}l r@{.}l  r@{.}l r@{.}l r@{.}l r@{.}l r@{.}l r@{.}l r@{.}l r@{.}l r@{.}l r@{.}l r@{.}l r@{.}l @{}}
\toprule 
\multicolumn{4}{c}{} & \multicolumn{8}{c}{$N_e = 100$} & \multicolumn{8}{c}{$N_e = 10^3$} & \multicolumn{8}{c}{$N_e = 10^4$} \\
\cmidrule(lr){5-12}
\cmidrule(lr){13-20}
\cmidrule(l){21-28}
\multicolumn{2}{c}{$r$} & \multicolumn{2}{c}{$m$} & \multicolumn{2}{c}{$q_c = 0$} & \multicolumn{2}{c}{$q_c = 0.2$} & \multicolumn{2}{c}{$q_c = 0.5$} & \multicolumn{2}{c}{$q_c = 0.8$} &  \multicolumn{2}{c}{$q_c = 0$} & \multicolumn{2}{c}{$q_c = 0.2$} & \multicolumn{2}{c}{$q_c = 0.5$} & \multicolumn{2}{c}{$q_c = 0.8$} & \multicolumn{2}{c}{$q_c = 0$} & \multicolumn{2}{c}{$q_c = 0.2$} & \multicolumn{2}{c}{$q_c = 0.5$} & \multicolumn{2}{c}{$q_c = 0.8$}\\
\midrule
0&05 & 0&006 & 3&818 & 1&995 & 0&770 & 0&211 & 6&508$\times$10\textsuperscript{5} & 1&879$\times$10\textsuperscript{4} & 241&353 & 6&201 & 1&336$\times$10\textsuperscript{57} & 1&441$\times$10\textsuperscript{42} & 4&175$\times$10\textsuperscript{23} & 3&344$\times$10\textsuperscript{8}\\
0&05 & 0&012 & 1&391 & 0&901 & 0&382 & 0&106 & 1&278$\times$10\textsuperscript{4} & 2&064$\times$10\textsuperscript{3} & 77&230 & 3&392 & 7&871$\times$10\textsuperscript{40} & 1&579$\times$10\textsuperscript{33} & 1&430$\times$10\textsuperscript{19} & 4&120$\times$10\textsuperscript{6}\\
0&05 & 0&018 & 0&280 & 0&299 & 0&167 & 0&054 & 26&941 & 41&405 & 5&751 & 0&457 & 6&210$\times$10\textsuperscript{14} & 1&430$\times$10\textsuperscript{17} & 1&010$\times$10\textsuperscript{10} & 419&102\\
0&05 & 0&024 & 0&025 & 0&118 & 0&092 & 0&035 & $-$0&011 & 0&830 & 0&347 & 0&086 & $-$0&393 & 1&274$\times$10\textsuperscript{4} & 4&193 & 0&192\\ [0.8ex]
0&10 & 0&006 & 0&404 & 0&301 & 0&169 & 0&060 & 22&709 & 11&552 & 3&765 & 0&842 & 3&840$\times$10\textsuperscript{13} & 7&486$\times$10\textsuperscript{10} & 5&327$\times$10\textsuperscript{6} & 433&553\\
0&10 & 0&012 & 0&149 & 0&129 & 0&080 & 0&030 & 4&909 & 4&137 & 1&987 & 0&537 & 5&790$\times$10\textsuperscript{7} & 1&559$\times$10\textsuperscript{7} & 7&196$\times$10\textsuperscript{4} & 85&052\\
0&10 & 0&018 & 0&033 & 0&048 & 0&037 & 0&015 & 0&388 & 0&611 & 0&342 & 0&092 & 48&191 & 472&445 & 85&111 & 3&239\\
0&10 & 0&024 & 0&003 & 0&022 & 0&022 & 0&010 & $-$0&010 & 0&056 & 0&056 & 0&023 & $-$0&058 & 0&174 & 0&131 & 0&046\\ [0.8ex]
0&20 & 0&006 & 0&083 & 0&067 & 0&042 & 0&017 & 1&179 & 0&914 & 0&528 & 0&191 & 2&233$\times$10\textsuperscript{3} & 620&925 & 66&145 & 4&670\\
0&20 & 0&012 & 0&027 & 0&027 & 0&019 & 0&008 & 0&438 & 0&441 & 0&312 & 0&125 & 39&267 & 40&862 & 15&423 & 2&404\\
0&20 & 0&018 & 0&006 & 0&010 & 0&009 & 0&004 & 0&034 & 0&067 & 0&055 & 0&022 & 0&640 & 1&758 & 1&391 & 0&401\\
0&20 & 0&024 & 0&000 & 0&005 & 0&006 & 0&003 & $-$0&002 & 0&010 & 0&012 & 0&006 & $-$0&006 & 0&020 & 0&024 & 0&012\\
\bottomrule  
\end{tabular}
\begin{tablenotes}[para, flushleft]
The relative error is computed as $\tbarQLErho/\tbarQLE - 1$. It quantifies the effect of assuming $\rho$ very large in $M(p)$ when deriving the diffusion approximation of the mean absorption time. Other parameters are $a = 0.02$ and $b = 0.04$. For a graphical representation, see Figure \ref{fig:relErrDiffApproxComb}E.
\end{tablenotes}
\end{threeparttable}
\end{sidewaystable}

\begin{sidewaystable}[!ht]
\centering 
\begin{threeparttable}
\topcaption{
\bf{The error of $\tilde{\bar{t}}_{\mathrm{QLE}\rho \gg 0}$ relative to $\bar{t}_{\mathrm{QLE}, \rho \gg 0}$ as in Table \ref{tab:relErrorp0SmallGivenrhoLarge}, but for $p_0 = 0.005$ fixed instead of $p_0 = 1/(2N)$.}
\label{tab:relErrorp0SmallGivenrhoLargep0Fixed}}
\footnotesize
\begin{tabular}{@{\extracolsep{\fill}} r@{.}l r@{.}l  r@{.}l r@{.}l r@{.}l r@{.}l r@{.}l r@{.}l r@{.}l r@{.}l r@{.}l r@{.}l r@{.}l r@{.}l @{}}
\toprule 
\multicolumn{4}{c}{} & \multicolumn{8}{c}{$N_e = 100$} & \multicolumn{8}{c}{$N_e = 10^3$} & \multicolumn{8}{c}{$N_e = 10^4$} \\
\cmidrule(lr){5-12}
\cmidrule(lr){13-20}
\cmidrule(l){21-28}
\multicolumn{2}{c}{$r$} & \multicolumn{2}{c}{$m$} & \multicolumn{2}{c}{$q_c = 0$} & \multicolumn{2}{c}{$q_c = 0.2$} & \multicolumn{2}{c}{$q_c = 0.5$} & \multicolumn{2}{c}{$q_c = 0.8$} &  \multicolumn{2}{c}{$q_c = 0$} & \multicolumn{2}{c}{$q_c = 0.2$} & \multicolumn{2}{c}{$q_c = 0.5$} & \multicolumn{2}{c}{$q_c = 0.8$} & \multicolumn{2}{c}{$q_c = 0$} & \multicolumn{2}{c}{$q_c = 0.2$} & \multicolumn{2}{c}{$q_c = 0.5$} & \multicolumn{2}{c}{$q_c = 0.8$}\\
\midrule
0&05 & 0&006 & 0&018 & 0&017 & 0&016 & 0&015 & 0&192 & 0&183 & 0&169 & 0&156 & 2&716 & 2&569 & 2&347 & 2&126\\
0&05 & 0&012 & 0&015 & 0&013 & 0&011 & 0&009 & 0&154 & 0&141 & 0&120 & 0&097 & 2&107 & 1&900 & 1&566 & 1&226\\
0&05 & 0&018 & 0&010 & 0&009 & 0&006 & 0&004 & 0&102 & 0&092 & 0&068 & 0&038 & 1&298 & 1&145 & 0&806 & 0&443\\
0&05 & 0&024 & 0&003 & 0&003 & 0&001 & $-$0&002 & 0&035 & 0&035 & 0&013 & $-$0&015 & 0&410 & 0&410 & 0&141 & $-$0&120\\ [0.8ex]
0&10 & 0&006 & 0&016 & 0&016 & 0&015 & 0&014 & 0&169 & 0&164 & 0&158 & 0&151 & 2&342 & 2&271 & 2&162 & 2&053\\
0&10 & 0&012 & 0&011 & 0&010 & 0&009 & 0&008 & 0&118 & 0&111 & 0&101 & 0&090 & 1&531 & 1&435 & 1&277 & 1&113\\
0&10 & 0&018 & 0&006 & 0&005 & 0&004 & 0&003 & 0&060 & 0&055 & 0&042 & 0&028 & 0&709 & 0&642 & 0&491 & 0&323\\
0&10 & 0&024 & 0&000 & 0&000 & $-$0&001 & $-$0&003 & $-$0&002 & $-$0&002 & $-$0&011 & $-$0&024 & $-$0&016 & $-$0&016 & $-$0&096 & $-$0&183\\ [0.8ex]
0&20 & 0&006 & 0&015 & 0&015 & 0&014 & 0&014 & 0&158 & 0&155 & 0&152 & 0&149 & 2&160 & 2&124 & 2&071 & 2&016\\
0&20 & 0&012 & 0&009 & 0&009 & 0&009 & 0&008 & 0&100 & 0&097 & 0&091 & 0&086 & 1&260 & 1&214 & 1&138 & 1&057\\
0&20 & 0&018 & 0&004 & 0&003 & 0&003 & 0&002 & 0&038 & 0&036 & 0&030 & 0&023 & 0&447 & 0&416 & 0&346 & 0&266\\
0&20 & 0&024 & $-$0&002 & $-$0&002 & $-$0&002 & $-$0&003 & $-$0&018 & $-$0&018 & $-$0&023 & $-$0&029 & $-$0&143 & $-$0&143 & $-$0&173 & $-$0&211\\
\bottomrule  
\end{tabular}
\begin{tablenotes}[para, flushleft]
The relative error is computed as $\tbartilQLErho/\bar{t}_{\mathrm{QLE}, \rho \gg 0} - 1$. It quantifies the effect of assuming $p_0$ small, given the assumption of $\rho$ large in $M(p)$ when deriving the diffusion approximation of the mean absorption time. Other parameters are $a = 0.02$ and $b = 0.04$. For a graphical representation, see Figure \ref{fig:relErrDiffApproxComb}F.
\end{tablenotes}
\end{threeparttable}
\end{sidewaystable}

\begin{sidewaystable}[!ht]
\centering 
\begin{threeparttable}
\topcaption{
\bf{The ratio $\tbarQLE / \bar{t}_{\mathrm{OLM}}$ as in Table \ref{tab:ratioTAbsQLEToTAbsOLM}, but for $p_0 = 0.005$ instead of $p_0 = 1/(2N)$.}
\label{tab:ratioTAbsQLEToTAbsOLMp0Fixed}}
\scriptsize
\begin{tabular}{@{\extracolsep{\fill}} r@{.}l r@{.}l  r@{.}l r@{.}l r@{.}l r@{.}l r@{.}l r@{.}l r@{.}l r@{.}l r@{.}l r@{.}l r@{.}l r@{.}l @{}}
\toprule 
\multicolumn{4}{c}{} & \multicolumn{8}{c}{$N_e = 100$} & \multicolumn{8}{c}{$N_e = 10^3$} & \multicolumn{8}{c}{$N_e = 10^4$} \\
\cmidrule(lr){5-12}
\cmidrule(lr){13-20}
\cmidrule(l){21-28}
\multicolumn{2}{c}{$r$} & \multicolumn{2}{c}{$m$} & \multicolumn{2}{c}{$q_c = 0$} & \multicolumn{2}{c}{$q_c = 0.2$} & \multicolumn{2}{c}{$q_c = 0.5$} & \multicolumn{2}{c}{$q_c = 0.8$} &  \multicolumn{2}{c}{$q_c = 0$} & \multicolumn{2}{c}{$q_c = 0.2$} & \multicolumn{2}{c}{$q_c = 0.5$} & \multicolumn{2}{c}{$q_c = 0.8$} & \multicolumn{2}{c}{$q_c = 0$} & \multicolumn{2}{c}{$q_c = 0.2$} & \multicolumn{2}{c}{$q_c = 0.5$} & \multicolumn{2}{c}{$q_c = 0.8$}\\
\midrule
0&05 & 0&006 & 3&887 & 2&738 & 1&763 & 1&228 & 1&340$\times$10\textsuperscript{6} & 4&191$\times$10\textsuperscript{4} & 463&139 & 9&790 & 2&964$\times$10\textsuperscript{180} & 3&124$\times$10\textsuperscript{165} & 9&765$\times$10\textsuperscript{145} & 1&828$\times$10\textsuperscript{129}\\
0&05 & 0&012 & 2&563 & 1&897 & 1&386 & 1&116 & 3&451$\times$10\textsuperscript{10} & 6&511$\times$10\textsuperscript{8} & 7&330$\times$10\textsuperscript{6} & 2&672$\times$10\textsuperscript{5} & 1&095$\times$10\textsuperscript{94} & 4&148$\times$10\textsuperscript{76} & 6&376$\times$10\textsuperscript{56} & 1&011$\times$10\textsuperscript{42}\\
0&05 & 0&018 & 1&679 & 1&407 & 1&187 & 1&059 & 2&634$\times$10\textsuperscript{4} & 1&542$\times$10\textsuperscript{3} & 141&905 & 49&620 & 2&614$\times$10\textsuperscript{38} & 1&786$\times$10\textsuperscript{25} & 7&849$\times$10\textsuperscript{12} & 1&006$\times$10\textsuperscript{6}\\
0&05 & 0&024 & 1&335 & 1&224 & 1&111 & 1&037 & 351&786 & 190&096 & 126&558 & 103&788 & \multicolumn{2}{c}{NA} & \multicolumn{2}{c}{NA} & \multicolumn{2}{c}{NA} & \multicolumn{2}{c}{NA}\\ [0.8ex]
0&10 & 0&006 & 2&183 & 1&815 & 1&418 & 1&140 & 3&971$\times$10\textsuperscript{3} & 601&729 & 44&963 & 4&276 & 1&858$\times$10\textsuperscript{155} & 1&269$\times$10\textsuperscript{147} & 7&351$\times$10\textsuperscript{135} & 4&574$\times$10\textsuperscript{125}\\
0&10 & 0&012 & 1&580 & 1&403 & 1&209 & 1&071 & 4&768$\times$10\textsuperscript{7} & 8&447$\times$10\textsuperscript{6} & 9&010$\times$10\textsuperscript{5} & 1&386$\times$10\textsuperscript{5} & 1&199$\times$10\textsuperscript{65} & 2&641$\times$10\textsuperscript{57} & 2&866$\times$10\textsuperscript{47} & 1&083$\times$10\textsuperscript{39}\\
0&10 & 0&018 & 1&256 & 1&189 & 1&104 & 1&037 & 280&052 & 142&014 & 67&583 & 42&921 & 4&907$\times$10\textsuperscript{16} & 7&793$\times$10\textsuperscript{12} & 1&521$\times$10\textsuperscript{8} & 8&927$\times$10\textsuperscript{4}\\
0&10 & 0&024 & 1&131 & 1&109 & 1&064 & 1&024 & 133&739 & 125&264 & 110&775 & 100&350 & \multicolumn{2}{c}{NA} & \multicolumn{2}{c}{NA} & \multicolumn{2}{c}{NA} & \multicolumn{2}{c}{NA}\\ [0.8ex]
0&20 & 0&006 & 1&519 & 1&386 & 1&218 & 1&079 & 91&196 & 34&687 & 8&664 & 2&321 & 8&305$\times$10\textsuperscript{138} & 5&399$\times$10\textsuperscript{134} & 5&247$\times$10\textsuperscript{128} & 1&014$\times$10\textsuperscript{123}\\
0&20 & 0&012 & 1&247 & 1&188 & 1&108 & 1&040 & 1&423$\times$10\textsuperscript{6} & 6&775$\times$10\textsuperscript{5} & 2&333$\times$10\textsuperscript{5} & 8&711$\times$10\textsuperscript{4} & 3&099$\times$10\textsuperscript{49} & 1&498$\times$10\textsuperscript{46} & 2&439$\times$10\textsuperscript{41} & 8&426$\times$10\textsuperscript{36}\\
0&20 & 0&018 & 1&112 & 1&091 & 1&055 & 1&021 & 71&310 & 60&791 & 47&766 & 39&139 & 3&490$\times$10\textsuperscript{8} & 2&761$\times$10\textsuperscript{7} & 5&223$\times$10\textsuperscript{5} & 1&997$\times$10\textsuperscript{4}\\
0&20 & 0&024 & 1&059 & 1&054 & 1&035 & 1&014 & 109&131 & 107&834 & 102&913 & 98&013 & \multicolumn{2}{c}{NA} & \multicolumn{2}{c}{NA} & \multicolumn{2}{c}{NA} & \multicolumn{2}{c}{NA}\\
\bottomrule  
\end{tabular}
\begin{tablenotes}[para, flushleft]
Here, $\tbarQLE$ is the mean absorption time assuming quasi-linkage equilibrium (QLE), and $\bar{t}_{\mathrm{OLM}}$ the one for the one-locus model (no linkage). Parameters are $a = 0.02$ and $b = 0.04$. NA denotes cases where $\bar{t}_{\mathrm{OLM}}$ is numerically indistinguishable from 0 and hence the ratio $\tbarQLE/\bar{t}_{\mathrm{OLM}}$ not defined.
\end{tablenotes}
\end{threeparttable}
\end{sidewaystable}
\clearpage
\renewcommand{\thesection}{S\arabic{section}}
\renewcommand{\thefile}{S\arabic{file}}
\setcounter{file}{0}

\begin{file}[h!]
	\topcaption[Additional Methods]{}
	\label{txt:SI}
\end{file}
\section{Supporting Information: Additional Methods}
\subsection{Details of the model}
We denote the frequencies on the island of haplotypes $A_1B_1$, $A_1B_2$, $A_2B_1$, and $A_2B_2$ by $x_1$, $x_2$, $x_3$, and $x_4$, respectively. The haplotype frequencies are related to the allele frequencies ($p$, $q$) and the linkage disequilibrium ($D$) as follows \cite[e.g.][]{Buerger:2000fk}. The frequencies of $A_1$ and $B_1$ on the island can be expressed as $p = x_1 + x_2$ and $q = x_1 + x_3$. Accordingly, the frequencies of $A_2$ and $B_2$ are $1-p = x_3 + x_4$ and $1-q = x_2 + x_4$. Moreover, $x_1 = p q + D$, $x_2 = p (1-q) - D$, $x_3 = (1-p) q - D$, and $x_4 = (1-p) (1-q) + D$, and the linkage disequilibrium can be expressed in terms of the haplotype frequencies as $D=x_1x_4 - x_2x_3$. Thereby, we must recall the constraints $x_i \ge 0$ $(i = 1,\dots,4)$ and $\sum_{i=1}^{4} x_i = 1$, which are equivalent to $0 \le p,q \le 1$ and
\begin{equation}
	\label{eq:condLD}
	-\min \left\{pq, (1-p)(1-q) \right\} \le D \le \min \left\{p(1-q), (1-p)q \right\}.
\end{equation}

The matrix of relative fitnesses on the island is
\begin{equation}
	\label{eq:fitMat}
	\mathbf{W} \; = \; \bordermatrix{
			~ & B_1B_1 & B_1B_2 & B_2B_2 \cr
                  		A_1A_1 & w_{11} & w_{12} & w_{22} \cr
                  		A_1A_2 & w_{13} & w_{14} = w_{23} & w_{24} \cr
			A_2A_2 & w_{33} & w_{34} & w_{44} \cr 
			},
\end{equation}
where $w_{ij}$ is the relative fitness of the genotype composed of haplotypes $i$ and $j$ ($i,j \in \{1,2,3,4\}$).
For additive fitnesses, we use Eq.\ \eqref{eq:fitMatAdd} in the main text. The marginal fitness of haplotype $i$ on the island is defined as $w_{i\sbt} = \sum_{j=1}^{4}w_{ij}x_j$ and the mean fitness of the island population as $\bar{\bar{w}} = \sum_{i,j}w_{ij}x_{i}x_{j} = \sum_{i=1}^{4}w_{i\sbt}x_i$.

Straightforward extension of two-locus models without migration (cf. \citealt{Lewontin:1960uq} or \citealt{Buerger:2000fk}, chap.\ 2) yields the recursion equations for the haplotype frequencies,
\begin{subequations}
	\label{eq:recEqFull}
	\begin{align}
		x_1^{\prime} & = (1-m)(x_1 w_{1\sbt} - r w_{14} D)/\bar{\bar{w}},\\
		x_2^{\prime} & = (1-m)(x_2 w_{2\sbt} + r w_{14} D)/\bar{\bar{w}},\\
		x_3^{\prime} & = (1-m)(x_3 w_{3\sbt} + r w_{14} D)/\bar{\bar{w}} + m q_c,\\
		x_4^{\prime} & = (1-m)(x_4 w_{4\sbt} - r w_{14} D)/\bar{\bar{w}} + m (1-q_c),
	\end{align}
\end{subequations}
where $r$ is the recombination rate, $m$ the migration rate, and $q_c$ the frequency of $B_1$ on the continent. For a monomorphic continent, $q_c = 0$. For this case, a continuous-time version of Eq.\ \eqref{eq:recEqFull} has been fully described \citep{Buerger:2011uq}.


\subsection{Approximating the dynamics for rare $A_1$}

Because $A_1$ arises as a novel mutation in our scenario (see main text), the haplotype frequencies $x_1$ and $x_2$ are initially small. We therefore ignore terms of order $x_ix_j$ ($i,j \in \{1,2\}$) and higher in Eq.\ \eqref{eq:recEqFull}. Moreover, we assume that, upon invasion of $A_1$, the frequency of $B_1$ stays constant at the one-locus migration--selection equilibrium ($q = \hat{q}_{\bloc}$). In principle, $q$ approaches an internal equilibrium $\hat{q}_{+}$, but the change is small compared to the change in $p$ \citep{Buerger:2011uq}. We then have $x_3 = q - x_1 \approx \qeqb$ and $x_4 = 1 - q - x_2 \approx 1- \qeqb$ for $x_1$ and $x_2$ small.
As a consequence, the dynamics in Eq.\ \eqref{eq:recEqFull} reduces to a system with only two equations in $x_1$ and $x_2$,
\begin{subequations}
	\label{eq:recEqApprox}
	\begin{align}
		x_1^{\prime} & = (1-m)\left[ w_1 x_1 + r w_{14} x_2 \qeqb - r w_{14} x_1 (1-\qeqb) \right] / \bar{w},\\
		x_2^{\prime} & = (1-m)\left[ w_2 x_2 - r w_{14} x_2 \qeqb + r w_{14} x_1 (1-\qeqb) \right] / \bar{w},
	\end{align}
\end{subequations}
where $w_1$ and $w_2$ are the marginal fitnesses of the $A_1B_1$ and $A_1B_2$ haplotypes, respectively. These are given by
\begin{subequations}
	\label{eq:margFitA1rare}
	\begin{align}
		w_1 & = w_{13} \qeqb + w_{14} (1-\qeqb) \label{eq:margFitA1rareA},\\
		w_2 & = w_{24} (1-\qeqb) + w_{14} \qeqb \label{eq:margFitA1rareB}.
	\end{align}
\end{subequations}
Moreover, $\bar{w}$ is the mean fitness of the resident population on the island, which is assumed to be monomorphic at locus $\aloc$:
\begin{equation}
	\label{eq:meanFitA1rare}
	\bar{w} = {\qeqb}^2 w_{33} + 2 \qeqb (1-\qeqb) w_{34} + (1-\qeqb)^2 w_{44}.
\end{equation}
This holds approximately if $A_1$ is rare on the island. Equation \eqref{eq:recEqApprox} can be written more compactly in matrix form as $\mathbf{x}^{\prime} = \mathbf{x}\mathbf{L}$,
where $\mathbf{x} = (x_1,x_2)$ is a row vector, and
\begin{equation}
	\mathbf{L} =
		\begin{pmatrix}
			\lambda_{11} & \lambda_{12}\\
			\lambda_{21} & \lambda_{22}
		\end{pmatrix},
	\label{eq:meanMatrixGeneric}
\end{equation}
with
\begin{subequations}
	\label{eq:recGEntries}
	\begin{align}
		\lambda_{11} & = (1-m)\left[w_1 -  r (1-\qeqb) w_{14}\right] / \bar{w},\\
		\lambda_{12} & = (1-m) r (1-\qeqb) w_{14} / \bar{w}, \\
		\lambda_{21} & = (1-m) r  \qeqb w_{14} / \bar{w}, \\
		\lambda_{22} & = (1-m) \left[w_2 - r \qeqb w_{14}\right] / \bar{w}.
	\end{align}
\end{subequations}
Setting $m = 0$, we recover the dynamics derived by \citet{Ewens:1967fk} for a panmictic population and a focal mutation occurring in linkage to a background locus at which overdominant selection maintains $B_1$ at frequency $\qeqb$. We note that Eqs.\ \eqref{eq:recEqApprox} to \eqref{eq:recGEntries} are valid for both a monomorphic and a polymorphic continent. The difference comes in only via $\qeqb$, which is derived in the following section. 
Matrix $\mathbf{L}$ will be encountered again as the \emph{mean matrix} of the two-type branching process used to study the invasion probability of $A_1$ (see also the following section).

Note the difference between $w_i$ and $w_{i\sbt}$: the former refers to the resident population under the assumption of the branching process (this section), whereas the latter applies to the island population in the general two-locus model (previous section). The same distinction holds for $\bar{w}$ and $\bar{\bar{w}}$. 


\subsection{Marginal one-locus migration--selection model}
We denote the marginal one-locus migration--selection equilibrium by $E_{\bloc} = (p = 0, q = \qeqb, D = 0)$. This equilibrium is assumed to be realised on the island before occurrence of the $A_1$ mutation. The equilibrium frequency $\qeqb$ of allele $B_1$ plays an important role. It determines the division of the resident island population into two genetic backgrounds and provides the weights for computing the average invasion probability of $A_1$ given the haplotype-specific invasion probabilities (see sections 2 and 4). 
Analysis of the one-locus dynamics (File \ref{prc:determDiscrNB}) shows that $\qeqb$ is obtained by solving
\begin{equation}
	\label{eq:equEB}
	q_{\bloc}^{\prime} = (1-m) \frac{\tilde{w}_1}{\tilde{\bar{w}}} q_{\bloc} + m q_c = q_{\bloc}
\end{equation}
for $q_{\bloc}$, where $\tilde{w}_1 = w_{33} q_{\bloc} + w_{34} (1-q_{\bloc})$ is the marginal relative fitness of the $B_1$ allele and
\begin{equation}
	\label{eq:margMeanFit}
	\tilde{\bar{w}} = q_{\bloc}^2 w_{33} + 2 q_{\bloc} (1- q_{\bloc}) w_{34} +(1- q_{\bloc})^2 w_{44}
\end{equation}
the mean fitness in the island population. From Eq.\ \eqref{eq:equEB}, one obtains
\begin{equation}
	\label{eq:qEqBGeneric}
	\qeqb = \frac{w_{34}(1-m) - \tilde{\bar{w}} + \sqrt{4(1-m)m q_c (w_{34}-w_{33}) \tilde{\bar{w}} + \left[ \tilde{\bar{w}} - (1-m) \right]^2}}{2(1-m)(w_{34} - w_{33})},
\end{equation}
which simplifies to $\qeqb = \left[w_{34} (1-m) - \tilde{\bar{w}}\right]/\left[(1-m)(w_{34} - w_{33})\right]$ for a monomorphic continent ($q_c = 0$). The equilibrium $E_{\bloc}$ is asymptotically stable if 
the migration rate is smaller than a critical value,
\begin{equation}
	\label{eq:mCritGeneric}
	m < \frac{w_{34} - \tilde{\bar{w}}}{w_{34}}.
\end{equation}

We note that $\tilde{\bar{w}}$ is a (non-linear) function of $q_{\bloc}$, and hence of $m$. Therefore, Eq.\ \eqref{eq:qEqBGeneric} is only an implicit solution and condition \eqref{eq:mCritGeneric} not immediately informative. However, for additive fitnesses (see Eq.\ \ref{eq:fitMatAdd} of the main text) and a monomorphic continent ($q_c = 0$) we find the explicit solution given in Eq.\ \eqref{eq:FreqB1OneLoc}. This is an admissible polymorphic equilibrium (i.e.\ $0 < \qeqb < 1$), if the migration rate is below a critical value,
\begin{equation}
	\label{eq:mCrit2}
	m < \frac{b}{1-a}=:m_{\bloc}.
\end{equation}
Because $a < 1$ was assumed, $m_{\bloc}$ is always positive. Straightforward calculations show that Eq.\ \eqref{eq:mCrit2} is also the condition for asymptotic stability of $E_{\bloc}$ within its marginal one-locus system. That is, under the marginal one-locus dynamics, $E_{\bloc}$ is stable whenever it is admissible (see File \ref{prc:determDiscrNB}, or \citealt{Nagylaki:1992fk}, chap.\ 6.1).

When the mutation $A_1$ occurs, there is a transition from one- to two-locus dynamics. It is therefore crucial to study the stability of $E_{\bloc}$ also under the full two-locus dynamics. We find that $E_{\bloc}$ is not hyperbolic if $m = m^{\ast}$ or if $m = m_{\bloc} > m^{\ast}$, with $m^{\ast}$ given in Eq.\ \eqref{eq:mCrit5Main}. In the first case, $E_{\bloc}$ changes stability from unstable to asymptotically stable as $m$ increases above $m^{\ast}$; in the second case, $E_{\bloc}$ leaves the state space as $m$ increases beyond $m_{\bloc}$. We do not have a complete stability and bifurcation analysis of $E_{\bloc}$. However, some numerical and analytical results suggest that the qualitative behaviour is the same as in the continuous-time model \citep{Buerger:2011uq}.
%
%
Then, the following holds. If $E_{\bloc}$ exists and is asymptotically stable under the one-locus dynamics, (i.e.\ $m < \min(b, m_{\bloc}$)), but unstable under the two-locus dynamics (i.e.\ $m < m^{\ast}$), then a fully-polymorphic internal equilibrium $E_{+}$ ($0 < \hat{p}_{+}, \hat{q}_{+} < 1$ and $\hat{D}_{+} > 0$) exists and is asymptotically stable. Therefore, if $m < m^{\ast}$, a novel mutation $A_1$ can invade via $E_{\bloc}$. Presumably, the internal equilibrium $E_{+}$ is reached.
%
%
Comprehensive numerical computations under the discrete-time dynamics corroborate this conjecture (see File \ref{prc:determDiscrNB} and Figure \ref{fig:stabilityEBMonomCont}). 

With a polymorphic continent ($0 < q_c < 1$) and additive fitnesses, the frequency of $B_1$ at the marginal one-locus migration--selection polymorphism ($E_\bloc$) is 
\begin{equation}
	\label{eq:FreqB1OneLocPolymCont}
	\qeqb = \frac{b-(1-a)m + 2b m q_c + \sqrt{R}}{2b(1+m)},
\end{equation}
where
\begin{equation}
	\label{eq:radincandR}
	R = 4b(1-a-b)m(1+m)q_c + \left[b-(1-a)m + 2b m q_c\right]^2 \ge 0.
\end{equation}
In contrast to the case of a monomorphic continent, where $E_\bloc$ exists only if $m < m_{\bloc}$, with a polymorphic continent, both alleles $B_1$ and $B_2$ are introduced by migration and hence $E_\bloc$ always exists and is always asymptotically stable under the one-locus dynamics if $0 < q_c < 1$ and $0 < m < 1$.

A comprehensive analysis of the stability of $E_\bloc$ involves solving a complicated cubic equation, which results in expressions that are not informative. We could not accomplish a complete analytical treatment, but a combination of analytical, numerical and graphical approaches suggests the following. Upon occurrence of $A_1$ at locus $\aloc$, $E_\bloc$ may either become unstable, in which case $A_1$ can invade and a fully-polymorphic internal equilibrium $E_+$ is reached, or $E_\bloc$ may stay asymptotically stable, in which case $A_1$ cannot invade. The transition between these two scenarios occurs at a critical recombination rate
\begin{equation}
	\label{eq:rCritPolymCont}
	r^{\ast} = \left\{ \begin{array}{ll}
 	\frac{1}{2} & \textrm{if $m \leq m_{r^{\ast}}$},\\
 	\tilde{r}^{\ast}(m) & \textrm{otherwise},
  	\end{array} \right.
\end{equation}
where $\tilde{r}^{\ast}(m)$ is a complicated function of $m$ that we do not present here (but see Eq.\ 3 in File \ref{prc:determDiscrNB}, and Eq.\ \ref{eq:rCritPolymContCont} in section 6), and $m_{r^{\ast}}$ is the migration rate at which $\tilde{r}^{\ast}(m)$ has a pole. Then, for a given combination of values for $a$, $b$, $m$ and $q_c$, $A_1$ can invade if and only if $r < r^{\ast}$ (Figure \ref{fig:stabilityEBPolymCont}).
A similar argument holds for a critical continental frequency $q_c^{\ast}$ of $B_1$, such that for a given combination of values for $a$, $b$, $m$ and $r$, $A_1$ can invade if and only if $q_c < q_c^{\ast}$ (see File \ref{prc:determDiscrNB} for details). We were not able to find an explicit expression for a critical migration rate $m^{\ast}$ with an interpretation analogous to that of $r^{\ast}$ or $q_c^{\ast}$. However, $m^{\ast}$ is implicitly defined by $r^{\ast}$ or $q_c^{\ast}$ and can be computed numerically.

As a final remark, we note that for weak evolutionary forces, Eqs.\ \eqref{eq:FreqB1OneLoc}, \eqref{eq:mCrit2} and \eqref{eq:mCrit5Main} can be approximated by the corresponding equations derived by \cite{Buerger:2011uq} for the continuous-time model with a monomorphic continent. Specifically, scaling $a$, $b$, $m$ and $r$ by $\epsilon$ and expanding Eqs.\ \eqref{eq:FreqB1OneLoc}, \eqref{eq:mCrit2} and \eqref{eq:mCrit5Main} into a Taylor series around $\epsilon = 0$ yields
\begin{equation}
	\label{eq:FreqB1OneLocApprox}
	\qeqb \approx 1 - \frac{m}{b},
\end{equation}
\begin{equation}
	\label{eq:mCrit2Approx}
	m_{\bloc} \approx b,
\end{equation}
and
\begin{equation}
	\label{eq:mCrit5Approx}
	m^{\ast} \approx a\left(1 + \frac{b-a}{r}\right)
\end{equation}
to first order of $\epsilon$ and after rescaling. Equations \eqref{eq:FreqB1OneLocApprox} and \eqref{eq:mCrit5Approx} correspond to Eqs.\ (3.9) and (3.11) in \cite{Buerger:2011uq}.


\subsection{Branching-process approximation to the invasion probability}

For a proper stochastic treatment, the evolution of haplotype frequencies has to be modelled by a Markov process. In the context of invasion of novel mutations, particularly useful approximations can be obtained using branching processes \citep{Fisher:1922fk} and diffusion processes \citep{Kimura:1962fk}.
Both approaches deal with the probabilistic effect due to the initially small absolute number of copies of the mutant allele. The effect of finite population size is only accounted for by the diffusion approximation, however. In the first part of the main paper, we are concerned only with initial rareness of the mutation.

We employ a two-type branching process \citep{Harris:1963dq,Ewens:1968fk,Ewens:1967fk} 
to study the dynamics of the two haplotypes of interest, $A_1B_1$ (type 1) and $A_1B_2$ (type 2) after occurrence of  mutation $A_1$ (see section 2 above). 
Let $\lambda_{ij}$ be the mean number of $j$-type offspring produced by an $i$-type parent each generation, and $x_i$ the proportion of type $i$ in the island population. Then the expected proportion of types $A_{1}B_{1}$ and $A_{1}B_{2}$ in the next generation is
\begin{subequations}
	\label{eq:recEqApproxStoc}
	\begin{align}
		\mathbb{E}\left[x_1^{\prime}\right] &= \lambda_{11} x_1 + \lambda_{21} x_2,\\
		\mathbb{E}\left[x_2^{\prime}\right] &= \lambda_{12} x_1 + \lambda_{22} x_2,
	\end{align}	
\end{subequations}
or, in matrix form
\begin{equation}
	\label{eq:recEqApproxStocMatr}
	\mathbb{E}\left[\mathbf{x}^{\prime}\right] = \mathbf{x}\mathbf{L},
\end{equation}
where $\mathbf{x} = (x_1,x_2)$, and $\mathbf{L} = (\lambda_{ij})$, ${i,j \in \{1,2\}}$, is called the \emph{mean matrix} (cf.\ Eq.\ \ref{eq:meanMatrixGeneric} in section 2). 
The leading eigenvalue $\nu$ of $\mathbf{L}$ determines whether the branching process is supercritical ($\nu > 1$) and $A_1$ has a strictly positive invasion probability, or subcritical ($\nu < 1$), in which case $A_1$ goes extinct with probability 1. 
Expressions for the $\lambda_{ij}$ were given in Eq.\ \eqref{eq:recGEntries}.

The leading eigenvalue of $\mathbf{L}$ is
\begin{equation}
	\label{eq:leadEvalL}
	\nu = \frac{1-m}{2 \bar{w}} \left[w_1 + w_2 - r w_{14} + \sqrt{(w_1 - w_2)^2 + 2 r w_{14} (2 \qeqb - 1)(w_1 - w_2) + r^2 w_{14}^2}\right],
\end{equation}
where $w_1$ and $w_2$ are the marginal fitnesses of type 1 and type 2 defined in Eq.\ \eqref{eq:margFitA1rare}, and $\bar{w}$ is the mean fitness of the resident population on the island as defined in Eq.\ \eqref{eq:meanFitA1rare} (section 2). 
After some algebra (see File \ref{prc:stochDiscrNB}), the condition for invasion of $A_1$, $\nu > 1$, is found to be equivalent to Eq.\ \eqref{eq:condNonExtGener} in the main text. Equations \eqref{eq:leadEvalL} and \eqref{eq:condNonExtGener} hold for both a monomorphic and a polymorphic continent.

\rev{Let $\zeta_{ij}$ be the random number of $j$-type offspring produced by a single $i$-type parent. We assume that $\zeta_{i1}$ and $\zeta_{i2}$ are independent and Poisson-distributed with mean $\lambda_{i1}$ and $\lambda_{i2}$, respectively ($i \in \left\{1,2\right\}$). Then, the probability-generating function (pgf) of $\zeta_{ij}$ is\begin{equation}
	\label{eq:pgf}
	f_{ij}(s_j) = \mathbb{E}[s_{j}^{\zeta_{ij}}] = \sum_{k=0}^{\infty} p_{k} s_{j}^{k} =  e^{-\lambda_{ij}(1-s_j)}, \quad i,j \in \{ 1,2 \},
\end{equation}
where $p_{k} = \mathbb{P}[\zeta_{ij} = k]$ is the probability that an $i$-type parent has $k$ offspring of type $j$. The first two equalities follow from the definition of the pgf \cite[e.g.][]{Harris:1963dq}, and the third from the properties of the Poisson distribution.}
Because of independent offspring distributions for each type, the pgf for the number of offspring (of any type) produced by an $i$-type parent is given by
\begin{equation}
	\label{eq:pgfJoint}
	f_{i}(s_1, s_2) = \prod_{j=1}^{2} f_{ij}(s_{j}).
\end{equation}
Inserting Eq.\ \eqref{eq:pgf} into Eq.\ \eqref{eq:pgfJoint}, we obtain
\begin{subequations}
	\label{eq:pgfExpl}
	\begin{align}
		f_{1}(s_1, s_2) & = e^{-\lambda_{11}(1-s_1)} \cdot e^{-\lambda_{12}(1-s_2)} \label{eq:pgfExpl1},\\
		f_{2}(s_1, s_2) & = e^{-\lambda_{21}(1-s_1)} \cdot e^{-\lambda_{22}(1-s_2)} \label{eq:pgfExpl2}.
	\end{align}
\end{subequations}

We use $Q_{i}$ for the extinction probability of allele $A_1$ conditional on initial occurrence on background $B_i$, and $\pi_i = 1 - Q_i$ for the respective probability of invasion. The extinction probabilities $Q_i$ are found as the smallest positive solution to Eq.\ \eqref{eq:bpSol} in the main text. The average invasion probability $\bar{\pi}$ is found as the weighted average of $\pi_{1}$ and $\pi_{2}$ (see Eq.\ \ref{eq:invProbOverall} in the main text).
%
As the problem stated in Eq.\ \eqref{eq:bpSol} amounts to solving a system of transcendental equations, an explicit solution cannot be found in general. Numerical solutions can be obtained, however (see File \ref{prc:stochDiscrNB}).

We proceed by assuming additive fitnesses as defined in Eq.\ \eqref{eq:fitMatAdd} of the main text. The entries $\lambda_{ij}$ of the mean matrix $\mathbf{L}$ in Eq.\ \eqref{eq:meanMatrixGeneric} are then given by
\begin{subequations}
	\label{eq:meanMatrixAdd}
	\begin{align}
		\lambda_{11} &= E + F r,\\
		\lambda_{12} &= -F r,\\
		\lambda_{21} &= H r,\\
		\lambda_{22} & = J - Hr,
	\end{align}
\end{subequations}
where
\begin{subequations}
	\label{eq:coefsAdd}
	\begin{align}
		E & = \frac{1+b+a m}{1-a+b},\\
		F & = -\frac{m}{b},\\
		H & = \frac{b-(1-a)m}{b(1-a+b)},\\
		J & = \frac{1+m(a-b)}{1-a+b}.
	\end{align}
\end{subequations}

Assuming weak evolutionary forces, i.e.\ replacing $a$, $b$, $m$ and $r$ by $\alpha\epsilon$, $\beta\epsilon$, $\mu\epsilon$ and $\rho\epsilon$, respectively, and expanding into a Taylor series around $\epsilon = 0$, the terms in Eq.\ \eqref{eq:meanMatrixAdd} are approximated to first order in $\epsilon$ by
\begin{align*}
	\lambda_{11} & \approx 1 + a - \frac{m}{b} r, & \lambda_{12} & \approx \frac{m}{b} r, \nonumber \\
	\lambda_{21} & \approx \left(1 - \frac{m}{b}\right) r, & \lambda_{22} & \approx 1 + a - b - \left(1 - \frac{m}{b}\right) r \nonumber,
\end{align*}
after resubstituting $\alpha \rightarrow a/\epsilon$, $\beta \rightarrow b/\epsilon$, $\mu \rightarrow m/\epsilon$ and $\rho \rightarrow r/\epsilon$.

With additive fitnesses and a monomorphic continent, the dominant eigenvalue of $\mathbf{L}$ is
\begin{equation}
	\label{eq:domEigenVecAdd}
	\nu = \frac{2+b-r+m(2a-b-r)+\sqrt{R_1}}{2(1-a+b)},
\end{equation}
where
\begin{equation}
	\label{eq:radicandR1}
	R_1 = (1+m)\left\{ b^2 (1 + m) + 2 b (1 - m) r + r \left[r - m (4 - 4 a - r)\right]\right\}.
\end{equation}	
The branching process is supercritical ($\nu > 1$) if $m < m^{\ast}$ or, alternatively, if $r < r^{\ast}$, with $m^{\ast}$ and $r^{\ast}$ the critical migration and recombination rates defined in Eqs.\ \eqref{eq:mCrit5Main} and \eqref{eq:rCrit} of the main text, respectively (see File \ref{prc:stochDiscrNB} for details). Assuming weak evolutionary forces, $\nu$ simplifies to
\begin{equation*}
	\nu \approx 1 + \frac{1}{2}\left(2 a - b - r + \sqrt{R_2}\right),
\end{equation*}
where
\begin{equation}
	\label{eq:radicandR2}
	R_2 = b^2 + 2br - 4mr + r^2.
\end{equation}
Then, $m^{\ast}$ is approximated by Eq.\ \eqref{eq:mCrit5Approx}  
and
\begin{equation}
	\label{eq:rCritApprox}
	r^{\ast} \approx \tilde{r}^{\ast} = \left\{ \begin{array}{ll}
 	\infty & \textrm{if $m \leq a$},\\
 	\frac{a(b-a)}{m-a} & \textrm{otherwise}
  	\end{array} \right.
\end{equation}
(see File \ref{prc:stochDiscrNB}). Note that the critical migration and recombination rates for invasion of $A_1
$ obtained under the deterministic model (section 3) 
and the corresponding two-type branching process are identical. In File \ref{prc:stochCompareJacobianVsMeanMatrixNB} we show that this agreement is generically expected.

To obtain the extinction probabilities of $A_1$ given initial occurrence on background $B_1$ or $B_2$, we plug Eq.\ \eqref{eq:meanMatrixAdd} into \eqref{eq:pgfExpl} and solve
\begin{subequations}
	\label{eq:pgfExplAdd}
	\begin{align}
		f_1\left(s_1,s_2\right)&=e^{\left(E + Fr\right)s_1 - F r s_2 - E}=s_1 \label{eq:pgfExplAdd1}\\
		f_2\left(s_1,s_2 \right)&=e^{H r s_1 + (J - H r) s_2 - J} = s_2 \label{eq:pgfExplAdd2}
	\end{align}
\end{subequations}
for $s_1$ and $s_2$. The smallest solutions between 0 and 1 are the extinction probabilities $Q_1 = 1 - \pi_1$ and $Q_2 = 1 - \pi_2$ (cf.\ Eq.\ \ref{eq:bpSol} in the main text). An explicit solution is not available and we need to use numerical methods to obtain exact results (File \ref{prc:stochDiscrNB}). 

We now turn to the case of a polymorphic continent ($0 < q_c < 1$), still assuming additive fitnesses. Then,
\begin{subequations}
	\label{eq:meanMatrixAddPolymCont}
	\begin{align}
		\lambda_{11} &= \tilde{E} + \tilde{F} r,\\
		\lambda_{12} &= \tilde{G} r,\\
		\lambda_{21} &= \tilde{H} r,\\
		\lambda_{22} & = \tilde{J} + \tilde{I}r,
	\end{align}
\end{subequations}
with
\begin{align*}
		\tilde{E} & = \frac{(1 - m) \left(2 + b + m + a m + 2 b m q_c + \sqrt{R}\right)}{2 \left[1 - a - b m (1 - 2 q_c) + \sqrt{R}\right]},\\
		\tilde{F} & = -\frac{(1 - m)  \left[b + (1 - a) m + 2 b m (1 - q_c) - \sqrt{R}\right]}{2 b \left[1 - a - b m (1 - 2 q_c) + \sqrt{R}\right]},\\
		\tilde{G} & = \frac{b + m \left[1 - a - 2 b (1 - q_c)\right] - \sqrt{R}}{2 b (1 - a - b)},\\
		\tilde{H} & = \frac{b - (1 - a) m - 2 b m q_c + \sqrt{R}}{2 b (1 - a + b)},\\
		\tilde{I} & = - \frac{(1 - m) \left[b - (1 - a) m + 2 b m q_c + \sqrt{R}\right]}{2 b \left[1 - a - b m (1 - 2 q_c) + \sqrt{R}\right]},\\
		\tilde{J} & = \frac{(1 - m) \left[2 + m + a m - b (1 + 2 m (1 - q_c)) + \sqrt{R}\right]}{2 \left[1 - a - b m (1 - 2 q_c) + \sqrt{R}\right]}.
\end{align*}
Here, $R$ is as defined in Eq.\ \eqref{eq:radincandR}. 
Assuming weak evolutionary forces, i.e.\ scaling $a$, $b$, $m$ and $r$ by $\epsilon$ and expanding into a Taylor series around $\epsilon = 0$, Eq.\ \eqref{eq:meanMatrixAddPolymCont} is approximated to first order in $\epsilon$ by
\begin{align*}
	\lambda_{11} & \approx \frac{1}{2} \left(2 + 2 a + b - m - \sqrt{R_3}\right) - \frac{b+m-\sqrt{R_3}}{2b}r, & \lambda_{12} & \approx \frac{b+m-\sqrt{R_3}}{2b} r, \nonumber \\
	\lambda_{21} & \approx \frac{b-m+\sqrt{R_3}}{2b}r, & \lambda_{22} & \approx \frac{1}{2}\left(2+2a-b-m-\sqrt{R_3}\right) - \frac{b-m+\sqrt{R_3}}{2b}r \nonumber,
\end{align*}
where
\begin{equation}
	\label{eq:radicandR3}
	R_3 = (b-m)^2 + 4 b m q_c > 0.
\end{equation}

Note that the continental frequency $q_c$ of $B_1$ enters these equations only via $4 b m q_c$ in the radicand $R_3$. For a polymorphic continent, the eigenvalues of $\mathbf{L}$ are complicated expressions, which we do not show here (but see File \ref{prc:stochDiscrNB}). The leading eigenvalue can be identified, though. For weak evolutionary forces, and to first order in $\epsilon$, it is approximately
\begin{equation}
	\label{eq:leadEvalPolymContApprox}
	\nu \approx 1 + \frac{1}{2}\left[2 a - m - r - \sqrt{R_3} + \sqrt{b^2 - r\left(2m - r - 2 \sqrt{R_3} \right)} \right]
\end{equation}
(see File \ref{prc:stochDiscrSlightlySupercritBPNB}). Finally, the system of transcendental equations to be solved in order to obtain the extinction probabilities of $A_1$ becomes
\begin{subequations}
	\label{eq:pgfExplAddPolymCont}
	\begin{align}
		f_1\left(s_1,s_2\right)&=e^{-(\tilde{E} + \tilde{F} r)(1-s_1)-\tilde{G}r(1-s_2)}=s_1 \label{eq:pgfExplAddPolymCont1}\\
		f_2\left(s_1,s_2 \right)&=e^{-\tilde{H}r(1-s_1)-(\tilde{J}+\tilde{I}r)(1-s_2)} = s_2 \label{eq:pgfExplAddPolymCont2}
	\end{align}
\end{subequations}
(cf.\ Eq.\ \ref{eq:bpSol} of main text).

To obtain analytical approximations to the invasion probability of $A_1$, we follow Haccou (\citeyear{Haccou:2005bh}, pp.\ 127--128) and assume that the branching process is slightly supercritical \cite[see also][]{Eshel:1984qf,Hoppe:1992ys,Athreya:1992zr,Athreya:1993ve}. This means that the leading eigenvalue of the mean matrix $\mathbf{L}$ is of the form
\begin{equation}
	\label{eq:ansatzSlightlySupercritBP}
	\nu = \nu(\xi) = 1 + \xi,
\end{equation}	
where $\xi$ is small and positive. 
To make explicit the dependence on $\xi$, we write $Q_i = Q_i(\xi)$ and $\pi_i = \pi_i(\xi)$ for the extinction and invasion probabilities, respectively ($i \in {1,2}$). Using the Ansatz in Eq.\ \eqref{eq:ansatzSlightlySupercritBP}, Haccou et al.\ state in their Theorem 5.6 that, as $\xi \rightarrow 0$, $q_{i}(\xi)$ converges to 1 and
\begin{equation}
	\label{eq:invProbSuperCritBPGeneric}
	\pi_{i}(\xi) = 1 - q_{i}(\xi) = \frac{2\left[ \nu(\xi) - 1 \right]}{\mathbf{B}(\xi)} v_{i}(\xi) + o(\xi).
\end{equation}
Here, $v_k=i$ is the $i$th entry of the right eigenvector $\mathbf{v} = (v_1, v_2)^{\top}$ pertaining to the leading eigenvalue $\nu$ of the mean matrix $\mathbf{L}$. The matrix $\mathbf{B}(\xi)$ is defined as
\begin{equation}
	\label{eq:limitMatrixGeneric}
	\mathbf{B}(\xi) = \sum_{i=1}^{2}u_i \sum_{j=1}^{2}v_j \lambda_{ij} + \nu(\xi) \left[1-\nu(\xi) \right] \sum_{j=1}^{2}u_j v_{j}^{2},
\end{equation}
where $u_i$ is the $i$th entry of the normalised left eigenvector $\mathbf{u} = (u_1, u_2)$ associated with the leading eigenvalue $\nu$ of $\mathbf{L}$. By normalised we mean that $\sum_{k}^{2} u_k = 1$. For Eq.\ \eqref{eq:invProbSuperCritBPGeneric} to hold, $\mathbf{u}$ and $\mathbf{v}$ must in addition fulfill $\sum_{k=1}^{2}u_k v_k = 1$.

For additive fitnesses (Eq.\ \ref{eq:fitMatAdd}) and a monomorphic continent ($q_c=0$), we combine Eqs.\ \eqref{eq:domEigenVecAdd} and \eqref{eq:ansatzSlightlySupercritBP} to identify $\xi$ as
\begin{equation}
	\label{eq:xiAdd}
	\xi = \frac{2a(1+m) - b - r - m(b+r) + \sqrt{R_1}}{2(1 - a + b)},
\end{equation}
where $R_1$ is defined in Eq.\ \eqref{eq:radicandR1}. Therefore, the assumption of a slightly supercritical branching process will hold for all parameter combinations that result in a small positive $\xi$ in Eq.\ \eqref{eq:xiAdd}. For weak evolutionary forces, Eq.\ \eqref{eq:xiAdd} is approximated by the simpler expression below Eq.\ \eqref{eq:approxInvProbAdd} in the maint text. After some algebra using \emph{Mathematica} (File \ref{prc:stochDiscrSlightlySupercritBPNB}), we obtain the appropriately normalised left and right eigenvectors of $\mathbf{L}$ as
\begin{equation}
	\label{eq:leftEigenVecAdd}
	\mathbf{u} = \begin{pmatrix}
		\frac{b(1+m) - (1+m)r + \sqrt{R_1}}{2b(1+m)}\\
		\frac{2(1-a+b)mr}{b\left[ b + r + m(b+r) + \sqrt{R_1}\right]}
	\end{pmatrix}^\top
\end{equation}
and
\begin{equation}
	\label{eq:rightEigenVecAdd}
	\mathbf{v} = \begin{pmatrix}
		\frac{b^2(1+m) - 2(1-a)m r + b\left(r - mr + \sqrt{R_1}\right)}{(b+r)^2 + m \left[ (b-r)^2 - 4(1-a) r\right] + (b-r) \sqrt{R_1}}\\
		\frac{2\left[b-(1-a)m\right]r}{(b+r)^2 + m \left[ (b-r)^2 - 4(1-a) r\right] + (b-r) \sqrt{R_1}}
	\end{pmatrix},
\end{equation}
respectively. Combining Eqs.\ \eqref{eq:meanMatrixAdd}, \eqref{eq:domEigenVecAdd}, \eqref{eq:leftEigenVecAdd}, \eqref{eq:rightEigenVecAdd}, and \eqref{eq:limitMatrixGeneric}, we find analytical expressions for the conditional invasion probabilities $\pi_{1}(\xi)$ and $\pi_{2}(\xi)$ under a slightly supercritical branching process. The weighted average invasion probability $\bar{\pi}(\xi)$ is obtained according to Eq.\ \eqref{eq:invProbOverall} with $\qeqb$ given in Eq.\ \eqref{eq:FreqB1OneLoc}. The resulting expressions are long and not very informative (see File \ref{prc:stochDiscrSlightlySupercritBPNB} for details and Figure \ref{fig:invProbFuncRApprox} for a graphical comparison to numerical solutions). However, if we assume weak evolutionary forces, we obtain the analytical approximations $\tilde{\pi}_{1}(\xi)$ and $\tilde{\pi}_{2}(\xi)$ given in Eq.\ \eqref{eq:approxInvProbAdd} of the main text. The corresponding average invasion probability $\bar{\tilde{\pi}}(\xi)$ is obtained by insertion of Eqs.\ \eqref{eq:approxInvProbAdd} and \eqref{eq:FreqB1OneLoc} into Eq.\ \eqref{eq:invProbOverall} (see main text).

For a polymorphic continent ($0 < q_c < 1$), the procedure is analogous to the one outlined above. Intermediate and final expressions are more complicated as those obtained for the monomorphic continent, though. We therefore refer to File \ref{prc:stochDiscrSlightlySupercritBPNB} for details and to Figures \ref{fig:invProbFuncMPolymCont} and \ref{fig:invProbFuncqCPolymCont} for a graphical comparison to numerical solutions. The approximations $\tilde{\pi}_{1}(\xi)$, $\tilde{\pi}_{1}(\xi)$ and $\bar{\tilde{\pi}}(\xi)$ given in Eqs.\ (7)--(9) in File \ref{prc:stochDiscrSlightlySupercritBPNB} for weak evolutionary forces and $0<q_c<1$ are accurate if $\xi$ is small, where
\begin{equation*}
	\xi \approx \frac{1}{2}\left[ m + r \sqrt{R_3} - \sqrt{b^2 - r \left( 2m - r - 2 \sqrt{R_3} \right)} \right]
\end{equation*}
and $R_3$ is defined in Eq.\ \eqref{eq:radicandR3}. Then, the branching process is slightly supercritical (cf.\ Eq.\ \ref{eq:ansatzSlightlySupercritBP}). In practice, the approximations derived for a polymorphic continent are useful for efficient plotting, but otherwise not very intuitive. Letting $q_c \rightarrow 0$ and assuming $m < m_\bloc$ (cf.\ Eq.\ \ref{eq:mCrit2} in section 3), 
we recover the respective analytical expressions for the case of a monomorphic continent.


\subsection{Condition for a non-zero optimal recombination rate}
Observation of the mean invasion probability $\bar{\pi}$ of allele $A_1$ as a function of the recombination rate $r$ suggests that $\bar{\pi}(r)$ may have a maximum at a non-zero recombination rate ($r_\mathrm{opt} > 0$) in some cases, whereas it is maximised at $r_{\mathrm{opt}} = 0$ in other cases (Figures \ref{fig:invProbFuncR}A and \ref{fig:invProbFuncR}B). To distinguish between these two regimes, we note that $r_{\mathrm{opt}} > 0$ holds whenever the derivative of $\bar{\pi}(r)$ with respect to $r$, evaluated at $r=0$, is positive. This is because $\bar{\pi}(r)$ will always decay for sufficiently large $r$. We denote the derivative of interest by
\begin{equation}
	\label{eq:derPiAvDef}
	\bar{\pi}^{\prime}(0) := \frac{d}{dr} \left[ \qeqb \pi_{1}(r) + (1-\qeqb) \pi_{2}(r) \right] \Big|_{r = 0} = \qeqb \frac{d\pi_{1}(r)}{dr} \bigg|_{r=0} + (1-\qeqb) \frac{d\pi_{2}(r)}{dr} \bigg|_{r=0},
\end{equation}
where $\pi_1$ and $\pi_2$ are the invasion probabilities of $A_1$ conditional on initial occurrence on the $B_1$ and $B_2$ background, respectively, and $\qeqb$ is the equilibrium frequency of $B_1$ before invasion of $A_1$. In the following, we obtain $\bar{\pi}^{\prime}(0)$ via implicit differentiation. We will first derive a general, implicit condition for $\bar{\pi}^{\prime}(0) > 0$, and then proceed by assuming additive fitnesses to obtain explicit conditions. We will do so first for a monomorphic ($q_c = 0$) and then for a polymorphic ($0 < q_c < 1$) continent.

We start from Eq.\ \eqref{eq:bpSol} of the main text with probability generating functions $f_i({s_1,s_2})$ ($i \in {1,2}$) as defined in Eq.\ \eqref{eq:pgfExpl} in section 4. 
Recall that the extinction probabilities $Q_i = 1 - \pi_i$ are the smallest positive solutions to Eq.\ \eqref{eq:bpSol}. Assuming that these solutions have been identified, we know that the invasion probabilities $\pi_i$ satisfy
\begin{align*}
	1 - \pi_1 & = e^{-\lambda_{11} \pi_1} \cdot e^{-\lambda_{12} \pi_2}\\ 
	1 - \pi_2 & = e^{-\lambda_{21} \pi_1} \cdot e^{-\lambda_{22} \pi_2}. 
\end{align*}
Taking the logarithm on both sides and making the dependence of both $\pi_{i}$ and $\lambda_{ij}$ on $r$ explicit, we have
\begin{subequations}
	\label{eq:pgfEqExpl}
	\begin{align}
		\ln\left[1 - \pi_{1}(r)\right] & = -\lambda_{11}(r) \pi_{1}(r) - \lambda_{12}(r) \pi_{2}(r) \label{eq:pgfEqExpl1}\\
		\ln\left[1 - \pi_{2}(r)\right] & =  -\lambda_{21}(r) \pi_{1}(r) - \lambda_{22}(r) \pi_{2}(r) \label{eq:pgfEqExpl2}.
	\end{align}
\end{subequations}
Applying the formulae for the $\lambda_{ij}(r)$ given in Eq.\ \eqref{eq:recGEntries}, Eq.\ \eqref{eq:pgfEqExpl} becomes
\begin{subequations}
	\label{eq:pgfEqExplFull}
	\begin{align}
		\ln\left[1 - \pi_{1}(r)\right] & = - \frac{1-m}{\bar{w}} \Big\{ \big[ w_1 - (1-\qeqb) r w_{14}\big] \pi_{1}(r) + (1-\qeqb) r w_{14} \pi_{2}(r) \Big\} \label{eq:pgfEqExplFull1}\\
		\ln\left[1 - \pi_{2}(r)\right] & = - \frac{1-m}{\bar{w}} \Big\{ \qeqb r w_{14} \pi_{1}(r) + \left( w_2 - \qeqb r w_{14}\right) \pi_{2}(r) \Big\} \label{eq:pgfEqExplFull2}.
	\end{align}
\end{subequations}
Differentiating both sides with respect to $r$, and setting $r=0$ yields
\begin{subequations}
	\label{eq:pgfEqExplD}
	\begin{align}
		\frac{\pi_{1}^{\prime}(0)}{1-\pi_{1}^{\circ}} & = (1-m) \frac{w_{1} \pi_{1}^{\prime}(0) - (1-\qeqb) w_{14} \left(\pi_{1}^{\circ} - \pi_{2}^{\circ}\right)}{\bar{w}} \label{eq:pgfEqExplD1}\\
		\frac{\pi_{2}^{\prime}(0)}{1-\pi_{2}^{\circ}} & = (1-m) \frac{w_{2} \pi_{2}^{\prime}(0) + \qeqb w_{14} \left(\pi_{1}^{\circ} - \pi_{2}^{\circ}\right)}{\bar{w}} \label{eq:pgfEqExplD2},
	\end{align}
\end{subequations}
where $\pi_{i}^{\prime}(0) = \left.\frac{d\pi_{i}(r)}{dr} \right|_{r=0}$ for $i \in \{1,2\}$. Moreover, $\pi_{1}^{\circ} = \pi_{1}(0)$ and $\pi_{2}^{\circ} = \pi_{2}(0)$ are the conditional invasion probabilities of $A_1$ if it initially occurs on background $B_1$ and $B_2$, respectively, and if there is no recombination ($r=0$). 
Solving the system in Eq.\ \eqref{eq:pgfEqExplD} for $\pi_{1}^{\prime}(0)$ and $\pi_{2}^{\prime}(0)$, and plugging the solutions into Eq.\ \eqref{eq:derPiAvDef}, we find after some algebra
\begin{equation}
	\label{eq:derPiAvImpl}
	\bar{\pi}^{\prime}(0) = (1-m) \qeqb(1-\qeqb)(\pi_{2}^{\circ} - \pi_{1}^{\circ}) \frac{w_{14}}{\bar{w}} \left( \frac{1-\pi_{1}^{\circ}}{1-(1-m)(1-\pi_{1}^{\circ}) w_{1}/\bar{w}} - \frac{1-\pi_{2}^{\circ}}{1-(1-m)(1-\pi_{2}^{\circ}) w_{2}/\bar{w}} \right).
\end{equation}

Setting $r=0$ in Eq.\ \eqref{eq:pgfEqExplFull} and rearranging, we obtain
\begin{equation}
	\label{eq:pgfEqImplR0}
	(1-m) \frac{w_i}{\bar{w}} = - \ln{\left( 1 - \pi_{i}^{\circ}\right)} / \pi_{i}^{\circ} \quad \quad i \in \{1,2\}.
\end{equation}
Insertion of Eq.\ \eqref{eq:pgfEqImplR0} into Eq.\ \eqref{eq:derPiAvImpl} yields
\begin{equation}
	\label{eq:derPiAvImplRA1}
	\bar{\pi}^{\prime}(0) = (1-m) \qeqb(1-\qeqb)(\pi_{2}^{\circ} - \pi_{1}^{\circ})  \frac{w_{14}}{\bar{w}} \left( \frac{1-\pi_{1}^{\circ}}{1 + \ln{\left( 1-\pi_{1}^{\circ}\right)} (1-\pi_{1}^{\circ}) / \pi_{1}^{\circ}} - \frac{1-\pi_{2}^{\circ}}{1 + \ln{\left(1-\pi_{2}^{\circ}\right)} (1-\pi_{2}^{\circ}) / \pi_{2}^{\circ}} \right).
\end{equation}

At this point, a closer inspection of Eq. \eqref{eq:pgfEqImplR0} is worthwhile. Straightforward rearrangement leads to
\begin{equation}
	\label{eq:pgfEqImplR0RA1}
	1 - \pi_{i}^{\circ} = \exp\left[{-(1-m) \frac{w_{i}}{\bar{w}} \pi_{i}^{\circ}}\right] \quad \quad i \in \{1,2\},
\end{equation}
which has a solution $\pi_{i}^{\circ}$ in $(0,1]$ if and only if $(1-m)w_{i}/\bar{w} > 0$. Otherwise, the only solution is $\pi_{i}^{\circ} = 0$. In our setting, we always assumed that when $A_1$ occurs on the deleterious background ($B_2$), it will form a suboptimal haplotype ($A_1B_2$ less fit on the island than $A_1B_1$) and go extinct in the absence recombination. This assumption translates into $w_2 < \bar{w}$. As $0<m<1$, we immediately note that for $i=2$, the only possible solution of Eq.\ \eqref{eq:pgfEqImplR0RA1} is $\pi_{2}^{\circ} = 0$. Therefore, whenever $w_1 > \bar{w}/(1-m)$ holds, the derivative of interest in Eq.\ \eqref{eq:derPiAvImpl} simplifies to
\begin{equation}
	\label{eq:derPiAvImplRA2}
	\bar{\pi}^{\prime}(0) = (1-m) \qeqb(1-\qeqb) \pi_{1}^{\circ}  \frac{w_{14}}{\bar{w}} \left( \frac{\bar{w}}{\bar{w} - (1-m)w_2} - \frac{1-\pi_{1}^{\circ}}{1-(1-m)(1-\pi_{1}^{\circ}) w_{1}/\bar{w}} \right).
\end{equation}
After some algebra (File \ref{prc:stochDiscrDerivativeZeroRecombRateNB}), we find that $\bar{\pi}^{\prime}(0) > 0$, and hence $r_{\mathrm{opt}} > 0$, is equivalent to Eq.\ \eqref{eq:aCritOptRecombRateGeneric} in the main text. Again, if we set $m=0$ in the derivation above, we obtain expressions previously derived by Ewens for a panmictic population in which the background locus is maintained polymorphic by heterozygote superiority \citep{Ewens:1967fk}.

To obtain more explicit conditions, we assume additive fitnesses (Eq.\ \ref{eq:fitMatAdd}). We start directly from Eq.\ \eqref{eq:pgfExplAdd}, replacing $s_i$ by the smallest solution $Q_i$ between 0 and 1. Taking the logarithm on both sides and making the dependence of $Q_i$ on $r$ explicit, we find
\begin{subequations}
	\label{eq:pgfExplAddLog}
	\begin{align}
		\ln{Q_1(r)} & = \left(E + Fr\right)Q_1(r) - F r Q_2(r) - E \label{eq:pgfExplAddLog1}\\
		\ln{Q_2(r)} & = H r Q_1(r) + (J - H r) Q_2(r) - J \label{eq:pgfExplAddLog2},
	\end{align}
\end{subequations}
where $E$, $F$, $J$ and $H$ are independent of $r$ and as defined in Eq.\ \eqref{eq:coefsAdd}. 
Differentiating Eq.\ \eqref{eq:pgfExplAddLog} on both sides, setting $r=0$ and rearranging, we obtain
	\begin{align*}
		\frac{Q_{1}^{\prime}(0)}{Q_{1}^{\circ}} & = F \left(Q_{1}^{\circ} - Q_{2}^{\circ}\right) + E Q_{1}^{\prime}(0)\\ 
		\frac{Q_{2}^{\prime}(0)}{Q_{2}^{\circ}} & = H \left(Q_{1}^{\circ} - Q_{2}^{\circ} \right) + J Q_{2}^{\prime}(0),
	\end{align*}
with $Q_{i}^{\prime}(0) = \frac{dQ_{1}(r) }{dr} \big|_{r=0}$. Here, we used $Q_{i}^{\circ} = Q_{i}(0)$ for the extinction probability of $A_1$ condidtional on initial occurrence on background $B_i$ ($i \in \{1,2\}$). Solving for $Q_{1}^{\prime}(0)$ and $Q_{2}^{\prime}(0)$ yields
\begin{subequations}
	\label{eq:siSolAdd}
	\begin{align}
		Q_{1}^{\prime}(0) & = \frac{F Q_{1}^{\circ} (Q_{1}^{\circ} - Q_{2}^{\circ})}{1-E Q_{1}^{\circ}} \label{eq:siSolAdd1}\\
		Q_{2}^{\prime}(0) & = \frac{H Q_{2}^{\circ} (Q_{1}^{\circ} - Q_{2}^{\circ})}{1-J Q_{2}^{\circ}}.\label{eq:siSolAdd2}
	\end{align}
\end{subequations}

To obtain an explicit solution, we aim at approximating the $Q_{i}^{\circ}$ in the following. Going back to Eq.\ \eqref{eq:pgfExplAdd} again, but setting $r=0$ directly, we find
\begin{equation}
	\label{eq:pgfEqDecoupled}
		Q_{i}^{\circ} = e^{-Z_i(1-Q_{i}^{\circ})} \quad \quad i \in \{1,2 \},
\end{equation}
where
\begin{subequations}
	\label{eq:defZi}
	\begin{align}
		Z_1 & := E = \frac{1+b+am}{1-a+b} \label{eq:defZi1},\\
		Z_2 & := J = \frac{1+m(a-b)}{1-a+b} \label{eq:defZi2}.
	\end{align}
\end{subequations}
Importantly, the equations for $Q_{1}^{\circ}$ and $Q_{2}^{\circ}$ in \eqref{eq:pgfEqDecoupled} are now decoupled. 
Moreover, we note that Eq.\ \eqref{eq:pgfEqDecoupled} has a solution $Q_{i}^{\circ}$ in $[0,1)$ if and only if $Z_i > 1$; if $Z_i \le 1$, the solution is $Q_{i}^{\circ} = 1$. In other words, in the case of complete linkage ($r=0$), type $i$ has a non-zero invasion probability if and only if $Z_i > 1$ (recall that $\pi_{i}^{\circ} = 1-Q_{i}^{\circ}$). Closer inspection of Eq.\ \eqref{eq:defZi} shows that, given our assumptions of $a<b$ and $0<m<1$, $Z_1 > 1$ and $Z_2 < 1$ hold always. Hence, we have $\pi_{2}^{\circ} = 1 - Q_{2}^{\circ} =  0$, and we are left with finding an approximate solution of Eq.\ \eqref{eq:pgfEqDecoupled} for $i=1$. For this purpose, we focus on the case where invasion is just possible, i.e.\ $\pi_{1}^{\circ}$ is close to 0 and hence $Q_{1}^{\circ}$ close to 1. This is equivalent to $Z_1$ being close to, but larger than, 1. We therefore use the Ansatz
\begin{equation}
	\label{eq:ansatzZ}
	Z_{1} = 1 + \epsilon
\end{equation}
with $\epsilon > 0$ small. We then have $Q_{1}^{\circ} = e^{-(1+\epsilon)(1-Q_{1}^{\circ})}$. Noting that $Q_{1}^{\circ}(\epsilon)$ must be close to 1 for $\epsilon$ small, we expand the right-hand side into a Taylor series around $Q_{1}^{\circ} = 1$, which results in
\begin{equation}
	\label{eq:Qi1AddEps}
	Q_{1}^{\circ} = 1 - (1 - Q_{1}^{\circ})(1 + \epsilon) + \frac{1}{2} (1 - Q_{1}^{\circ})^{2} (1 + \epsilon)^{2} + \mathcal{O}(Q_{1}^{\circ})^{3}
\end{equation}
Neglecting terms beyond $\mathcal{O}(Q_{1}^{\circ})^{2}$ and solving for $Q_{1}^{\circ}$, we obtain $Q_{1}^{\circ} = (1+\epsilon^2) / (1+\epsilon)^2$ (excluding the trivial solution $Q_{1}^{\circ} = 1$). To first order in $\epsilon$, this is approximated by
\begin{equation}
	\label{eq:Qi1AddEpsApprox}
	Q_{1}^{\circ} = 1 - \pi_{1}^{\circ} \approx 1 - 2 \epsilon.
\end{equation}

We identify $\epsilon$ by inserting Eq.\ \eqref{eq:defZi1} into Eq.\ \eqref{eq:ansatzZ} and solving for $\epsilon$. To first order in $a$, this yields $\epsilon \approx a(1+m)/(1+b)$ and hence, from Eq.\ \eqref{eq:Qi1AddEpsApprox}, we find
\begin{equation}
	\label{eq:Qi1AddApprox}
	Q_{1}^{\circ} =  1 - 2 \frac{a(1+m)}{(1+b)} + \mathcal{O}(a)^2.
\end{equation}
Note that if we set $m=0$ (no migration) and $b=0$ (no background selection), we recover Haldane's \citeyearpar{Haldane:1927fk} well-known approximation $\pi \approx 2a$.

Comparison of Eqs.\ \eqref{eq:Qi1AddEpsApprox} and \eqref{eq:Qi1AddApprox} suggests that the invasion probability $\pi_{1}^{\circ}$ increases with the migration rate $m$. This may seem counterintuitive. However, with complete linkage ($r=0$), the cases of $A_1$ occurring on background $B_1$ or $B_2$ can be considered separately. If $A_1$ occurs on background $B_1$, it forms haplotype $A_1B_1$. From then on it competes against the resident population consisting of haplotypes $A_2B_1$ and $A_2B_2$ at frequencies $\qeqb$ and $1 - \qeqb$, respectively. Because, initially, $A_1B_1$ types do not interfere nor contribute to the resident population,  what matters is the ratio of the marginal fitness $w_1$ of $A_1B_1$ to the mean fitness $\bar{w}$ of the resident population. This follows directly from Eq.\ \eqref{eq:pgfEqImplR0}. Equations \eqref{eq:margFitA1rareA} and \eqref{eq:meanFitA1rare} in section 2 show that both $w_1$ and $\bar{w}$ depend on $\qeqb$. For additive fitnesses, $\qeqb$ is given by Eq.\ \eqref{eq:FreqB1OneLoc} in the main text; it depends on $m$. Therefore, to understand the apparently paradoxical increase of $\pi_{1}^{\circ}$ on $m$, we must compare the dependence on $m$ of $w_1$ and $\bar{w}$. We have $w_1 = (1 + b + am)/(1+m)$ and $\bar{w} = (1-m)(1-a+b)/(1+m)$. Both decrease with $m$, but $\bar{w}$ does so faster. The ratio $w_1/\bar{w} = (1+b+am)/\left[(1-a+b)(1-m)\right]$ increases quickly with $m$ (File \ref{prc:stochDiscrDerivativeZeroRecombRateNB}). This explains why $\pi_{1}^{\circ}$ increaes with $m$. It also explains why $\pi_1$ increases with small $m$ in Figure \ref{fig:invProbFuncM}D for very weak recombination. If recombination is too strong, the effect vanishes (Figures \ref{fig:invProbFuncM}E and \ref{fig:invProbFuncM}F).

Finally, plugging Eq.\ \eqref{eq:coefsAdd} from section 4 
and Eq.\ \eqref{eq:Qi1AddApprox} into Eq.\ \eqref{eq:siSolAdd}, we obtain the explicit approximations
\begin{subequations}
	\label{eq:siSolAddFinal}
	\begin{align}
		Q_{1}^{\prime}(0) & \approx \frac{2m(1-a+b)\left[ 1+b-2a(1+m)\right]}{b(1+b)(1+b+2am)} \label{eq:siSolAddFinal1},\\
		Q_{2}^{\prime}(0) & \approx \frac{2a\left[ b-(1-a)m\right]}{b(1+b)(a-b)} \label{eq:siSolAddFinal2},
	\end{align}
\end{subequations}
valid for $a$ small relative to $m$ and $b$. Noting that $\bar{\pi}^{\prime}(0)= - \left[ \qeqb Q_{1}^{\prime}(0) + (1 - \qeqb) Q_{2}^{\prime}(0) \right]$ and using $\qeqb$ from Eq.\ \eqref{eq:FreqB1OneLoc} of the main text for additive fitnesses and a monomorphic continent, we find the approximate derivative of the mean invasion probability $\bar{\pi}$ at $r=0$ as
\begin{equation}
	\label{eq:piAvDerAddApprox}
	\bar{\pi}^{\prime}(0) \approx  \frac{2m(1-a+b) \left[ b - (1-a)m \right] \left\{ 2a^2 + b + b^2 - 2a\left[ 1 + b(2+m) \right] \right\}}{b^2 (1+b)(a-b)(1+m)(1+b+2am)}.
\end{equation}
After some algebra, one can show that $\bar{\pi}^{\prime}(0) > 0$, and hence $r_{\mathrm{opt}} > 0$, if $a > 1 - b/m$ and $a > a^{\ast}$, with $a^{\ast}$ defined in Eq.\ \eqref{eq:aCritOptRecombRate} of the main text. Combination of Eq.\ \eqref{eq:aCritOptRecombRate} with our assumption $a<b$ and the condition for existence of the marginal one-locus equilibrium $E_\bloc$ ($a > 1-b/m$, from Eq.\ \ref{eq:mCrit2} in section 3) yields a sufficient condition for $r_{\mathrm{opt}}>0$ (Figure \ref{fig:optRecombRate}). For further details, we refer to File \ref{prc:stochDiscrDerivativeZeroRecombRateNB}.

For the case of a polymorphic continent ($q_c > 0$), we were not able to derive informative analytical conditions for $r_{\mathrm{opt}} > 0$. Analytical and numerical computations in File \ref{prc:stochDiscrDerivativeZeroRecombRateNB} suggest that if we start with a monomorphic continent ($q_c$ = 0) in a constellation where $r_{\mathrm{opt}}>0$ holds, and then increase $q_c$, the maximum in $\bar{\pi}(r)$ shifts to 0 ($r_{\mathrm{opt}} \rightarrow 0$). There must be a critical value of $q_c$ at which the shift from $r_{\mathrm{opt}}>0$ to $r_{\mathrm{opt}}=0$ occurs, but we could not determine it analytically.


\subsection{Analysis of the deterministic model in continuous time}

For the diffusion approximation in the following section 
 we will need a continuous-time version of our model as a starting point. Here, we derive this model from the discrete-time version. We will analyse some properties of interest in the context of invasion and survival of a weakly beneficial mutation arising in linkage to a migration--selection polymorphism. The continuous-time version with a monomorphic continent ($q_c = 0$) has been completely analysed by \citet{Buerger:2011uq}. Therefore, we only summarise some of their results and focus on the extension to a polymorphic continent ($0 < q_c < 1$). We use a tilde ($\sim$) to distinguish continuous-time expressions from their analogous terms in discrete time. For ease of typing, though, this distinction is not made in all \emph{Mathematica} Notebooks provided in the Supporting Information.

We start from the recursion equations for the haplotype frequencies given in Eq.\ \eqref{eq:recEqFull} of this text, with relative fitnesses $w_{ij}$ according to Eq.\ \eqref{eq:fitMatAdd}. As we will assume quasi-linkage equilibrium (QLE) in the following section, 
it is more convenient to express the dynamics in terms of allele frequencies ($p, q$) and linkage disequilibrium ($D$), rather than haplotype frequencies. This is achieved by recalling the relationships between $D$, $p$, $q$, and the $x_i$ ($i=1,\dots,4$) given in section 1.
The resulting difference equations are complicated and only shown in File \ref{prc:stochDiffApproxQLENB}.
We obtain the differential equations by assuming that the changes due to selection, migration and recombination are small during a short time interval $\Delta t$. Scaling $a$, $b$, $m$ and $r$ by $\Delta t$ and taking the limit $
\lim_{\Delta t \rightarrow 0} \frac{\Delta x}{\Delta t}$ for $x \in \{p, q, D \}$ results in
\begin{subequations}
	\label{eq:diffEqs}
	\begin{align}
		\dot{p} &= \frac{dp}{dt} = a p (1-p) - mp + bD\label{eq:diffEqsa},\\
		\dot{q} &= \frac{dq}{dt} = b q (1-q) - m(q-q_c) + aD \label{eq:diffEqsb},\\
		\dot{D} &= \frac{dD}{dt} = \left[a(1-2p)+b(1-2q)\right]D + m\left[ p(q-q_c) - D\right] - rD \label{eq:diffEqsc}.
	\end{align}
\end{subequations}

For a monomorphic continent ($q_c = 0$), one finds the marginal one-locus migration--selection equilibrium $\tilde{E}_\bloc$ for locus $\bloc$ by setting $p = D = 0$ and solving $\dot{q} = 0$ for $q$, which yields
\begin{equation}
	\label{eq:FreqB1OneLocCont}
	\qeqbc = 1 - \frac{m}{b}
\end{equation}
as the solution of interest (cf.\ Eq.\ \ref{eq:FreqB1OneLocApprox}). \citet{Buerger:2011uq}  have shown that this equilibrium is asymptotically stable in its one-locus dynamics whenever it exists, i.e.\ when $m < b = \tilde{m}_{\bloc}$. Moreover, it is asymptotically stable under the two-locus dynamics if and only if $\tilde{m}^\ast < m < b$, where $\tilde{m}^\ast = a \left(1+\frac{b-a}{r}\right)$ (cf.\ Eq.\ \ref{eq:mCrit5Approx} in section 3, and Eq.\ 3.13 in \citealt{Buerger:2011uq}). Note that B\"{u}rger and Akerman used $m_\bloc$ for what we call $\tilde{m}^\ast$. 
Invasion of $A_1$ via $\tilde{E}_{\bloc}$ requires $m < \min(b, \tilde{m}^\ast)$. After invasion, the system reaches an asymptotically stable, fully-polymorphic equilibrium $\tilde{E}_{+}$. There may exist a second fully-polymorphic equilibrium $\tilde{E}_{-}$, but this is never stable and does not exist when $\tilde{E}_{\bloc}$ is unstable. It is therefore of limited interest to us. B\"{u}rger and Akerman give the coordinates of these equilibria in their Eq.\ (3.15).

For a polymorphic continent ($0 < q_c < 1$), we find the frequency $\qeqbc$ of $B_1$ at the marginal one-locus migration--selection equilibrium $\tilde{E}_\bloc$ as
\begin{equation}
	\label{eq:FreqB1OneLocPolymContCont}
	\qeqbc = \frac{b - m + \sqrt{R_3}}{2b},
\end{equation}
with $R_3 = (b-m)^2 + 4bmq_c > 0$ as previously encountered in Eq.\ \eqref{eq:radicandR3} in section 4. 
Equilibrium $\tilde{E}_\bloc$ always exists and is always asymptotically stable under its one-locus dynamics (File \ref{prc:stochDiffApproxQLENB}). To know when a weakly beneficial mutation at locus $\aloc$ can invade, we investigate the stability properties of $\tilde{E}_\bloc$ under the two-locus dynamics. The Jacobian matrix evaluated at $\tilde{E}_\bloc = (p = 0, q = \qeqbc, D = 0)$ 
is
\begin{equation}
	\mathbf{J}_{\tilde{E}_{\bloc}} = 
		\begin{pmatrix}
			a-m & 0 & b \\
			0 & - \sqrt{R_3} & a \\
			m \left(b-m-2bq_c+\sqrt{R_3}\right)/(2b) & 0 & a - r - \sqrt{R_3}
		\end{pmatrix}
\end{equation}
and its leading eigenvalue is
\begin{equation}
	\label{eq:leadEvalPolymContCont}
	\tilde{\nu} = \frac{1}{2}\left[2 a - m - r - \sqrt{R_3} + \sqrt{b^2 - r\left(2m - r - 2 \sqrt{R_3} \right)} \right]
\end{equation}
(cf.\ Eq.\ \ref{eq:leadEvalPolymContApprox}). 
Equilibrium $\tilde{E}_\bloc$ is unstable if and only if $\tilde{\nu}>0$. To obtain explicit conditions, we determine values of $r$ and $q_c$ at which $\tilde{E}_{\bloc}$ is not hyperbolic (i.e.\ $\tilde{\nu} = 0$) and may therefore enter or leave the state space, or change its stability.
%
%
Equilibrium $\tilde{E}_\bloc$ is not hyberbolic if the recombination rate is equal to
\begin{equation}
	\label{eq:rCritPolymContCont}
	\tilde{r}^{\ast\ast} = \frac{2a^2 - 2a\left(m + \sqrt{R_3}\right) + m\left[m-b(1-2q_c) + \sqrt{R_3}\right]}{2(a-m)}
\end{equation}
(File \ref{prc:stochDiffApproxQLENB}). As a function of $m$, $\tilde{r}^{\ast\ast}$ has a pole at $m = a$, and $\tilde{r}^{\ast\ast} = 0$ if $m = a(a+b)/(a + b q_c)$. This holds for $a < b$, which is one of our general assumptions. We conclude that $\tilde{E}_\bloc$ is unstable and $A_1$ can invade whenever $r < \tilde{r}_{\bloc}$, where
\begin{equation}
	\label{eq:rBPolymContCont}
	\tilde{r}_{\bloc} = \left\{
				\begin{array}{ll}
					\infty & \textrm{if $0 \le m \le a$},\\
					\tilde{r}^{\ast\ast} & \textrm{if $m > a$}.
				\end{array}
			\right.
\end{equation}
Figure \ref{fig:stabilityEBPolymMR} shows the division of the $(m,r)$-parameter space into areas where $\tilde{E}_\bloc$ is asymptotically stable (blue) and unstable (orange), respectively.

By solving $\tilde{\nu} = 0$ for $q_c$, we obtain two critical continental frequencies of $B_1$ at which $\tilde{E}_\bloc$ is not hyperbolic. These are given by
\begin{equation}
	\label{eq:qCCritPolymContCont}
	\tilde{q}_{c\pm}^{\ast\ast} = \frac{1}{2} + \frac{(a-m)(a+r)}{bm} \pm \frac{(2a-m) \sqrt{R_4}}{2bm},
\end{equation}
where $R_4 = 4r(a-m)+b^2$. We first investigate the properties of $\tilde{q}_{\pm}^{\ast\ast}$ as a function of the migration rate $m$. A combination of algebra and graphical exploration given in File \ref{prc:stochDiffApproxQLENB} suggests that the following cases must be distinguished:
\begin{description}
	\item[Case 1] $2a \le b$ and $\left( r \le a\ \textrm{or}\ b-a \le r\right)$. Then $\tilde{E}_\bloc$ is unstable if $q_c < \tilde{q}_{c,\bloc}$, with $\tilde{q}_{c,\bloc}$ defined as
		\begin{equation}
			\label{eq:qCBPolymCont1}
			\tilde{q}_{c,\bloc} = \left\{
				\begin{array}{ll}
					\infty & \textrm{if $m < a$},\\
					\tilde{q}_{c+}^{\ast\ast} & \textrm{if $a \le m < a+b-r$},\\
					0 & \textrm{if $a+b-r \le m$}.
				\end{array}
			\right.
		\end{equation} 

	\item[Case 2] $\left(2a < b\ \textrm{and}\ a<r<b-a\right)$ or $\left(2a>b\ \textrm{and}\ b-a<r<a \right)$. Then $\tilde{E}_\bloc$ is unstable if $q_c < \tilde{q}_{c,\bloc}$, with $\tilde{q}_{c,\bloc}$ defined as
		\begin{equation}
			\label{eq:qCBPolymCont2}
			\tilde{q}_{c,\bloc} = \left\{
				\begin{array}{ll}
					\infty & \textrm{if $m < a$},\\
					\tilde{q}_{c+}^{\ast\ast} & \textrm{if $a \le m < a(b-a+r)/r$},\\
					0 & \textrm{if $a(b-a+r)/r \le m$}.
				\end{array}
			\right.
		\end{equation}
	\item[Case 3] $2a > b$ and $2r > b$ and $a \le r$ $\Leftrightarrow$ $2a > b$ and $a \le r$. We distinguish four subcases:
		\begin{description}
			\item[3a] $m<a$. Then $\tilde{E}_\bloc$ is always unstable.
			
			\item[3b] $a \le m \le a(b-a+r)/r$. Then $\tilde{E}_\bloc$ is unstable if $q_c < \tilde{q}_{c+}^{\ast\ast}$.
			
			\item[3c] $a(b-a+r)/r < m < a + b^2/(4r)$. Then $\tilde{E}_\bloc$ is unstable if $\tilde{q}_{c-}^{\ast\ast} < q_c < \tilde{q}_{c+}^{\ast\ast}$.
			
			\item[3d] $a+b^2/(4r) \le m$. Then $\tilde{E}_\bloc$ is asymptotically stable.
		\end{description}
	\item[Case 4] $2a > b$ and $2r > b$ and $a>r$ $\Leftrightarrow$ $2r>b$ and $a>r$. We distinguish four subcases:
		\begin{description}
			\item[4a] $m<a$. Then $\tilde{E}_\bloc$ is always unstable.
			
			\item[4b] $a \le m \le a+b-r$. Then $\tilde{E}_\bloc$ is unstable if $q_c < \tilde{q}_{c+}^{\ast\ast}$.
			
			\item[4c] $a+b-r<m<a+b^2/(4r)$. Then $\tilde{E}_\bloc$ is unstable if $\tilde{q}_{c-}^{\ast\ast} < q_c < \tilde{q}_{c+}^{\ast\ast}$.
			
			\item[4d] $a+b^2/(4r) \le m$. Then $\tilde{E}_\bloc$ is asymptotically stable.
		\end{description}
\end{description}
Figure \ref{fig:stabilityEBPolymMqC} shows the partition of the $(m,q_c)$-parameter space into areas where $\tilde{E}_\bloc$ is asymptotically stable (blue) and unstable (orange), respectively. There are parameter combinations such that $\tilde{E}_\bloc$ is asymptotically stable for very low and for high values of $q_c$, but unstable for intermediate $q_c$ (Figures \ref{fig:stabilityEBPolymMqC}B and \ref{fig:stabilityEBPolymMqC}C). This effect is weak and constrained to a small proportion of the parameter space ($q_c$ small).

Alternatively, we assess the properties of $\tilde{q}_{\pm}^{\ast\ast}$ as a function of the recombination rate $r$. Graphical exploration (File \ref{prc:stochDiffApproxQLENB}) suggests the following, provided that $a<\min(m,b)$ holds. If recombination is weak, i.e.\ $r < a(b-a)/(m-a) = \tilde{r}^{\ast}$, then $\tilde{E}_\bloc$ is unstable if $q_c < \tilde{q}_{c+}^{\ast\ast}$. If recombination is intermediate, i.e.\ $\tilde{r}^{\ast}<r<b^2/\left[ 4(m-a)\right]$, then $\tilde{E}_\bloc$ is unstable if $\tilde{q}_{c-}^{\ast\ast}<q_c<\tilde{q}_{c+}^{\ast\ast}$. Last, if recombination is strong, i.e.\ $r \ge b^2/\left[ 4(m-a)\right]$, then $\tilde{E}_\bloc$ is asymptotically stable. Note that $\tilde{r}^{\ast}$ was previously encountered in Eq.\ \eqref{eq:rCritApprox} 
in the context of the branching process. Figure \ref{fig:stabilityEBPolymRqC} shows the division of the $(r,q_c)$-parameter space into areas where $\tilde{E}_\bloc$ is asymptotically stable (blue) and unstable (orange), respectively. As just shown, there are parameter combinations such that $\tilde{E}_\bloc$ is asymptotically stable for very low and for high values of $q_c$, but unstable for intermediate $q_c$ (Figures \ref{fig:stabilityEBPolymRqC}A--\ref{fig:stabilityEBPolymRqC}C).

In principle, analogous conditions for asymptotic stability of $\tilde{E}_\bloc$ under the two-locus dynamics could be obtained in terms of a critical migrationr rate $m^{\ast\ast}$ at which $\tilde{E}_\bloc$ is not hyperbolic ($\tilde{\nu} = 0$). However, we were not able to derive informative explicit conditions (see File \ref{prc:stochDiffApproxQLENB} for a graphical exploration).

So far, we have described the conditions for instability of the marginal one-locus migration--selection equilibrium $\tilde{E}_\bloc$ under the two-locus dynamics, both for a monomorphic ($q_c = 0$) and a polymorphic ($0 < q_c < 1$) continent. In both cases, there is no other stable equilibrium on the boundary for $0<m<1$.
%
As mentioned above, for the case of a monomorphic continent, the coordinates of the fully-polymorphic equilibria can be found \citep{Buerger:2011uq} and asymptotic stability proved \citep{Bank:2012vn}. For a polymorphic continent, simple explicit expressions are not available, but we could show analytically that at most three candidates for a fully-polymorphic equilibrium exist. Numerical and graphical explorations suggest that if $\tilde{E}_{\bloc}$ is unstable, at most one of these candidates is an admissible equilibrium, and it is asymptotically stable (see File \ref{prc:stochDiffApproxQLENB} for details).
%
%
Figures \ref{fig:stabilityEBPolymMR}--\ref{fig:stabilityEBPolymRqC} therefore directly tell us when $A_1$ can be established if introduced near $\tilde{E}_\bloc$ (orange areas).

In the following section, 
we will derive a diffusion approximation of sojourn and absorption times under the assumption of quasi-linkage equilibrium (QLE), i.e.\ for $r \gg \max(m,b)$. Therefore, we briefly discuss the properties under the QLE assumption of the fully-polymorphic, asymptotically stable, equilibria mentioned in the previous paragraphs. For a monomorphic continent, $\tilde{E}_{+}$ is approximated to first order in $1/r$ by
\begin{subequations}
	\label{eq:EPlusMonomContQLE}
		\begin{align}
			\hat{\tilde{p}}_{+} & = \frac{bm + ar - m(m+r)}{ar} = 1 - \frac{m}{a} + \frac{m}{r} \frac{(b-m)}{a}, \\
			\hat{\tilde{q}}_{+} & = \frac{am + br - m(m+r)}{br} = 1 - \frac{m}{b} + \frac{m}{r} \frac{(a-m)}{b}, \\
			\hat{\tilde{D}}_{+} & = \frac{(a-m)(b-m)m}{abr} = \frac{m}{r} \left(1 - \frac{m}{a}\right)\left(1 - \frac{m}{b}\right),
		\end{align}
\end{subequations}
(cf.\ Eq.\ 4.3 in \citealt{Buerger:2011uq}). As $r \rightarrow \infty$, Eq.\ \eqref{eq:EPlusMonomContQLE} converges to the case of no linkage, where $\hat{\tilde{p}}_{+} = 1 - m/a$, $\hat{\tilde{q}}_{+} = 1 - m/b$, and $\hat{\tilde{D}}_{+} = 0$.
%
%
Turning to the case of a polymorphic continent, we recall from above that there is at most one admissible fully-polymorphic equilibrium. 
To first order in $1/r$, its coordinates are
\begin{subequations}
	\label{eq:EPlusPolymContQLE}
		\begin{align}
			\hat{\tilde{p}}_{+} & = \frac{2ar + m(b - 2b q_c - m - 2r + \sqrt{R_3})}{2ar} = 1 - \frac{bmq_c}{ar} + \frac{m(b-m)}{2ar} + \frac{m \sqrt{R_3}}{2ar} - \frac{m}{a},\\
			\hat{\tilde{q}}_{+} & = \frac{1}{2} - \frac{am(2bq_c - b + m - \sqrt{R_3})}{2br\sqrt{R_3}} + \frac{m(m+r)(m-\sqrt{R_3})}{2br\sqrt{R_3}} + \frac{b}{2\sqrt{R_3}} + \frac{m(2q_c - 1)(m + 2r)}{2r\sqrt{R_3}}, \\
			\hat{\tilde{D}}_{+} & = \frac{m(a-m)\left[b(1-2q_c) - m + \sqrt{R_3}\right]}{2abr}.
		\end{align}
\end{subequations}
Setting $q_c = 0$ and recalling that $m < \tilde{m}^{\ast} = a \left(1+\frac{b-a}{r}\right)$ must hold for invasion in this case (section 3), 
it is easy to verify that Eq.\ \eqref{eq:EPlusPolymContQLE} coincides with Eq.\ \eqref{eq:EPlusMonomContQLE}. This is why we call the equilibrium in Eq.\ \eqref{eq:EPlusPolymContQLE} $\tilde{E}_{+\mathrm{QLE}}$. Graphical exploration in File \ref{prc:stochDiffApproxQLENB} confirms that $\tilde{E}_{+\mathrm{QLE}}$ is asymptotically stable whenever it exists under the QLE regime.

Finally, we ask when $\tilde{E}_{+\mathrm{QLE}}$ exists in the admissible state space. We note that $\hat{\tilde{p}}_{+\mathrm{QLE}}$ is a strictly decreasing function of the recombination rate $r$, independently of the migration rate $m$. In contrast, $\hat{\tilde{q}}_{+\mathrm{QLE}}$ is a strictly decreasing function of $r$ if and only if $m \le a$, which is of limited interest, because $A_1$ can then be established in any case. We denote by $r_{\hat{\tilde{p}}^{0}_{+\mathrm{QLE}}}$ and $r_{\hat{\tilde{p}}^{1}_{+\mathrm{QLE}}}$ the recombination rates at which $\hat{\tilde{p}}_{+\mathrm{QLE}}$ equals 0 and 1, respectively. Analogously,  we use $r_{\hat{\tilde{q}}^{0}_{+\mathrm{QLE}}}$ and $r_{\hat{\tilde{q}}^{1}_{+\mathrm{QLE}}}$ for the recombination rates at which $\hat{\tilde{q}}_{+\mathrm{QLE}}$ equals 0 and 1, respectively. These critical recombination rates are found to be
\begin{subequations}
	\label{eq:rp0QLE}
		\begin{align}
			r_{\hat{\tilde{p}}^{0}_{+\mathrm{QLE}}} & = m\; \frac{m-b(1-2q_c) - \sqrt{R_3}}{2(a-m)}, \\
			r_{\hat{\tilde{p}}^{1}_{+\mathrm{QLE}}} & = \frac{1}{2} \left( b - m - 2bq_c + \sqrt{R_3}\right),
		\end{align}
\end{subequations}
and
\begin{subequations}
	\label{eq:rp0QLE}
		\begin{align}
			r_{\hat{\tilde{q}}^{0}_{+\mathrm{QLE}}} & = (m-a)\; \frac{b+m-\sqrt{R_3}}{2\sqrt{R_3}}, \\
			r_{\hat{\tilde{q}}^{1}_{+\mathrm{QLE}}} & = (a-m)\; \frac{b-m+\sqrt{R_3}}{2\sqrt{R_3}}.
		\end{align}
\end{subequations}
As shown in File \ref{prc:stochDiffApproxQLENB}, if $m<a$, $\tilde{E}_{+\mathrm{QLE}}$ exists in the admissible state space if and only if $r > \max\left(r_{\hat{\tilde{p}}^{1}_{+\mathrm{QLE}}}, r_{\hat{\tilde{q}}^{1}_{+\mathrm{QLE}}} \right)$. If $m \ge a$, $\tilde{E}_{+\mathrm{QLE}}$ exists in the admissible state space if and only if $\max\left(r_{\hat{\tilde{p}}^{1}_{+\mathrm{QLE}}}, r_{\hat{\tilde{q}}^{1}_{+\mathrm{QLE}}} \right) < r < r_{\hat{\tilde{p}}^{0}_{+\mathrm{QLE}}}$. At a first glance, it may seem surprising to obtain an upper limit on $r$. However, as is easily verified, $r_{\hat{\tilde{p}}^{0}_{+\mathrm{QLE}}}$ is also the critical value at which $\tilde{E}_{+\mathrm{QLE}}$ coincides with the QLE approximation of $\tilde{E}_{\bloc}$, which becomes asymptotically stable. Thus, with looser linkage, allele $A_1$ is lost.

\subsection{Diffusion approximation to sojourn and absorption times assuming quasi-linkage equilibrium}

Although some two-locus diffusion theory has been developped \citep{Ewens:1979hc,Ethier:1989qf,Ethier:1988bh,Ethier:1980dq}, explicit calculation of quantities of interest, such as absorption probabilities or times, seems difficult. Substantial progress can be made, though, by assuming that recombination is much stronger compared to selection (and migration). Then, linkage disequilibrium decays on a fast time scale, whereas allele frequencies evolve on a slow time scale under quasi-linkage equilibrium (QLE) \citep{Kimura:1965fk,Nagylaki:1999uq,Kirkpatrick:2002uq}. Here, we employ the QLE assumption to approximate the expected amount of time the focal allele $A_1$ spends in a certain range of allele frequencies (the sojourn times), as well as the expected time to extinction (the mean absorption time). We do so in detail for a monomorphic continent ($q_c = 0$) first. For a polymorphic continent ($0 < q_c < 1$), we will only give a brief outline and refer to File \ref{prc:stochDiffApproxQLENB} for details. Throughout, we closely follow \citet{Ewens:1979hc} in our application of diffusion theory.

We start from the continuous-time dynamics of the allele frequencies ($p,q$) and the linkage disequilibrium ($D$) in Eq.\ \eqref{eq:diffEqs}, setting $q_c = 0$ for a monomorphic continent. Given that recombination is strong compared to selection and migration, $D$ will be close to an equilibrium, so that $\dot{D} = dD/dt \approx 0$ may be assumed. Moreover, we assume that the frequency of the beneficial background allele $B_1$ is not affected by establishment of $A_1$. Specifically, $q = \qeqbc$ constant, where $\qeqbc = 1 - m/b$ is the frequency of $B_1$ at the one-locus migration--selection equilibrium in continuous time (Eq.\ \ref{eq:FreqB1OneLocCont}). Equation \eqref{eq:diffEqs} is therefore approximated by
\begin{subequations}
	\label{eq:diffEqsQLE}
	\begin{align}
		\dot{p} &= \frac{dp}{dt} = a p (1-p) - mp + bD \label{eq:diffEqsQLEa},\\
		\dot{q} &= \frac{dq}{dt} = 0 \label{eq:diffEqsQLEb},\\
		\dot{D} &= \frac{dD}{dt} = \left[a(1-2p)+b(1-2q)\right]D + m\left( p q - D\right) - rD = 0 \label{eq:diffEqsQLEc}.
	\end{align}
\end{subequations}
Solving Eq.\ \eqref{eq:diffEqsQLEc} for $D$, plugging the solution into Eq.\ \eqref{eq:diffEqsQLEa} and setting $q = \qeqbc$, we obtain a single differential equation in $p$:
\begin{equation}
	\label{eq:diffEqQLEa}
	\dot{p} = ap(1-p) - mp + \frac{m(b-m)}{b - m - a(1-2p) + r}\,p.
\end{equation}
In the limit of $r \rightarrow \infty$, we recover the one-locus migration-selecion dynamics for the continent--island model, $\dot{p} = ap(1-p)-mp$. 

We now consider the diffusion process obtained from the Wright--Fisher model \citep{Fisher:1930fk,Wright:1931rr}. More precisely, we measure time in units of $2N_e$ generations, where $N_e$ is the effective population size, and use $T$ for time on the diffusion scale. Further, we introduce the scaled selection coefficients $\alpha = 2N_e a$ and $\beta = 2N_e b$, the scaled recombination rate $\rho = 2 N_e r$, and the scaled migration rate $\mu = 2N_e m$. Equation \eqref{eq:diffEqQLEa} yields the infinitesimal mean
\begin{equation*}
	M(p)  =  \alpha p(1-p) - \mu p + \frac{\mu(\beta-\mu)}{\beta - \mu - \alpha(1-2p) + \rho}\,p
\end{equation*}
(cf.\ Eq.\ \ref{eq:Mp} in the main text). It expresses the mean change in $p$ per unit of time on the diffusion scale. The infinitesimal variance is
\begin{equation}
	\label{eq:Vp}
	V(p) = p(1-p)
\end{equation}
\cite[p.\ 159]{Karlin:1981fk}.

Later, we will need the ratio of $M(p)$ to $V(p)$, which is
\begin{equation}
	\label{eq:MpToVp}
	\frac{M(p)}{V(p)} = \alpha - \frac{\mu}{1-\rho} \left(1 - \frac{\beta - \mu}{\beta - \alpha (1-2p) - \mu + \rho}\right).
\end{equation}
We define the function $\psi(p)$ according to Eq.\ (4.16) in \cite{Ewens:1979hc} as
\begin{equation}
	\label{eq:psiDef}
	\psi(p) := \exp\left[-2\int_{0}^{p} \frac{M(z)}{V(z)} dz\right].
\end{equation}
Inserting Eq.\ \eqref{eq:MpToVp}, we find,
\begin{equation}
	\label{eq:psiQLE}
	\psi(p) = e^{-2 \alpha p} (1-p)^{-\frac{2\mu(\alpha+\rho)}{\alpha+\beta-\mu+\rho}} (\beta-\alpha-\mu+\rho)^{\frac{2\mu(\beta-\mu)}{\alpha+\beta-\mu+\rho}} \left[ \beta-(1-2p)\alpha-\mu+\rho \right]^{\frac{2\mu(\mu-\beta)}{\alpha+\beta-\mu+\rho}}.
\end{equation}
The derivation assumes that $(\alpha - \beta + \mu - \rho)/(\alpha p) < 0$ holds. Recalling from section 3 
that, for instability of the marginal one-locus equilibrium $\tilde{E}_\bloc$, it is required that $m < \tilde{m}^{\ast} = a\left(1 + \frac{b-a}{r} \right)$ and that then $a < \min(b,r)$, one can show that $(\alpha - \beta + \mu - \rho)/(\alpha p) < 0$ holds indeed (see File \ref{prc:stochDiffApproxQLENB}).

We now turn to the sojourn times as defined in Ewens (\citeyear{Ewens:1979hc}, pp.\ 141--144). We denote the initial frequency of the focal mutation $A_1$ by $p_0$ and introduce the function $t(p; p_0)$ to describe the sojourn-time density (STD). The interpretation of $t(p; p_0)$ is the following. The integral
\begin{equation*}
	\int_{p_1}^{p_2} t (p; p_0) dp
\end{equation*}
approximates the mean time in units of $2N_e$ generations allele $A_1$ spends at a frequency in the interval $(p_1, p_2)$, conditional on the initial frequency $p_0$. According to Eqs.\ (4.38) and (4.39) in \cite{Ewens:1979hc}, we define
\begin{equation}
	\label{eq:STDQLEDef}
	t(p; p_0) = \left\{ \begin{array}{ll}
		t_{1}(p; p_0) & \textrm{if $0 \le p \le p_0$},\\
		t_{2}(p; p_0) & \textrm{if $p_0 \le p \le 1$}.
	\end{array} \right.
\end{equation}
To make the assumption of quasi-linkage equilibrium explicit, we will add the subscript QLE to relevant quantities from now on. The densities $t_{i,\mathrm{QLE}}(p; p_0)$ are given by Eq.\ \eqref{eq:STDQLE} in the main text, with $\psi(y)$ as in Eq.\ \eqref{eq:psiDef}. 
The integral $\int_{0}^{x}\psi(y)dy$ cannot be found explicitly. However, because Eq.\ \eqref{eq:STD1QLE} takes the form $\toneQLE = 2 \psi(y)^{-1} (1-p)^{-1} p^{-1} \int_{0}^{p}  \psi(y) dy$ and $p^{-1} \int_{0}^{p} \psi(y) dy \rightarrow 1$ as $p \rightarrow 0$ (File \ref{prc:stochDiffApproxQLENB}), we approximate $\toneQLE$ by
\begin{equation}
	\label{eq:STD1QLEApprox}
	\tonetilQLE = \frac{2p}{V(p)\psi(p)}
\end{equation}
whenever $p$ is small. Recall from Eq.\ \eqref{eq:STDQLEDef} that $t_{1}(p; p_0)$ is needed only if $0 \le p \le p_0$. We are in general interested in a de-novo mutation, i.e.\ $p_0 = 1/(2N)$, with population size $N$ at least about 100. Hence, $p \le p_0$ automatically implies that $p$ is small whenever $\toneQLE$ is employed. The approximation in Eq.\ \eqref{eq:STD1QLEApprox} is therefore valid for our purpose.

Similarly, we may multiply $\ttwoQLE$ by $p_0$ and $1/p_0$ 
and write 
\begin{equation*}
	\ttwoQLE = 2p_0 \psi(y)^{-1} (1-p)^{-1} p^{-1} p_{0}^{-1} \int_{0}^{p_0}\psi(y)dy.
\end{equation*}
	Again, $p_0^{-1} \int_{0}^{p_0} \psi(y) dy \rightarrow 1$ as $p_{0} \rightarrow 0$ (File \ref{prc:stochDiffApproxQLENB}). We therefore approximate $\ttwoQLE$ by
\begin{equation}
	\label{eq:STD2QLEApprox}
	\ttwotilQLE = \frac{2p_0}{V(p)\psi(p)}
\end{equation}
whenever $p_0$ is small. In the following, we use a tilde ($\sim$) to denote the assumption of small $p_0$.

The expected time to extinction of allele $A_1$ in our model is identical to the mean absorption time, because extinction is the only absorbing state. For arbitrary initial frequency $p_0$, the approximate mean absorption time under the QLE approximation is obtained from the sojourn-time densities as shown in Eq.\ \eqref{eq:meanAbsTimeQLE} of the main text. Assuming small $p_0$, this simplifies to
\begin{equation}
	\label{eq:meanAbsTimeQLEApprox}
	\tbartilQLE = \int_{0}^{p_0} \tonetilQLE dp + \int_{p_0}^{1} \ttwotilQLE dp.
\end{equation}
In both cases, the integrals must be computed numerically. As a further approximation for very small $p_0$, one may omit the first integral on the right-hand side of Eq.\ \eqref{eq:meanAbsTimeQLEApprox}, as its contribution becomes negligible when $p_0 \rightarrow 0$.

The predictions for the sojourn-time densities (STDs) and the mean absorption time derived above are accurate if the QLE assumption holds (Figures \ref{fig:meanabstimeQLEOLM}, \ref{fig:STDCompSimMonomCont}  and \ref{fig:STDCompSimPolymCont}). However, the analytical expressions for the STDs in Eqs.\ \eqref{eq:STD1QLEApprox} and \eqref{eq:STD2QLEApprox} are not very informative once we plug in explicit formulae for $V(p)$ and $\psi(p)$ (see File \ref{prc:stochDiffApproxQLENB}). In the following, we will gain more insight by making an additional assumption.

We assume that recombination is much stronger than selection and migration, and expand $M(p)$ from Eq.\ \eqref{eq:Mp} as a function of $\rho^{-1}$ to first order into a Taylor series. This yields
\begin{equation*}
	M(p) \approx M_{\rho\gg0}(p) = \alpha p(1-p) - \mu p + \frac{\mu(\beta - \mu)}{\rho} p
\end{equation*}
and hence Eq.\ \eqref{eq:MpRhoLarge} in the main text. The infinitesimal variance $V(p)$ from Eq.\ \eqref{eq:Vp} remains unchanged, but the ratio of $M(p)$ to $V(p)$ simplifies to
\begin{equation}
	\label{eq:MpToVpRhoLarge}
	\frac{M_{\rho\gg0}(p)}{V(p)} = \alpha - \frac{\mu}{1-\rho}\left(1 - \frac{\beta-\mu}{\rho} \right).
\end{equation}
Insertion into Eq.\ \eqref{eq:psiDef}, integration and some algebra yields
\begin{equation}
	\label{eq:psiQLERhoLarge}
	\psi_{\rho \gg 0}(p) = e^{-2\alpha p} (1-p)^{-\frac{2\mu(\mu - \beta + \rho)}{\rho}}.
\end{equation}
The sojourn-time density (STD) is then given by
\begin{subequations}
	\label{eq:STDQLERhoLarge}
		\begin{align}
			\toneQLErho & = \frac{2}{V(p)\psi_{\rho \gg 0}(p)} \int_{0}^{p}\psi_{\rho \gg 0}(y)dy, \\
			\ttwoQLErho & = \frac{2}{V(p)\psi_{\rho \gg 0}(p)} \int_{0}^{p_0}\psi_{\rho \gg 0}(y)dy.
		\end{align}
\end{subequations}
As before, $x^{-1}\int_{0}^{x} \psi_{\rho \gg 0}(p) dp \rightarrow 1$ as $x \rightarrow 0$. Arguments analogous to those leading to Eqs.\ \eqref{eq:STD1QLEApprox} and \eqref{eq:STD2QLEApprox} show that, for a small initial frequency $p_0$, the STD is approximated by
	\begin{align*}
		\tilde{t}_{1,\mathrm{QLE},\rho \gg 0}(p; p_0) &= \frac{2p}{V(p)\psi_{\rho \gg 0}(p)} = 2 e^{2p\alpha}(1-p)^{\frac{2\mu(\mu - \beta + \rho)}{\rho} - 1},\\
		\tilde{t}_{2,\mathrm{QLE},\rho \gg 0}(p; p_0) &= \frac{2p_0}{V(p)\psi_{\rho \gg 0}(p)} = 2p_0 e^{2p\alpha} p^{-1}(1-p)^{\frac{2\mu(\mu - \beta + \rho)}{\rho} - 1}
	\end{align*}
(cf.\ Eq.\ \ref{eq:STDQLEApproxRhoLarge} of the main text). For details, we refer to File \ref{prc:stochDiffApproxQLENB}. The mean absorption time is again obtained as
\begin{equation}
	\label{eq:meanAbsTimeQLERhoLarge}
	\tbarQLErho = \int_{0}^{p_0} \toneQLErho dp + \int_{p_0}^{1} \ttwoQLErho dp
\end{equation}
using the STD in Eq.\ \eqref{eq:STDQLERhoLarge} for arbitrary initial frequency $p_0$, or as 
\begin{equation}
	\label{eq:meanAbsTimeQLEApproxRhoLarge}
	\tbartilQLErho = \int_{0}^{p_0} \tonetilQLErho dp + \int_{p_0}^{1} \ttwotilQLErho dp
\end{equation}
using the STD in Eq.\ \eqref{eq:STDQLEApproxRhoLarge} for small $p_0$. Figure \ref{fig:stdQLE} compares the various approximations to the STD derived under the QLE assumption for a monomorphic continent ($q_c$). It also includes a comparison to the STD for a one-locus model (OLM), which is specified by
		\begin{align*}
			\tonetilOLM &= 2 e^{2 p \alpha}(1-p)^{2\mu - 1} & \textrm{if}\ 0 \le p \le p_0,\\
			\ttwotilOLM &= 2 p_0 e^{2 p \alpha}p^{-1}(1-p)^{2\mu - 1} & \textrm{if}\ p_0 \le p \le 1
		\end{align*}
for small $p_0$ (cf.\ Eq.\ \ref{eq:STDOLM} in the main text).

A comparison of the STD given in Eq.\ \eqref{eq:STDQLEApproxRhoLarge} for two loci with large $\rho$ and small $p_0$ to the corresponding one-locus STD in Eq.\ \eqref{eq:STDOLM} is interesting. The difference is that $\mu$ in the one-locus model is replaced by $\mu (\mu - \beta + \rho)/ \rho$ to obtain the formulae for the two-locus model. Hence, for strong recombination, we may define an effective scaled migration rate
\begin{equation*}
	\mu_e = \mu \frac{\mu + \rho - \beta}{\rho}  = \mu - \frac{\beta \mu}{\rho} + \frac{\mu^2}{\rho} \approx \mu\left(1-\frac{\beta}{\rho}\right),
\end{equation*}
where the approximation holds for $\mu \ll \min(\beta, \rho)$. The interpretation is that $\mu_e$ denotes the scaled migration rate in a one-locus migration--selection model for which allele $A_1$ has the same sojourn-time properties as if it arose in a two-locus model with scaled migration rate $\mu$ and linkage to a previously established polymorphism that decays at a scaled recombination rate $\rho$. Transforming back from the diffusion to the natural scale, we obtain the invasion-effective migration rates $m_e$ and $\tilde{m}_e$ given in Eqs.\ \eqref{eq:mEffQLE} and \eqref{eq:mEffQLEweakMig} of the main text, respectively (see also Figure \ref{fig:mEffSojourn}A).

We now turn to the case of a polymorhpic continent ($0 < q_c < 1$). Derivations are analogous to those shown above for the monomorphic continent, but more cumbersome. We therefore give only a rough summary here and refer to File \ref{prc:stochDiffApproxQLENB} for details.

The mean change in $p$ per unit of time on the diffusion scale and under the assumption of quasi-linkage equilibrium (QLE) is
\begin{equation}
	\label{eq:MpPolymCont}
	M(p): = \frac{dp}{dT} =  \alpha p(1-p) - \mu p - \frac{\mu\left(\beta-\mu-2\beta q_c + \sqrt{R_5}\right)}{2\left[\alpha\left(1-2p\right) - \rho - \sqrt{R_5}\right]}\,p,
\end{equation}
where $R_5 = (\beta-\mu)^2 + 4 \beta \mu q_c > 0$. 

Equation \eqref{eq:MpPolymCont} can be used to numerically compute the sojourn-time densities (STDs) and the mean absorption time analogous to Eqs.\ \eqref{eq:STDQLE} and \eqref{eq:meanAbsTimeQLE} (see File \ref{prc:stochDiffApproxQLENB}). To obtain informative analytical results for the STDs, however, it is necessary to assume that recombination is strong compared to selection and migration, i.e.\ $\rho \gg \min(b,m)$. Then, the infinitesimal mean is approximated by
\begin{equation}
	\label{eq:MpPolymContApprox}
	M(p) \approx M_{\rho \gg 0}(p) =  \alpha p(1-p) - \mu p + \frac{\mu\left(\beta - \mu - 2\beta q_c + \sqrt{R_5}\right)}{2\rho}\,p
\end{equation}
The infinitesimal variance is the same as for a monomorphic continent, $V(p) = p(1-p)$. Inserting $M_{\rho \gg 0}(p)$ from Eq.\ \eqref{eq:MpPolymContApprox} and $V(p)$ into the definition of $\psi(p)$ in Eq.\ \eqref{eq:psiDef}, we obtain
\begin{equation}
	\label{eq:psiQLERhoLargePolymCont}
	\psi_{\rho \gg 0}(p) = e^{-2\alpha p}(1-p)^{\frac{\mu\left(\beta-\mu-2\rho-2\beta q_c + \sqrt{R_5}\right)}{\rho}}.
\end{equation}
The STDs $\toneQLErho$ and $\ttwoQLErho$ are found by insertion of $\psi_{\rho \gg 0}(p)$ from Eq.\ \eqref{eq:psiQLERhoLargePolymCont} into Eq.\ \eqref{eq:STDQLERhoLarge}. Exploiting the fact that $x^{-1}\int_{0}^{x}\psi_{\rho \gg 0}(p)dp$ converges to 1 as $x$ approaches 0, the STDs can be approximated by
\begin{subequations}
	\label{eq:STDQLEApproxRhoLargePolymCont}
	\begin{align}
		\tilde{t}_{1,\mathrm{QLE},\rho \gg 0}(p; p_0) &= 2 e^{2p\alpha}(1-p)^{\frac{\mu\left(\mu - \beta + 2\beta q_c + 2\rho - \sqrt{R_5}\right)}{\rho} - 1}, \label{eq:STD1QLEApproxRhoLargePolymCont}\\
		\tilde{t}_{2,\mathrm{QLE},\rho \gg 0}(p; p_0) &= 2p_0 e^{2p\alpha} p^{-1}(1-p)^{\frac{\mu\left(\mu - \beta + 2\beta q_c + 2\rho - \sqrt{R_5}\right)}{\rho} - 1} \label{eq:STD2QLEApproxRhoLargePolymCont}
	\end{align}
\end{subequations}
This approximation is valid if the initial frequency $p_0$ is small and $\rho$ is large.
The mean absorption time for arbitrary $p_0$ is found according to Eq.\ \eqref{eq:meanAbsTimeQLERhoLarge}. For small $p_0$, it is given by Eq.\ \eqref{eq:meanAbsTimeQLEApproxRhoLarge}, with $\tilde{t}_{i,\mathrm{QLE},\rho \gg 0}(p;p_0)$ from Eq.\ \eqref{eq:STDQLEApproxRhoLargePolymCont}.
%
%
%


\subsection{Effective migration rate at a neutral site linked to two migration--selection polymorphisms}

We derive the effective migration rate experienced by a neutral locus ($\cloc$) linked to two loci ($\aloc$ and $\bloc$) that are maintained polymorphic at migration--selection balance. Locus $\cloc$ has two alleles $C_1$ and $C_2$, which are assumed to segregate at constant frequencies $n_c$ and $1 - n_c$ on the continent. The frequency of $C_1$ on the island at time $t$ is denoted by $n(t)$. 
Loci $\aloc$ and $\bloc$ are as above, with alleles $A_1$ and $B_1$ segregating at frequencies $p$ and $q$ on the island, respectively. Without loss of generality, we assume that $\aloc$ is located to the left of $\bloc$ on the chromosome. We denote by $r_{XY}$ the recombination rate between loci $X$ and $Y$, where $r_{XY} = r_{YX}$. Because we consider a continuous-time model here, we may assume that the recombination rate increases additively with distance on the chromosome. For simplicity, we restrict the analysis to the case of a monomorphic continent, i.e.\ alleles $A_2$ and $B_2$ are fixed on the continent.

Following \citet{Buerger:2011uq}, we define the effective migration rate as the asymptotic rate of convergence of $n(t)$ to the fully-polymorphic three-locus equilibrium. This rate of convergence is defined by the leading eigenvalue $\lambda_N$ of the Jacobian of the system that describes the evolution of the frequency of $C_1$ and the linkage disequilibria associated with locus $\cloc$. Specifically, we define the effective migration rate as $m_e = -\lambda_N$ \cite[cf.][]{Kobayashi:2008fk}.

We start by assuming that the neutral locus is located between the two selected ones (configuration \configM). We denote by $D_{\aloc \bloc} = D$, $D_{\aloc \cloc}$ and $D_{\cloc \bloc}$ the linkage disequilibria between the indicated loci, and by $D_{\aloc \cloc \bloc} = y_1 - pqn - p D_{\cloc \bloc} - q D_{\aloc \cloc} - n D_{\aloc \bloc}$ the three-way linkage disequilibrium, where $y_1$ is the frequency of gamete $A_1 C_1 B_1$. The changes due to selection, migration and recombination in $p$, $q$, and $D_{\aloc \bloc}$ are given by Eq.\ \eqref{eq:diffEqs} of this text, with $r$ replaced by $r_{\aloc \bloc}$. The frequency of $C_1$ evolves according to
\begin{equation}
	\label{eq:diffEqNeutr}
	\dot{n} = m(n_c - n) + a D_{\aloc \cloc} + b D_{\cloc \bloc}
\end{equation}
and the differential equations for the linkage disequilibria associated with locus $\cloc$ are
\begin{subequations}
	\label{eq:LD3LocConfigM}
	\begin{align}
		\dot{D}_{\aloc \cloc} & = a (1 - 2 p) D_{\aloc \cloc} + b D_{\aloc \cloc \bloc} - m D_{\aloc \cloc} - m p (n_c - n) - r_{\aloc \cloc} D_{\aloc \cloc}, \\
		\dot{D}_{\cloc \bloc} & = a D_{\aloc \cloc \bloc} + b (1-2q) D_{\cloc \bloc} - m D_{\cloc \bloc} - m q (n_c - n) - r_{\cloc \bloc} D_{\cloc \bloc}, \\
		\dot{D}_{\aloc \cloc \bloc} & = \left[ a(1-2p) + b(1-2q)\right] D_{\aloc \cloc \bloc} - 2 (a D_{\aloc \cloc} + b D_{\cloc \bloc}) D_{\aloc \bloc} + m (p D_{\cloc \bloc} + q D_{\aloc \cloc} - D_{\aloc \cloc \bloc}) \notag \\
		& + m (pq - D_{\aloc \bloc}) (n_c - n) - r_{\aloc \bloc} D_{\aloc \cloc \bloc}
	\end{align}
\end{subequations}
(we use $\dot{x}$ for the differential of $x$ with respect to time, $dx/dt$). We refer to File \ref{prc:determEffMigRate} for the derivation. Recall that $r_{\aloc \bloc} = r_{\aloc \cloc} + r_{\cloc \bloc}$.
This system has an asymptotically stable equilibrium such that the selected loci are at the equilibrium $\tilde{E}_{+}$ (Eq.\ 3.15 in \citealt{Buerger:2011uq}), and $n = n_c$ and $D_{\aloc \cloc} = D_{\cloc \bloc} = D_{\aloc \cloc \bloc} = 0$ hold. The Jacobian at this equilibrium has the block structure
\begin{equation*}
	\mathbf{J} =
		\begin{pmatrix}
			\mathbf{J}_{S} & 0\\
			0 & \mathbf{J}_{N}
		\end{pmatrix},
	\label{eq:JacobianThreeLoc}
\end{equation*}
where $\mathbf{J}_{S}$ is the Jacobian approximating convergence of $(p,q,D_{\aloc \bloc})$ to $\tilde{E}_{+}$, and $\mathbf{J}_{N}$ is the Jacobian approximating convergence of $(n, D_{\aloc \cloc}, D_{\cloc \bloc}, D_{\aloc \cloc \bloc})$ to $(n_c, 0, 0, 0)$. In the limit of weak migration, i.e.\ $m \ll (a, b, r)$, the latter is given by
\begin{equation}
	\label{eq:JacobianSubNeutrConfigM}
		\mathbf{J}_{N}^{\aloc \cloc \bloc} =
		\begin{pmatrix}
			-m & a & b & 0\\
			m & -a -r_{\aloc \cloc} + \frac{m(a - b + r_{\aloc \bloc})}{a + b + r_{\aloc \bloc}} & 0 & b \\
			m & 0 & -b -r_{\cloc \bloc} + \frac{m(b - a + r_{\aloc \bloc})}{a + b + r_{\aloc \bloc}} & a\\
			-m & \frac{m(b - a + r_{\aloc \bloc})}{a + b + r_{\aloc \bloc}} & \frac{m(a - b + r_{\aloc \bloc})}{a + b + r_{\aloc \bloc}} & -a -b -r_{\aloc \bloc} +  \frac{m(a + b +3r_{\aloc \bloc})}{a + b + r_{\aloc \bloc}}
		\end{pmatrix}.
\end{equation}
As shown previously \citep{Buerger:2011uq}, to first order in $m$, the leading eigenvalue of $\mathbf{J}_{N}^{\aloc \cloc \bloc}$ is given by 
\begin{equation}
	\label{eq:leadEvalJNConfigM}
	\lambda_{N}^{\aloc \cloc \bloc} = m \frac{r_{\aloc \cloc} r_{\cloc \bloc}}{\left(a + r_{\aloc \cloc} \right) \left(b + r_{\cloc\bloc} \right)},
\end{equation}
and hence the approximation of the effective migration rate in Eq.\ \eqref{eq:effMigRateNeutrM} in the main text is obtained (see File \ref{prc:determEffMigRate} for details). We note that Eqs.\ \eqref{eq:diffEqNeutr}, \eqref{eq:LD3LocConfigM} and \eqref{eq:JacobianSubNeutrConfigM} correct errors in Eqs.\ (4.25), (4.26) and (4.28) of \citet{Buerger:2011uq}, respectively. The main results by \citet{Buerger:2011uq} were not affected, though.  

 If the neutral locus is located to the right of the two selected ones (configuration \configR), Eqs. \eqref{eq:diffEqNeutr} and \eqref{eq:LD3LocConfigM} remain the same (recall that $r_{XY} = r_{YX}$ and in this case $r_{\aloc \cloc} = r_{\aloc \bloc} + r_{\bloc \cloc}$). In Eq.\ \eqref{eq:diffEqsc}, $r$ must be replaced by $r_{\aloc \cloc}$. Then, the Jacobian $\mathbf{J}_{N}^{\aloc \bloc \cloc}$ approximating convergence of $(n, D_{\aloc \cloc}, D_{\bloc \cloc} = D_{\cloc \bloc}, D_{\aloc \bloc \cloc} = D_{\aloc \cloc \bloc})$ to $(n_c, 0, 0, 0)$ in the limit of weak migration is equal to $\mathbf{J}_{N}^{\aloc \cloc \bloc}$ with the last entry of the last row replaced by $-a -b -r_{\aloc \cloc} +  \frac{m(a + b +3r_{\aloc \bloc})}{a + b + r_{\aloc \bloc}}$.
%
To first order in $m$, the leading eigenvalue of $\mathbf{J}_{N}^{\aloc \bloc \cloc}$ is
 \begin{equation}
	\label{eq:leadEvalJNConfigR}
	\lambda_{N}^{\aloc \bloc \cloc} = m \frac{r_{\bloc\cloc} \left(b + r_{\aloc\cloc}\right)}{\left(b + r_{\bloc\cloc}\right)\left(a + b + r_{\aloc\cloc}\right)},
\end{equation}
and hence Eq.\ \eqref{eq:effMigRateNeutrR} in the main text. Details are given in File \ref{prc:determEffMigRate}.

Last, the leading eigenvalue for configuration \configL\ follows directly by symmetry,
 \begin{equation}
	\label{eq:leadEvalJNConfigL}
	\lambda_{N}^{\cloc \aloc \bloc} = m \frac{r_{\cloc\aloc} \left(a + r_{\cloc\bloc}\right)}{\left(a + r_{\cloc\aloc}\right)\left(a + b + r_{\cloc\bloc}\right)},
\end{equation}
and hence Eq.\ \eqref{eq:effMigRateNeutrL} in the main text.
 
Recall that the Jacobian matrices $\mathbf{J}_{N}^{\aloc \cloc \bloc}$ and $\mathbf{J}_{N}^{\aloc \bloc \cloc}$ hold under the assumption of weak migration. In File \ref{prc:determEffMigRate}, we derive analogous matrices under the assumption of weak recombination, i.e.\ $r \ll (a, b, m)$. These are too complicated to be shown here, but importantly, to first order in $m$, their leading eigenvalues are identical to Eqs.\ \eqref{eq:leadEvalJNConfigM} and \eqref{eq:leadEvalJNConfigR}, respectively. By symmetry, this also applies to the configuration \configL. Therefore, the approximate effective migration rates in Eq.\ \eqref{eq:effMigRatesNeutr} are valid also for tight linkage between the neutral locus and the selected loci.

To test the robustness of our results agaist violation of the assumption of weak migration, we numerically computed exact effective migration rates. In most cases, the deviation is very small; compare dashed to solid curves in Figures \ref{fig:statAllDistNeutr} and \ref{fig:meanabstimeNeutr}, and dots to curves in Figure \ref{fig:coalRateAndNeNeutr}.

\clearpage
\section*{Supporting Information: Procedures}
\vspace{0.5cm}
Files \ref{prc:determDiscrNB}--\ref{prc:stochDiffNeutrVar} are available upon request from the corresponding author (saeschbacher@mac.com).
\singlespacing
\begin{file}[!ht]
\topcaption{
{}Deterministic analysis of a diploid two-locus continent$-$island model in discrete time. (Mathematica Notebook)
}
\label{prc:determDiscrNB}
\end{file}

\begin{file}[!ht]
\topcaption{
{}Branching-process approximation of the invasion probability of a weakly beneficial mutation linked to an established polymorphism at migration$-$selection balance. (Mathematica Notebook)
}
\label{prc:stochDiscrNB}
\end{file}

\begin{file}[!ht]
\topcaption{
{}Comparison of the Jacobian of the marginal one-locus migration$-$selection equilibrium ($E_\bloc$) to the mean matrix of the corresponding branching process. (Mathematica Notebook)
}
\label{prc:stochCompareJacobianVsMeanMatrixNB}
\end{file}

\begin{file}[!ht]
\topcaption{
{}Analytical approximation of the invasion probability for a slightly supercritical branching process. (Mathematica Notebook)
}
\label{prc:stochDiscrSlightlySupercritBPNB}
\end{file}

\begin{file}[!ht]
\topcaption{
{}Derivative of the weighted mean invasion probability $\bar{\pi}$ at recombination rate $r=0$. (Mathematica Notebook)
}
\label{prc:stochDiscrDerivativeZeroRecombRateNB}
\end{file}


\begin{file}[!ht]
\topcaption{
{}Diffusion approximation of sojourn and absorption times assuming quasi-linkage disequilibrium. (Mathematica Notebook)
}
\label{prc:stochDiffApproxQLENB}
\end{file}

\begin{file}[!ht]
\topcaption{
{}The effective migration rate experienced by a neutral site linked to two loci at migration$-$selection balance. (Mathematica Notebook)
}
\label{prc:determEffMigRate}
\end{file}

\begin{file}[!ht]
\topcaption{
{}The effect on neutral variation of migration and selection at two linked sites. (Mathematica Notebook)
}
\label{prc:stochDiffNeutrVar}
\end{file}



\end{document}